\documentclass[preprint,showpacs]{revtex4}
\usepackage{graphicx}
\usepackage{amssymb}
\newcommand{\be}{\begin{eqnarray}}
\newcommand{\ee}{\end{eqnarray}}
\newcommand{\eq}{\begin{eqnarray}}
\newcommand{\en}{\end{eqnarray}}

\newcommand{\bfk}{{\bf k}_{\perp}}

\newcommand{\bfb}{{\bf b}_{\perp}}

\begin{document}

\title{Chiral-odd generalized parton distributions for proton in a light-front quark-diquark model}% in AdS/QCD }

\author{Dipankar Chakrabarti}
%\email{dipankar@iitk.ac.in}
\author{Chandan Mondal}
%\email{mchandan@iitk.ac.in}
\affiliation{ Department of Physics, Indian Institute of Technology Kanpur, Kanpur 208016, India}

\date{\today}

\begin{abstract}

We present a study of the chiral-odd generalized parton distributions (GPDs) for $u$ and $d$ quarks in a proton using the light front wave functions (LFWFs) of the scalar quark-diquark model for nucleon constructed from the soft-wall AdS/QCD correspondence. We obtain the GPDs  in terms of overlaps of the LFWFs. Numerical results for  chiral-odd GPDs in momentum as well as transverse position (impact) spaces considering both zero and nonzero skewness($\zeta$) are presented. For nonzero skewness,  the GPDs  are also evaluated in longitudinal position space. 

\end{abstract}

\maketitle

\section{Introduction}

Generalized parton distributions (GPDs) encode the informations about the three dimensional spatial structure of the proton as well as the spin and orbital angular momentum of the constituents.
% provide us a wealth of information about the nucleon structure and spin (see \cite{rev} for reviews on GPDs). 
The GPDs(see \cite{rev} for reviews on GPDs)  are off-forward matrix elements and appear in the exclusive processes like Deeply Virtual Compton Scattering (DVCS) or vector meson productions.  The GPDs being functions of three variables namely, longitudinal momentum faction $x$ of the parton, square of the total momentum transferred $t$  and  the longitudinal momentum transferred $\zeta$ so called skewness in the process contains more informations than the ordinary parton distribution functions(PDFs).
 %The GPDs contain interesting informations about the spin and orbital angular momentum of the constituents, as well as the spatial structure, of the nucleons. 
 The first moments of GPDs give the form factors accessible in exclusive processes whereas they reduce to PDFs in the forward limit. 
 At leading twist,  we can define three generalized distributions in parallel to 
  three PDFs, namely, the unpolarized,
 helicity, and transversity distributions.  Similar to transversity distribution, the generalized  transversity distribution  $F_T$ is also chiral-odd. In the most general way, $F_T$ is parametrized in terms of four chiral-odd GPDs, namely $H_T$ , $\widetilde{H}_T$ , $E_T$ , and $\widetilde{E}_T$~\cite{diehl01,Pasquini1,Diehl05,Burk3}. The chiral-odd GPDs give information on the correlation between the spin and angular momentum of quarks inside the proton.
 At zero skewness, by performing a Fourier transform (FT) of the GPDs with respect to the momentum transfer in the transverse direction $\Delta_{\perp}$, one obtains the impact parameter dependent parton distributions, which provide us the picture that how the partons of a given longitudinal momentum fraction ($x$) are distributed in impact parameter ($b_{\perp}$) or transverse position space. Unlike the GPDs themselves, impact parameter dependent parton distributions have probabilistic interpretation and satisfy the positivity condition~\cite{burk1,burk2,Burk3}. In the $t\to 0$ limit, the second 
moment of the GPDs are related to the angular momentum contribution to the nucleon by the quark or gluon~\cite{ji}. The impact parameter dependent PDFs are transversely distorted when one considers transversely polarized nucleons. The transverse distortion can also be connected with Ji's angular momentum relation. An interesting interpretation of Ji's angular momentum sum rule~\cite{ji} for transversely polarized state was obtained in terms of the impact parameter dependent PDFs in~\cite{Burk3}. For the unpolarized quark, transverse distortion arises due to the chiral-even GPD $E$ which is related to the anomalous magnetic moment of the quarks. As far as the transverse distortion of transversely polarized quark distributions is concerned, the linear combination of chiral-odd GPDs ($2\widetilde{H}_T+E_T$) plays a  role similar to the GPD $E$ as for the unpolarized quark distributions. In the forward limit, a relation between the transverse total angular momentum of the quarks and a combination of second 
moments 
of $H_T$, $\widetilde{H}_T$ and $E_T$ has been proposed in~\cite{Burk3}, in analogy with Ji's relation.  $\widetilde{E}_T$ being an odd function of  $\zeta$, does not contribute at $\zeta=0$. For nonzero skewness one can also represent the GPDs in the longitudinal position space by taking FT of the GPDs with respect to $\zeta$~\cite{CMM1,CMM2,Dahiya07,BDHAV,chandan,Kumar1}.

Unlike the chiral-even GPDs, it is very difficult to measure chiral-odd GPDs. In a  very recent  COMPASS experiment \cite{Adolph},  exclusive production of  $\rho^0$ mesons by scattering muons off transversely polarized proton was measured. The target spin asymmetries measured in the experiment agree well with GPD-based model calculations which indicate the first experimental evidence  of chiral-odd GPDs, especially the transversity GPD $H_T$. There has been  proposals to get access to the chiral-odd GPDs through diffractive double meson production~\cite{Yu,Enberg}. 
The role of transversity GPDs in leptoproduction of vector mesons~\cite{Goloskokov1} as well as in hard exclusive electroproduction of pseudoscalar mesons~\cite{Goloskokov2} have been investigated within the framework of the handbag approach. A simple model for the dominant transversity GPD $H_T$ based on the concept of double distribution has been proposed and has been used to estimate the unpolarized differential cross section for this process in the kinematics of the Jlab and COMPASS experiments in~\cite{Beiyad}. The chiral-odd GPDs in a constituent quark model have been studied for nonzero skewness using the overlap representation in terms of light-front wave functions (LFWFs) in~\cite{Pasquini1}. The general properties of the chiral-odd GPDs in a QED 
model have been investigated in both momentum and transverse position as well as longitudinal position spaces~\cite{CMM1}; the impact parameter representation of the GPDs have been studied in a QED model of a dressed electron~\cite{Dahiya07} and  in a quark-diquark model~\cite{kumar} for $\zeta=0$. The Mellin moments of the transverse GPDs have been evaluated on lattice~\cite{gock,Hagler:2009,Hagler:2004,Bratt:2010}. 

There have been numerous  attempts to gain insight into the  hadron structure by studying QCD inspired models as nonperturbative properties of hadrons are always very difficult to  evaluate from QCD first principle.% In the quark-diquark model, a nucleon is considered to be a bound state of a single quark and a scalar or vector diquark state. The quark-diquark model is proven to reproduce many interesting properties of nucleons and has been extensively used to investigate the proton structure.
 In this work, we consider  a phenomenological light front quark-diquark model recently proposed by Gutsche et. al~\cite{Gut} where
 the LFWFs  are modeled by the wave functions obtained from a soft-wall model in light front AdS/QCD correspondence~\cite{BT1,BT}. 
%The LFWFs are derived by matching the electromagnetic form factors of hadrons in the light front QCD and soft-wall model of AdS/QCD correspondence. 
This model is consistent with Drell-Yan-West relation which relates the high $Q^2$ behavior of the nucleon form factors and  the large $x$ behavior of the structure functions. 
% In this work we investigate the chiral-odd GPDs in the AdS/QCD inspried quark-diquark model~\cite{Gut} in the momentum space for both zero and nonzeo $\zeta$. We also study them in transverse impact parameter as well as longitudinal position spaces. 

The paper is organized in the following way. In Section \ref{model}, a brief introductions about the nucleon LFWFs of quark-diquark model has been given. We present the overlap formalism of the chiral-odd GPDs and show the results for proton GPDs of $u$ and $d$ quarks in momentum space in Section \ref{chiral_gpds}. The GPDs in the transverse as well as the longitudinal impact parameter space are presented in Sections \ref{chiral_impact} and \ref{chiral_longi_impact}. Finally we summarize all the results in Section \ref{summary}.

\section {Light-front quark-diquark model for the nucleon}\label{model}
%In quark-scalar diquark model, the three valence quarks of nucleon are considered as an effectively composite system composed of a fermion and a spectator bound state of diquark based on one loop quantum fluctuations. The spectator with spin-0 and spin-1 are called scalar and vector diquark model respectively. Therefore the nucleon state can be represented as 2-particle Fock-state expansion. In this article, we consider only the scalar diquark model.
Here, we consider the quark-diquark model with a scalar  diquark.
The 2-particle Fock-state expansion for $J^z = + \frac{1}{2}$ and   $J^z = - \frac{1}{2}$  are then written as
 \be
  |P;+\rangle %& = &\psi^{+(s)}_{2-Particle} (P^+,\textbf{0}_{\perp})\nonumber \\
& =& \sum_q \int \frac{dx~ d^2\textbf{k}_{\perp}}{2(2\pi)^3\sqrt{x(1-x)}} \bigg[ \psi^{+}_{+q}(x,\textbf{k}_{\perp})|+\frac{1}{2},0; xP^+,\textbf{k}_{\perp}\rangle \nonumber \\
 &+& \psi^{+}_{-q}(x,\textbf{k}_{\perp})|-\frac{1}{2},0; xP^+,\textbf{k}_{\perp}\rangle\bigg],\\
 % \ee
 % \be
  |P;-\rangle %& = &\psi^{-(s)}_{2-Particle} (P^+,\textbf{0}_T)\nonumber \\
& =& \sum_q \int \frac{dx~d^2\textbf{k}_{\perp}}{2(2\pi)^3\sqrt{x(1-x)}} \bigg[\psi^{-}_{+q}(x,\textbf{k}_{\perp})|+\frac{1}{2},0; xP^+,\textbf{k}_{\perp}\rangle \nonumber \\
 &+& \psi^{-}_{-q}(x,\textbf{k}_{\perp})|-\frac{1}{2},0; xP^+,\textbf{k}_{\perp}\rangle\bigg],
  \ee
 where the $|\lambda_q,\lambda_s ; xP^+, \textbf{k}_{\perp} \rangle $ represents a two particle state with a quark spin $\lambda_q = \pm $, longitudinal momentum $xP^+$ and a spectator of spin $\lambda_s=0$ (scalar diquark). The states are   normalize as: 
\be
\langle \lambda'_q,\lambda'_s; x'P^+,\textbf{k}'_{\perp}|\lambda_q \lambda_s;xP^+,\textbf{k}_{\perp}\rangle = \prod^2_{i=1} 16\pi^3 p^+_i \delta(p'^+_i-p^+_i)\delta^2(\textbf{k}'_{\perp i}-\textbf{k}_{\perp i}) \delta_{\lambda'_i \lambda_i},
\ee 
and $\psi^{\lambda_N}_{\lambda_q q}$ are the light-front wave functions with nucleon helicities $\lambda_N = \pm $ and quark helicities $\lambda_q = \pm $. We adopt the generic ansatz for the quark-diquark model of the valence Fock state of the nucleon LFWFs at an initial scale $\mu_0=313$~MeV as proposed in \cite{Gut} : 
\be\label{wf}
\psi^{+}_{+q}(x,\textbf{k}_{\perp})&=&\varphi^{(1)}_q(x,\textbf{k}_{\perp}),\nonumber\\
\psi^{+}_{-q}(x,\textbf{k}_{\perp})&=&-\frac{p^1+ip^2}{xM}\varphi^{(2)}_q(x,\textbf{k}_{\perp}),\nonumber\\
\psi^{-}_{+q}(x,\textbf{k}_{\perp})&=&\frac{p^1-ip^2}{xM}\varphi^{(2)}_q(x,\textbf{k}_{\perp}),\\
\psi^{-}_{-q}(x,\textbf{k}_{\perp})&=& \varphi^{(1)}_q(x,\textbf{k}_{\perp}),\nonumber
\ee
where $\varphi_q^{(1)}(x,\bfk) $ and $\varphi_q^{(2)}(x,\bfk) $ are the wave functions predicted by soft-wall AdS/QCD\cite{BT}
\be
\varphi_q^{(i)}(x,\bfk)=N_q^{(i)}\frac{4\pi}{\kappa}\sqrt{\frac{\log(1/x)}{1-x}}x^{a_q^{(i)}}(1-x)^{b_q^{(i)}}\exp\bigg[-\frac{\bfk^2}{2\kappa^2}\frac{\log(1/x)}{(1-x)^2}\bigg],
\ee
%%%%%%%%%%%%%%%%%%%%%%%%%%%%%%%%%%%%%%%%%%%%%%%%%%%%%%%%%%%%%%%%%%%%%%
%\begin{table}[ht]
%\centering % used for centering table 
%\begin{tabular}{| p{2cm} | p{1.7cm} | p{1.7cm} | p{1.7cm} | p{1.7cm} | p{1.7cm} | p{1.7cm} |}
 %   \hline
  %   parameters&  $a^{(1)}$ & $a^{(2)}$ & $b^{(1)}$ & $b^{(2)}$ & $N^{(1)}$ & $N^{(2)}$ \\ \hline
   % u quark & 0.035 & 0.750 & 0.080 & -0.6 & 29.18 & 1.459 \\ \hline
    %d quark & 0.2 & 1.25 & 1.0  & -0.2 & -33.918 & 1.413 \\ \hline
    %\end{tabular} 
%\caption{The parameters in the model for $\kappa=0.4066$ GeV} % title of Table 
%\label{table} % is used to refer this table in the text 
%\end{table} 
%%%%%%%%%%%%%%%%%%%%%%%%%%%%%%%%%%%%%%%%%%%%%%%%%%%%%%%%%%%%%%%%%%%%%%%
%%%%%%%%%%%%%%%%%%%%%%%%%%%%%%%%%%%%%%%%%%%%%%%%%%%%%%%%%%%%%%%%%%%%%%
%\begin{table}[ht]
%\centering % used for centering table 
%\begin{tabular}{ c c c } % centered columns (5 columns) 
%\hline\hline %inserts double horizontal lines 
%Parameters&~~~~~~~~~~~~$u$~~~~~~~~~~~~~& ~~~~~~~~~~~$d$~~~~~~~~~~~~  \\ [0.5ex] % inserts table 
%heading 
%\hline\hline % inserts single horizontal line 
%$a^{(1)}$ & 0.035 & 0.20\\ % inserting body of the table 
%$b^{(1)}$ & 0.080 & 1.00 \\
%\hline
%$a^{(2)}$ & 0.75 & 1.25 \\ 
%$b^{(2)}$ & -0.60 & -0.20 \\
%\hline
%$N^{(1)}$ & 29.180 & 33.918 \\
%$N^{(2)}$ & 1.459 & 1.413 \\
%\hline\hline%inserts single line 
%\end{tabular} 
%\caption{The parameters in the model for $\kappa=406$ MeV} % title of Table 
%\label{table} % is used to refer this table in the text 
%\end{table} 
%%%%%%%%%%%%%%%%%%%%%%%%%%%%%%%%%%%%%%%%%%%%%%%%%%%%%%%%%%%%%%%%%%%%%%%
where $\kappa$ is the AdS/QCD scale parameter which is taken to be $0.4$ GeV \cite{CM1,CM2}. 
%The wavefunctions involve four more parameters $a_q^{(i)}$ and $b_q^{(i)}$ (with $i=1,2$) for each quark.  We use the scale parameter  $\kappa=406$ MeV  which was obtained by fitting the nucleon form factors in AdS/QCD soft-wall model .
 The   parameters $a^{(i)}_q$ and $b^{(i)}_q$ with the constants $N^{(i)}_q$  are fixed by fitting the electromagnetic properties of the nucleons:  %\cite{chandan}. For completeness, we list the parameters:
$a^{(1)}_u  = 0.020,~  a^{(1)}_d= 0.10,~
b^{(1)}_u = 0.022,~b^{(1)}_d=0.38,~
a^{(2)}_u=  1.05,~ a^{(2)}_d=  1.07,~
b^{(2)}_u= -0.15, ~b^{(2)}_d= -0.20,
N^{(1)}_u = 2.055,~ N^{(1)}_d = 1.7618,
N^{(2)}_u= 1.322, N^{(2)}_d = -2.4827 $.
%%%%%%%%%%%%%%%%%%%%%%%%%%%%%%%%%%%%%%%%%%%%%%%%%%%%%%%%%%%%%%%%%%%%%%%
\section{Chiral-odd generalized parton distributions}\label{chiral_gpds}
%%%%%%%%%%%%%%%%%%%%%%%%%%%%%%%%%%%%%%%%%%%%%%%%%%%%%%%%%%%%%%%%%%%%%%%
The chiral-odd GPDs are defined as off-forward matrix elements of the bilocal operator of light-front correlation functions of the 
tensor current~\cite{diehl01}
\be
\label{gpd_eq}
& &\frac{1}{2}\int\frac{dz^-}{2\pi}e^{i \bar xP^+z^-}\langle
p', \lambda'|\bar{\psi}(-{z}/2)\sigma^{+i}\gamma_5
\psi({z}/2)|p, \lambda\rangle_{|_{z^+ = 0, \vec z_\perp=
0}}
\nonumber\\
& &= \frac{1}{2P^+}\bar{u}(p',
\lambda')\bigg[H^q_T\sigma^{+i}\gamma_5 +
\widetilde{H}^q_T\frac{\epsilon^{+i\alpha\beta}\Delta_\alpha
P_\beta}{M^2} +
E^q_T\frac{\epsilon^{+i\alpha\beta}\Delta_\alpha\gamma_\beta}{2M}
+
\widetilde{E}^q_T\frac{\epsilon^{+i\alpha\beta}P_\alpha\gamma_\beta}{M}\bigg]u(p,
\lambda),
\ee
where $i=1,2$ is a transverse index. $p$ $(p')$ and $\lambda$ $(\lambda')$ denote the proton momenta and the helicity of the initial (final) state of proton, respectively. 
%We use the light-front coordinates $v^\mu=[v^+,v^-,\vec v_\perp]$, where $v^\pm=(v^0\pm v^3)/\sqrt{2}$ and $\vec v_\perp=(v^1,v^2)$. 
In the symmetric frame, the kinematical variables are 
\be
P^\mu=\frac{(p+p')^\mu}{2}, \quad\quad \Delta^\mu=p'^\mu-p^\mu, \quad\quad \zeta=-\Delta^+/2P^+,
\ee
and $t=\Delta^2$. We choose the light-front gauge $A^+=0$, so that the gauge link appears in between the quark fields in Eq. (\ref{gpd_eq}) is unity. The GPDs which involve the quark helicity flip can be related to the following matrix elements \cite{diehl01,Pasquini1}
\be
  \label{amplitude}
A_{\lambda'+, \lambda-} &=&
\int \frac{d z^-}{2\pi}\, e^{i\bar x P^+ z^-}
  \langle p',\lambda'|\, {\cal O}_{+,-}(z)
  \,|p,\lambda \rangle \Big|_{z^+=0,\, \vec{z}_\perp=0} \, ,
\nonumber \\
A_{\lambda'-, \lambda+} &=&
\int \frac{d z^-}{2\pi}\, e^{i\bar x P^+ z^-}
  \langle p',\lambda'|\, {\cal O}_{-,+}(z)
  \,|p,\lambda \rangle \Big|_{z^+=0,\, \vec{z}_\perp=0} \, ,
\ee
with the operators $O_{+,-}$ and $O_{-,+}$ defined by
\be
{\cal O}_{+,-} 
&=& \frac{i}{4}\, 
  \bar{\psi}\, \sigma^{+1} (1-\gamma_5)\, \psi \,,\nonumber\\
{\cal O}_{-,+} 
&=& - \frac{i}{4}\, \bar{\psi}\, \sigma^{+1} (1+\gamma_5)\, \psi.
\hspace{2em}
\label{eq:op_o}
\ee
Using  the reference frame where  the momenta $\vec p$ and $\vec p\,'$ lie in the $x-z$ plane, one can explicitly derive the following relations~\cite{diehl01}
\be \label{relations}
A_{++,+-} &=&   \epsilon\, \frac{\sqrt{t_0-t}}{2m} \left( \widetilde{H}^q_T
              + (1-\zeta)\, \frac{E^q_T + \widetilde{E}^q_T}{2} \right) , 
\nonumber \\
A_{-+,--} &=&   \epsilon\, \frac{\sqrt{t_0-t}}{2m} \left( \widetilde{H}^q_T
              + (1+\zeta)\, \frac{E^q_T - \widetilde{E}^q_T}{2} \right) , 
\nonumber \\
A_{++,--} &=& \sqrt{1-\zeta^2} \left(H^q_T + \
              \frac{t_0-t}{4 m^2}\, \widetilde{H}^q_T -
              \frac{\zeta^2}{1-\zeta^2}\, E^q_T +
              \frac{\zeta}{1-\zeta^2}\, \widetilde{E}^q_T \right) , 
\nonumber \\
A_{-+,+-} &=& - \sqrt{1-\zeta^2}\; \frac{t_0-t}{4 m^2}\, \widetilde{H}^q_T,
\ee
where, $\epsilon =
\mathrm{sgn}(D^1)$, where $D^1$ is the $x$-component of $D^\alpha =
P^+ \Delta^\alpha - \Delta^+ P^\alpha$ and $D^1=0$ corresponds to $t=t_0$. The minimum value of $-t$ for given $\zeta$ is $- t_0 = 4 m^2 \zeta^2/(1-\zeta^2)$.  Due to parity invariance one has the relation 
$A_{-\lambda'-,-\lambda +}=(-1)^{\lambda'-\lambda}A_{\lambda'+,\lambda -}$.

%The GPDs with quark helicity flip can also be related with the matrix elements of the 
%transversity basis for the nucleon spin states 
%i.e.
%\be
%|p,\uparrow\rangle=\frac{1}{\sqrt{2}}(|p,+\rangle+|p,-\rangle)
%\nonumber\\
%|p,\downarrow\rangle=\frac{1}{\sqrt{2}}(|p,+\rangle -|p,-\rangle),
%\label{trans_states}
%\ee
The chiral-odd GPDs are off-diagonal in the quark helicity basis but they can also be calculated in the transversity basis~\cite{Pasquini1} which is more useful for the overlap formalism used in this work. Here we  briefly discuss the transformation of matrix elements defining chiral-odd GPDs from helicity basis to transversity basis \cite{Pasquini1}.
Consider  the operators $
%\mathcal{N}_T=
\mathcal{O}_{+,-} + \mathcal{O}_{-,+}=
-\frac{i}{2}\bar{\psi}\sigma^{+1}\gamma_5\psi$
and $
%\mathcal{F}_T=
\mathcal{O}_{+,-} - \mathcal{O}_{-,+} = \frac{i}{2}\bar{\psi}\sigma^{+1}\psi$  in the transversity basis  i.e.
\be
\label{T}
T^{q}_{\lambda'_t\lambda_t} &=& \langle p',
\lambda'_t|\int\frac{dz^-}{2\pi}e^{i \bar xP^+z^-}
\bar{\psi}(-{z}/2)\gamma^+\gamma^1\gamma_5
\psi({z}/2)|p, \lambda_t\rangle,
\\
\label{Ttilde}
\widetilde{T}^{q}_{\lambda'_t\lambda_t} &=&\langle p',
\lambda'_t|\int\frac{dz^-}{2\pi}e^{i\bar xP^+z^-}
\frac{i}{2}\bar{\psi}(-{z}/2)\sigma^{+1}\psi({z}/2)
|p, \lambda_t\rangle,
\ee
where $\lambda_t$ ($\lambda'_{t}$) labels the transverse polarization of the initial (final) nucleon polarized along +ve $x$($\uparrow$) or -ve $x$($\downarrow$) direction and
the transverse basis states are defined as 
\be
\mid p,\uparrow\rangle &=& \frac{1}{\sqrt{2}}(\mid p,+\rangle+\mid p,-\rangle),\\
\mid p,\downarrow\rangle &=&  \frac{1}{\sqrt{2}}(\mid p,+\rangle-\mid p,-\rangle).
\ee
These matrix elements obey the following relations as a result of  parity invariance
\be
T^{q}_{\uparrow\uparrow}&=&-T^{q}_{\downarrow\downarrow}\,,\quad\quad
T^{q}_{\uparrow\downarrow}=T^{q}_{\downarrow\uparrow}\,,\nonumber\\
\widetilde T^{q}_{\uparrow\uparrow}&=&\widetilde T^{q}_{\downarrow\downarrow}\, ,~~\quad\quad
\widetilde T^{q}_{\uparrow\downarrow}=-\widetilde T^{q}_{\downarrow\uparrow}\, .
\ee
We can now express them in terms of the matrix elements in the helicity basis as
\be
T^q_{\uparrow\uparrow} =A_{++,--} + A_{-+,+-}, &\quad&
T^q_{\uparrow\downarrow} =A_{++,+-} - A_{-+,--},\nonumber\\
\widetilde{T}^q_{\uparrow\uparrow} = 
A_{++,+-}+A_{-+,--}, &\quad&
\widetilde{T}^q_{\downarrow\uparrow} = 
A_{++,--}-A_{-+,+-}.
\ee
Finally, one can obtained the chiral-odd GPDs from the transverse matrix elements through the relations %~\cite{Pasquini1}
\be
H^{q}_T &=& \frac{1}{\sqrt{1 - \zeta^2}}T^q_{\uparrow\uparrow} -
\frac{2M\zeta} {\epsilon\sqrt{t_0 - t}(1 - \zeta^2)}T^q_{\uparrow\downarrow},\label{HT}
%\nonumber
\\
E_T^{q} &=& \frac{2M}{\epsilon\sqrt{t_0 - t}(1 -\zeta^2)}
\bigg(\zeta T^q_{\uparrow\downarrow}
+ \widetilde{T}^q_{\uparrow\uparrow}\bigg)
- \frac{4M^2}{(t_0 - t)\sqrt{1 - \zeta^2}(1 - \zeta^2)}
\bigg(\widetilde{T}^q_{\downarrow\uparrow} - T^q_{\uparrow\uparrow}\bigg).\label{ET}
\\\nonumber
\\
\widetilde{H}_T^{q} &=& \frac{2M^2}{(t_0 - t)\sqrt{1 -\zeta^2}}
(\widetilde{T}^q_{\downarrow\uparrow} - T^q_{\uparrow\uparrow}),\label{HtT}
%\nonumber
\\
\widetilde{E}^{q}_T &=& \frac{2M}{\epsilon\sqrt{t_0 - t}(1 -\zeta^2)}
\bigg(T^q_{\uparrow\downarrow}
+ \zeta\widetilde{T}^q_{\uparrow\uparrow}\bigg)
- \frac{4M^2\zeta}{(t_0 - t)\sqrt{1 - \zeta^2}(1 - \zeta^2)}
\bigg(\widetilde{T}^q_{\downarrow\uparrow} - T^q_{\uparrow\uparrow}\bigg).\label{EtT}
%\label{gpds_relations}
\ee

%%%%%%%%%%%%%%%%%%%%%%%%%%%%%%%%%%%%%%%%%
\subsection{Overlap formalism}
%%%%%%%%%%%%%%%%%%%%%%%%%%%%%%%%%%%%%%%%%
%%%%%%%%%%%%%%%%%%%%%%%%%%%%%%%%%%%%%%%%%%%%%%%%%%%
%%%%%%%%%%%%%%%%%%%
\begin{figure}[htbp]
\begin{minipage}[c]{0.98\textwidth}
\small{(a)}
\includegraphics[width=7.6cm,height=5.5cm,clip]{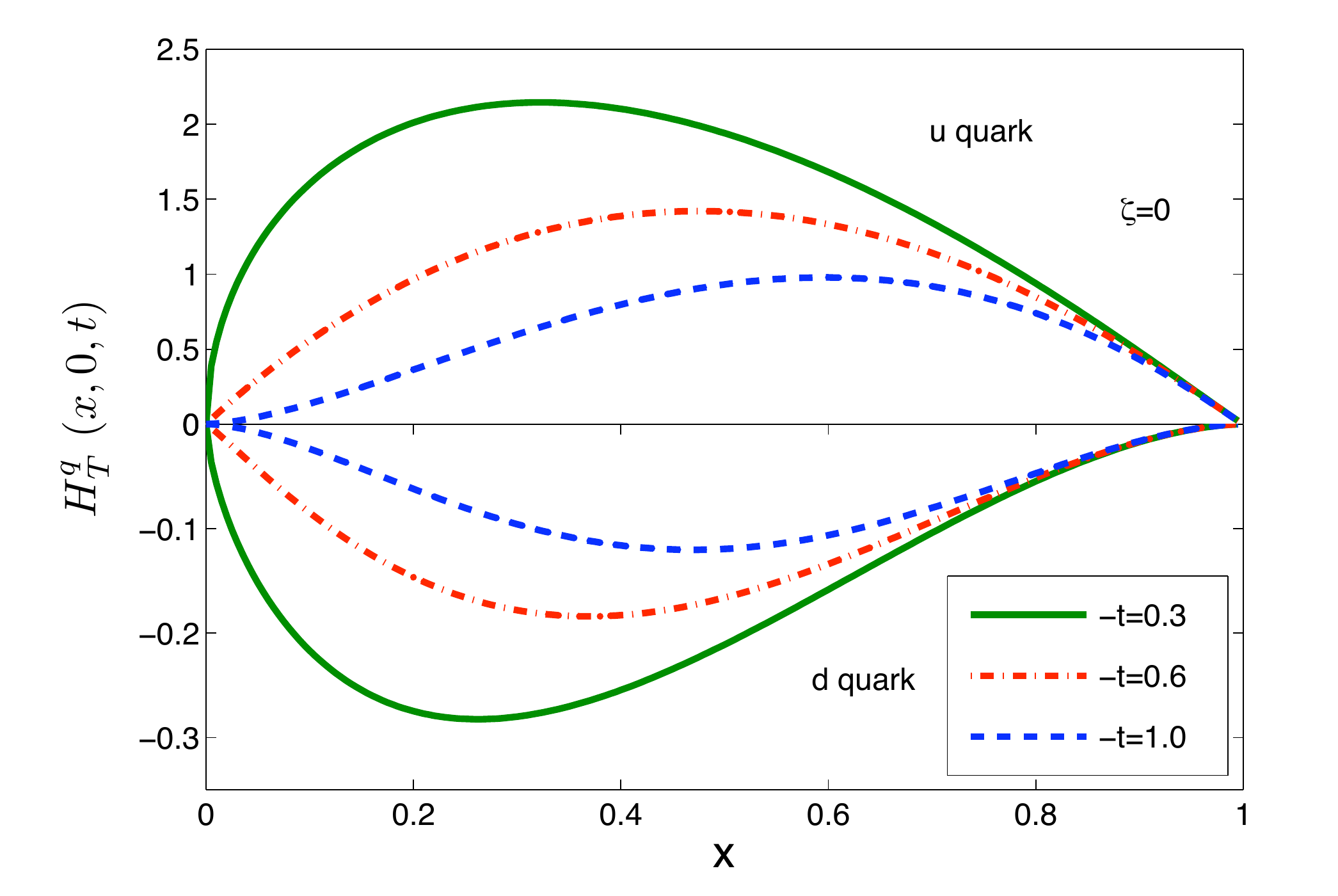}
\hspace{0.1cm}%
\small{(b)}\includegraphics[width=7.6cm,height=5.5cm,clip]{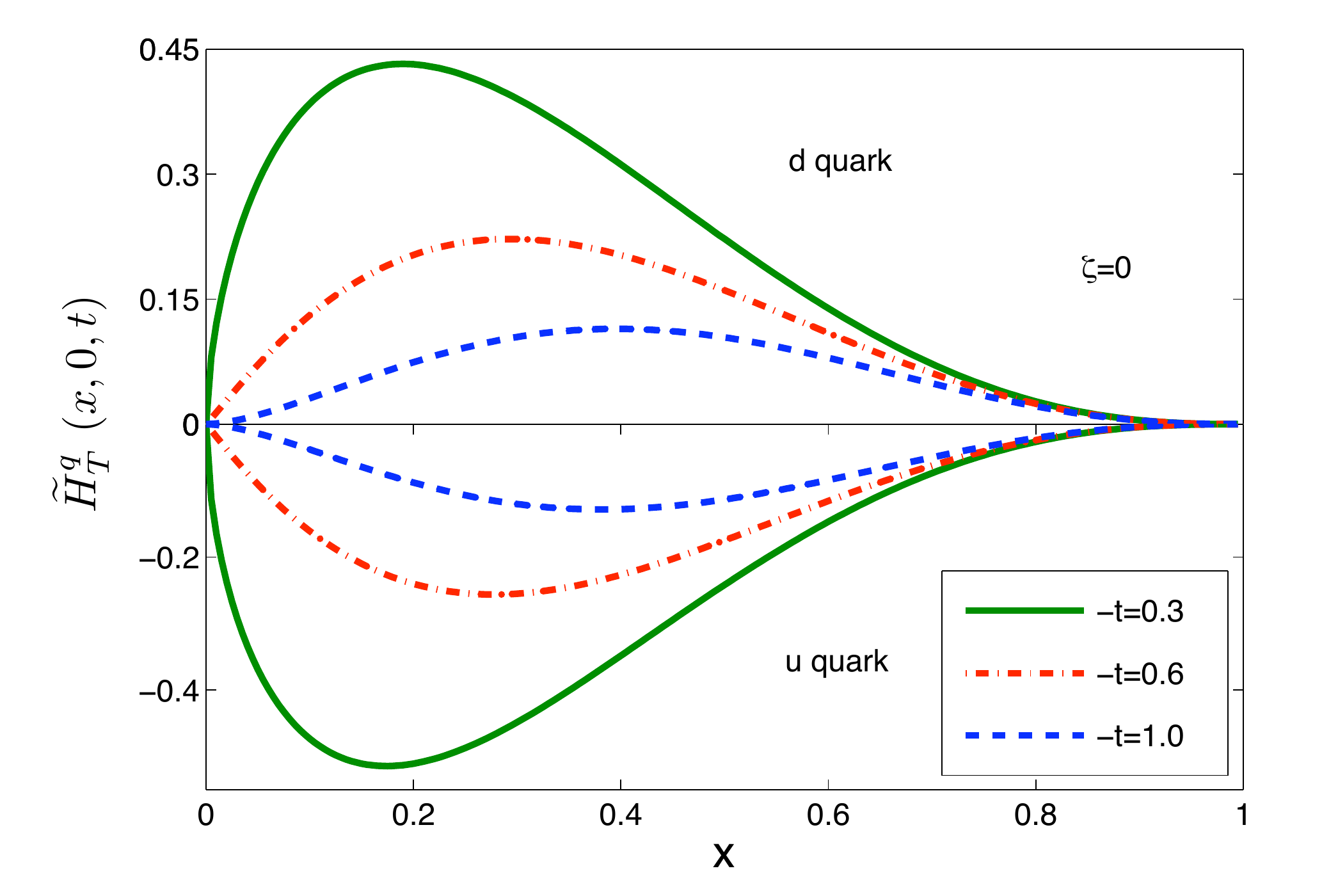}
\end{minipage}
\begin{minipage}[c]{0.98\textwidth}
\small{(c)}\includegraphics[width=7.6cm,height=5.5cm,clip]{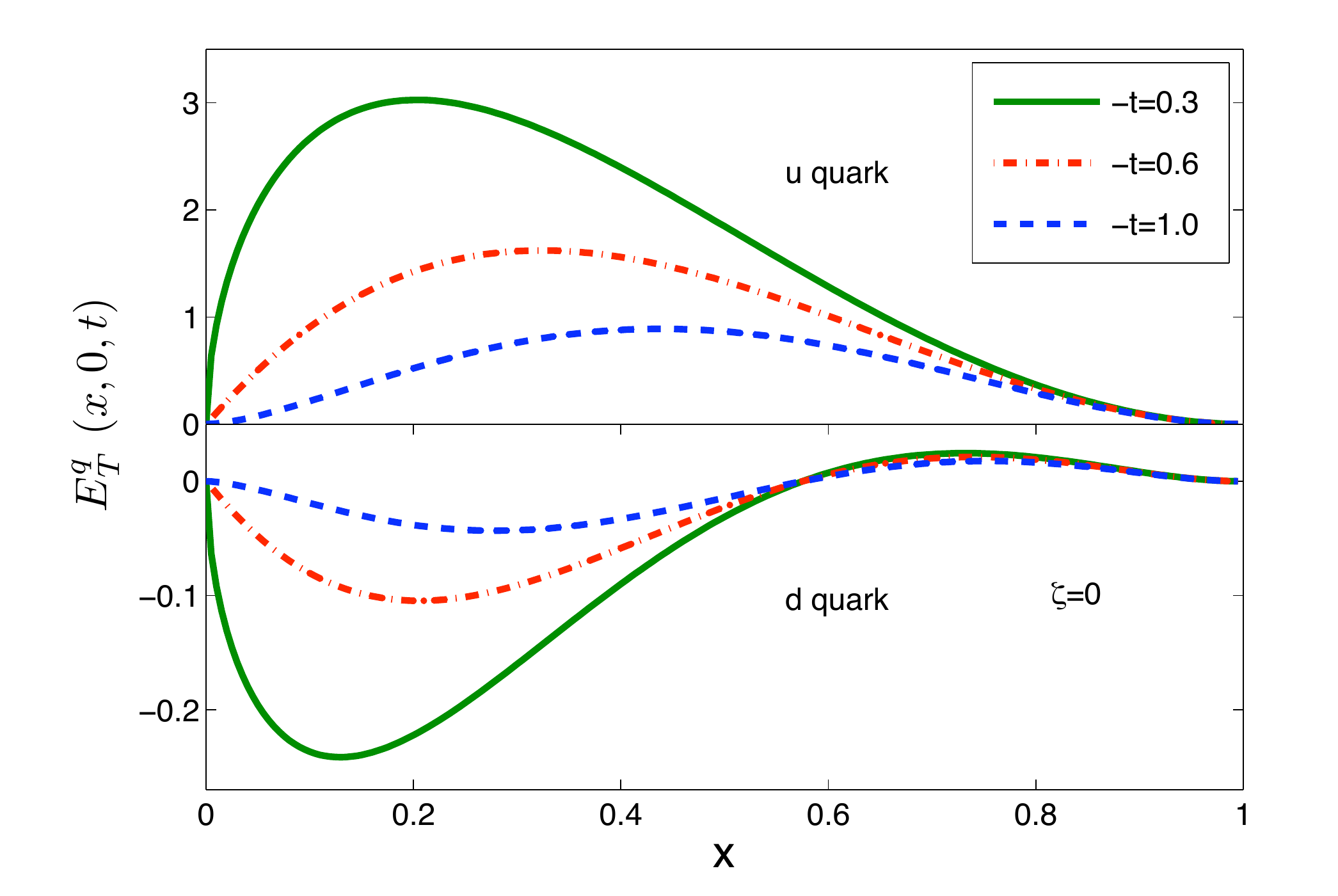}
%\hspace{0.1cm}%
%\small{(d)}\includegraphics[width=7.5cm,height=5.15cm,clip]{ETd_vs_x.pdf}
\end{minipage}
\caption{\label{gz0}(Color online) Plots of the chiral-odd GPDs for zero skewness vs $x$ and different values of $-t$ in $\rm{GeV}^2$ for $u$ and $d$ quarks.}
\end{figure}
%%%%%%%%%%%%%%%%%%%%%%%%%%%%%%%%%%%%%%%%%%%%%%%%%%
%%%%%%%%%%%%%%%%%%%...zeta dependent.xt..%%%%%%%%%
\begin{figure}[htbp]
\begin{minipage}[c]{0.98\textwidth}
\small{(a)}
\includegraphics[width=7.5cm,height=5.15cm,clip]{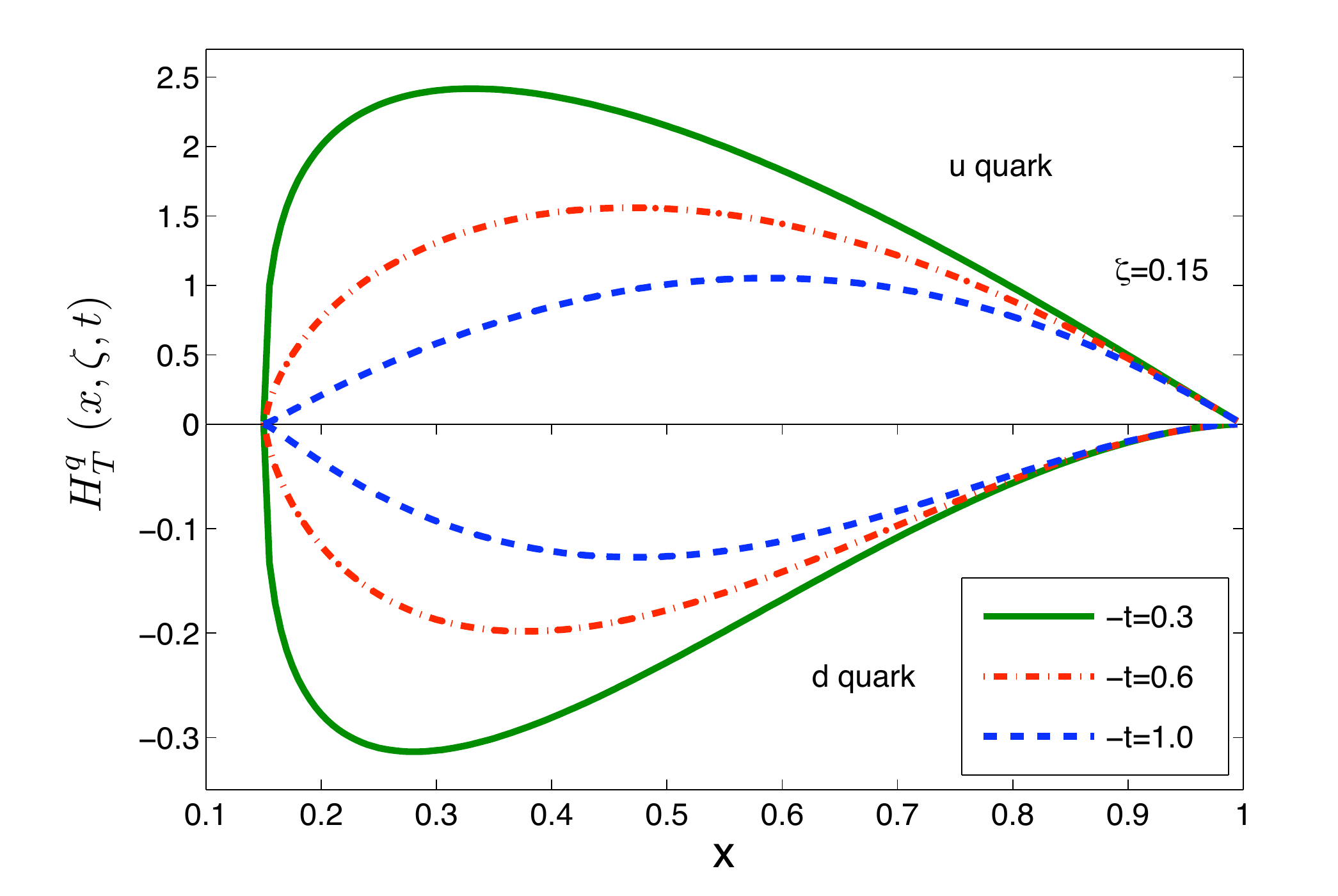}
\hspace{0.1cm}%
\small{(b)}\includegraphics[width=7.5cm,height=5.15cm,clip]{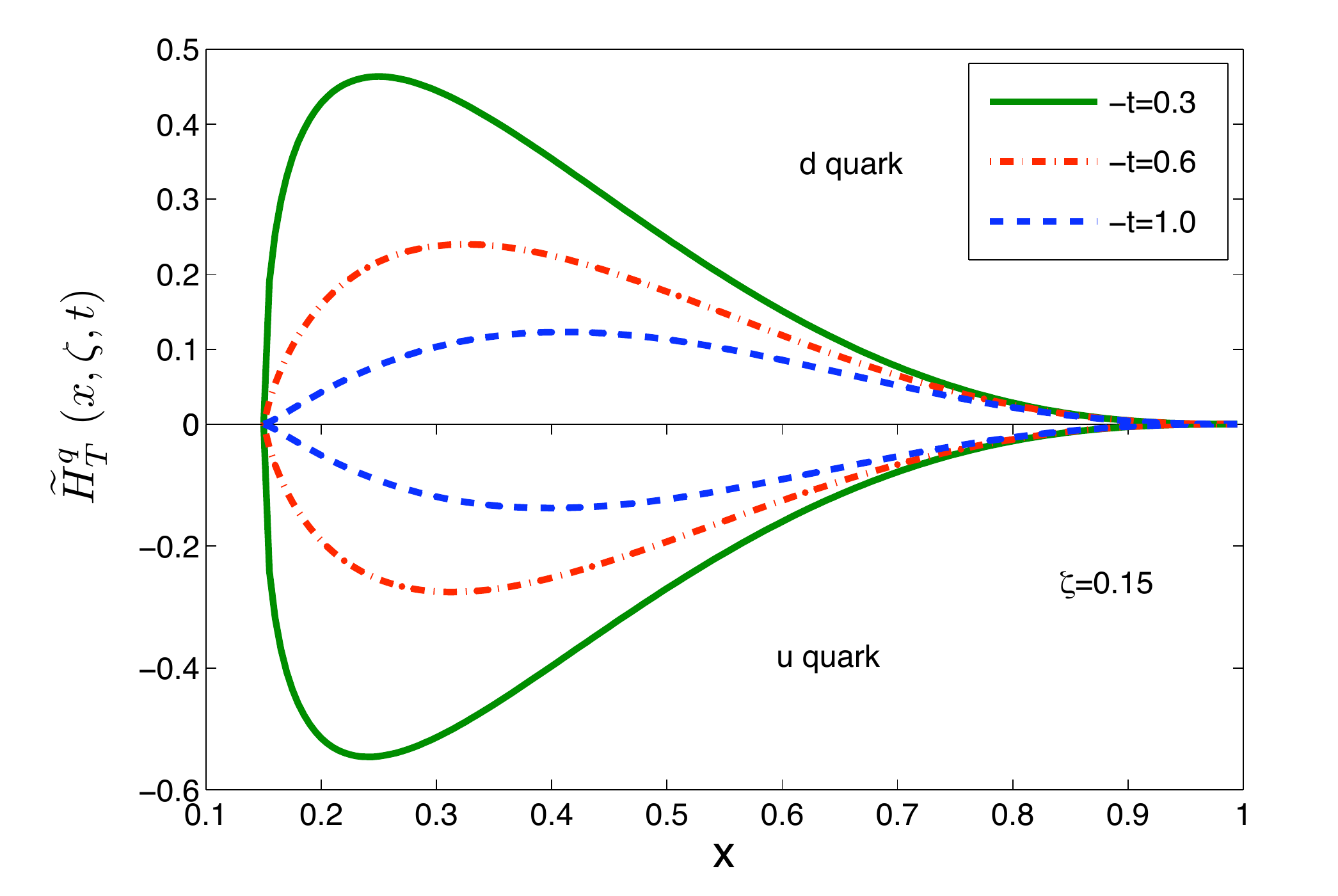}
\end{minipage}
\begin{minipage}[c]{0.98\textwidth}
\small{(c)}\includegraphics[width=7.5cm,height=5.15cm,clip]{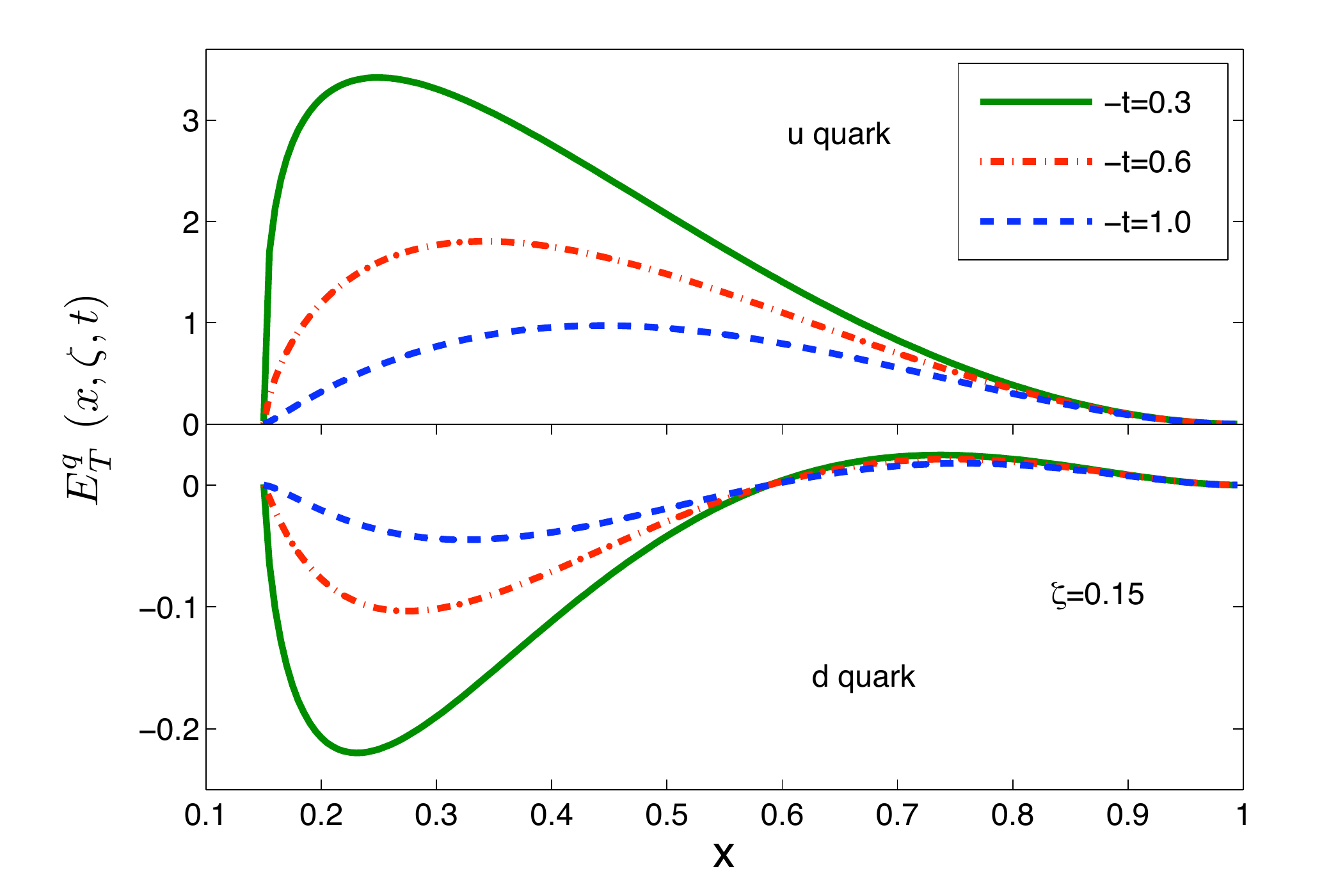}
\hspace{0.1cm}%
\small{(d)}\includegraphics[width=7.5cm,height=5.15cm,clip]{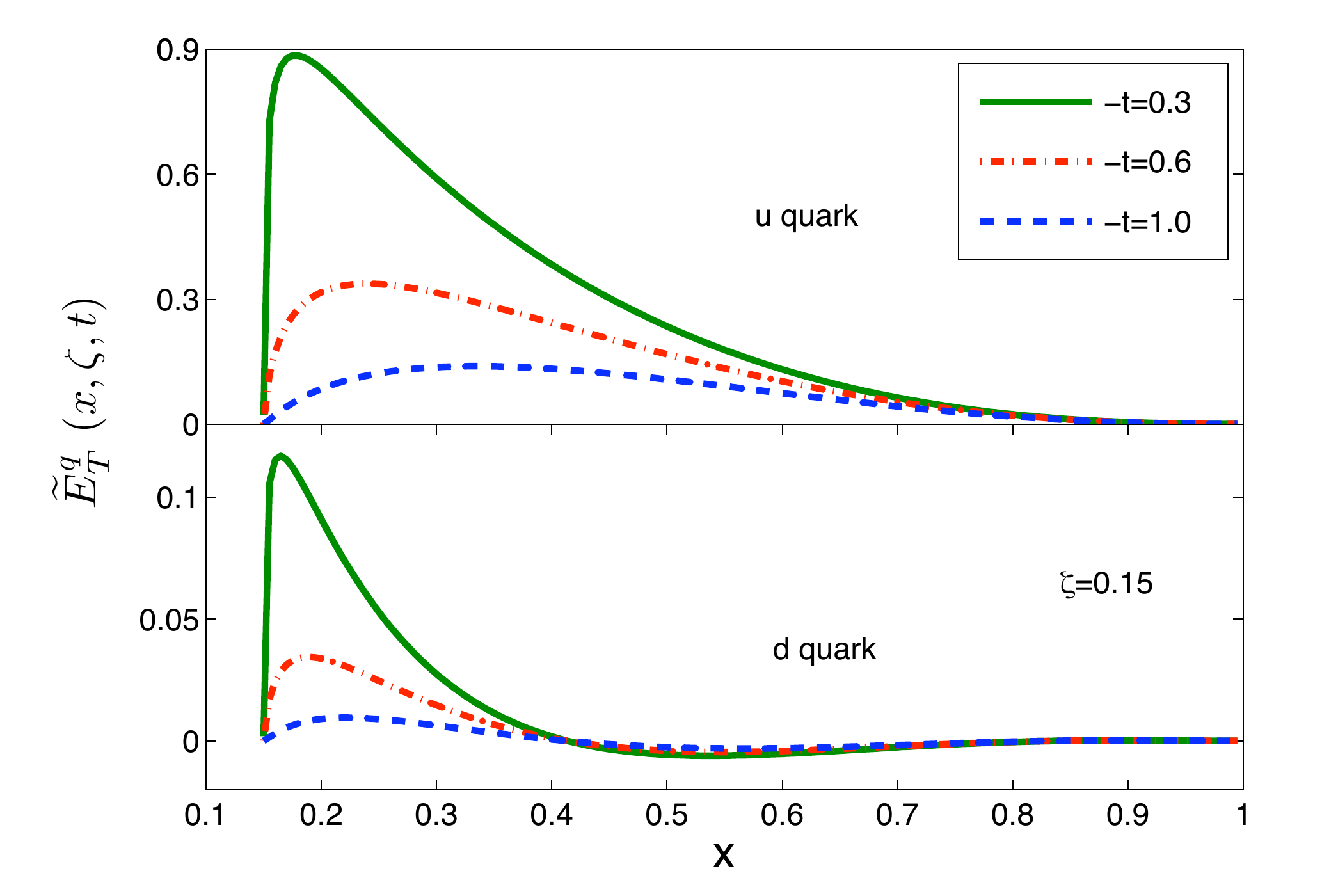}
\end{minipage}
\caption{\label{gz15}(Color online) Plots of chiral odd GPDs for the nonzero skewness vs $x$ and different values of $-t$ in $GeV^2$, for fixed value of $\zeta=0.15$. (a) $H^q_T$, (b) $\widetilde{H}^q_T$ and (c) $E^q_T$, (d) $\widetilde{E}^q_T$ ; $q$ stands for $u$ and $d$ quark.}
\end{figure}
%%%%%%%%%%%%%%%%%%%%%%%%%%%%%%%%%%%%%%%%%%%%%%%%%%%%%%%%%%%%%%%%%%%%%%%
%%%%%%%%%%%%%%%%%%%...zeta dependent.xt..%%%%%%%%%
\begin{figure}[htbp]
\begin{minipage}[c]{0.98\textwidth}
\small{(a)}
\includegraphics[width=7.5cm,height=5.15cm,clip]{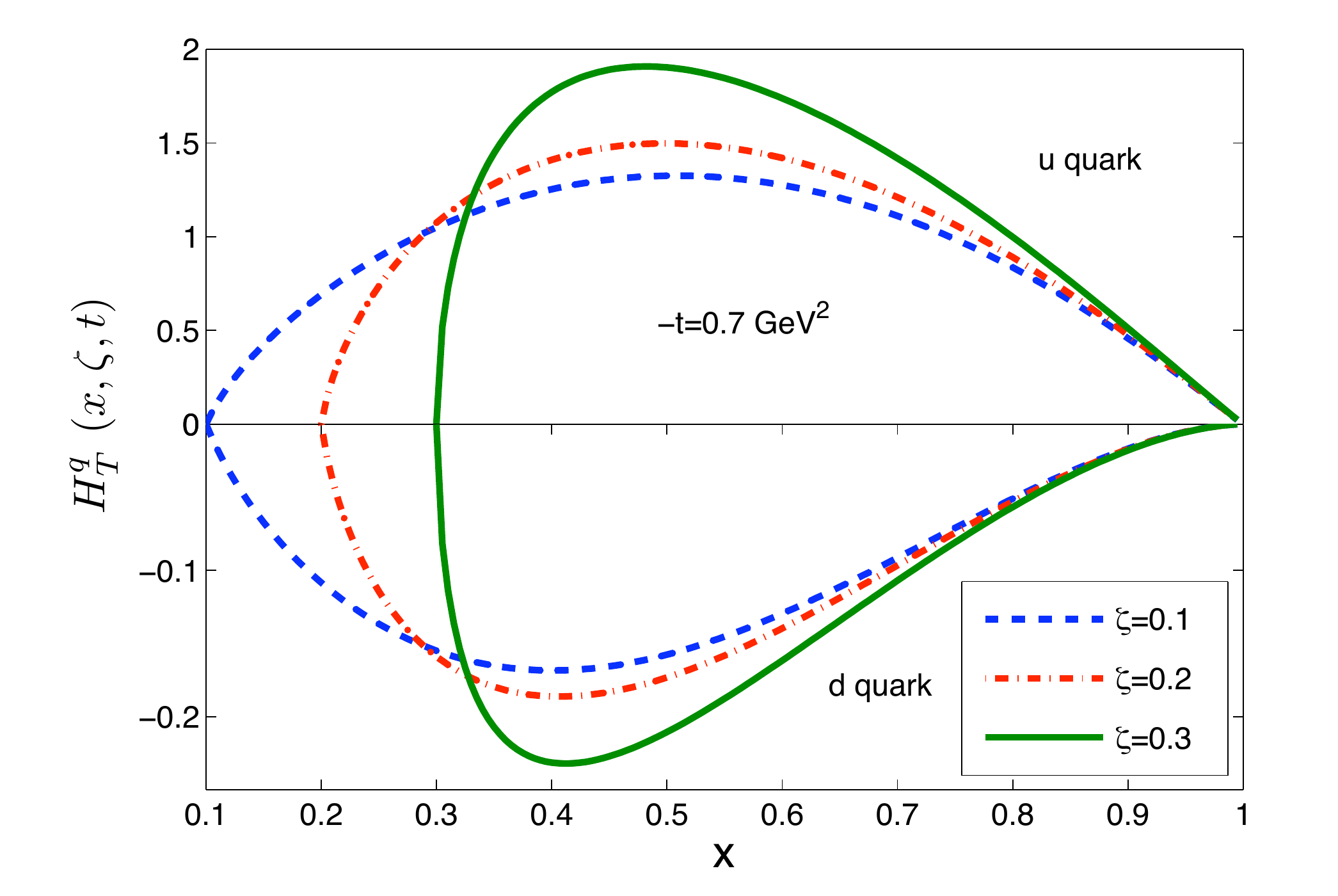}
\hspace{0.1cm}%
\small{(b)}\includegraphics[width=7.5cm,height=5.15cm,clip]{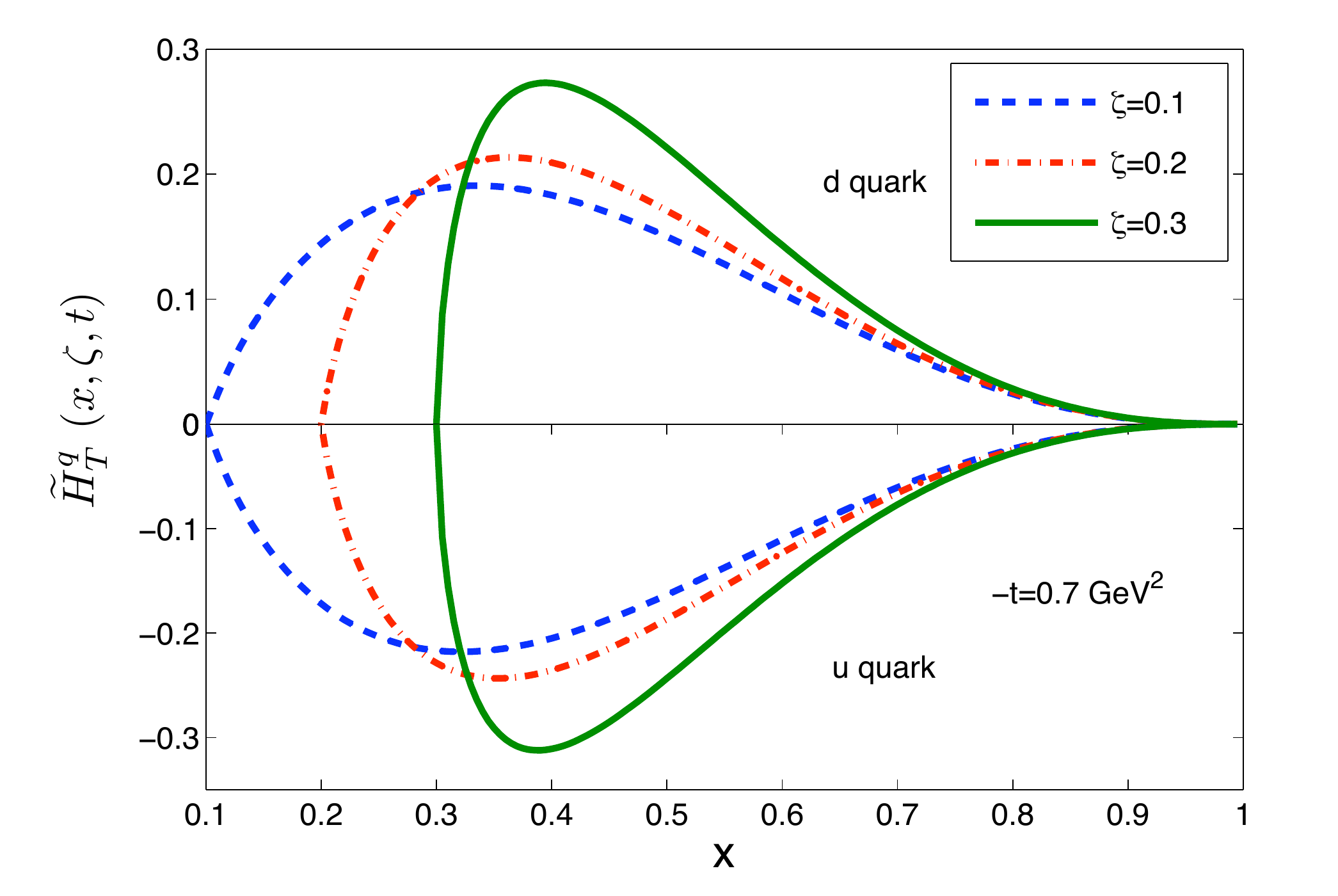}
\end{minipage}
\begin{minipage}[c]{0.98\textwidth}
\small{(c)}\includegraphics[width=7.5cm,height=5.15cm,clip]{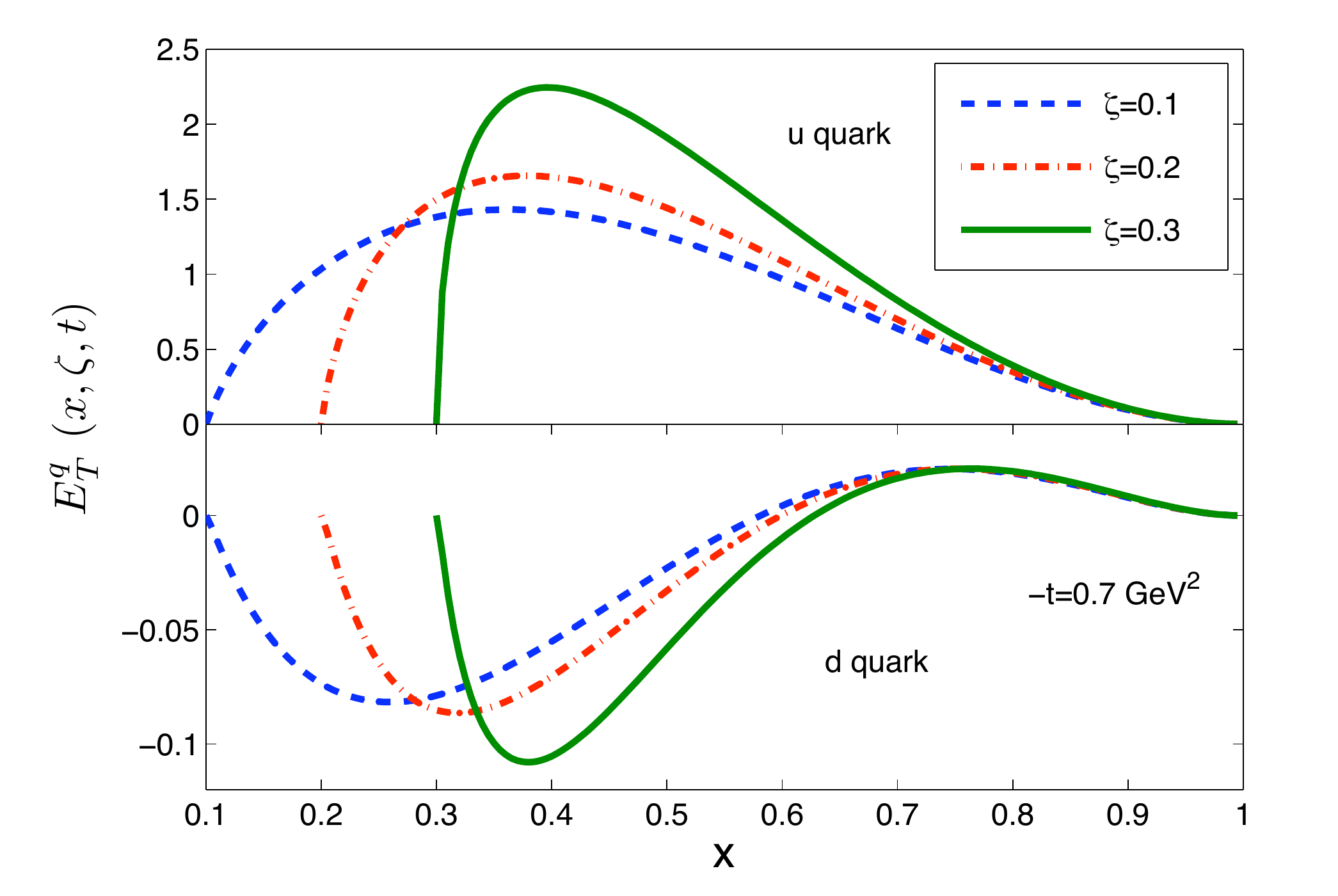}
\hspace{0.1cm}%
\small{(d)}\includegraphics[width=7.5cm,height=5.15cm,clip]{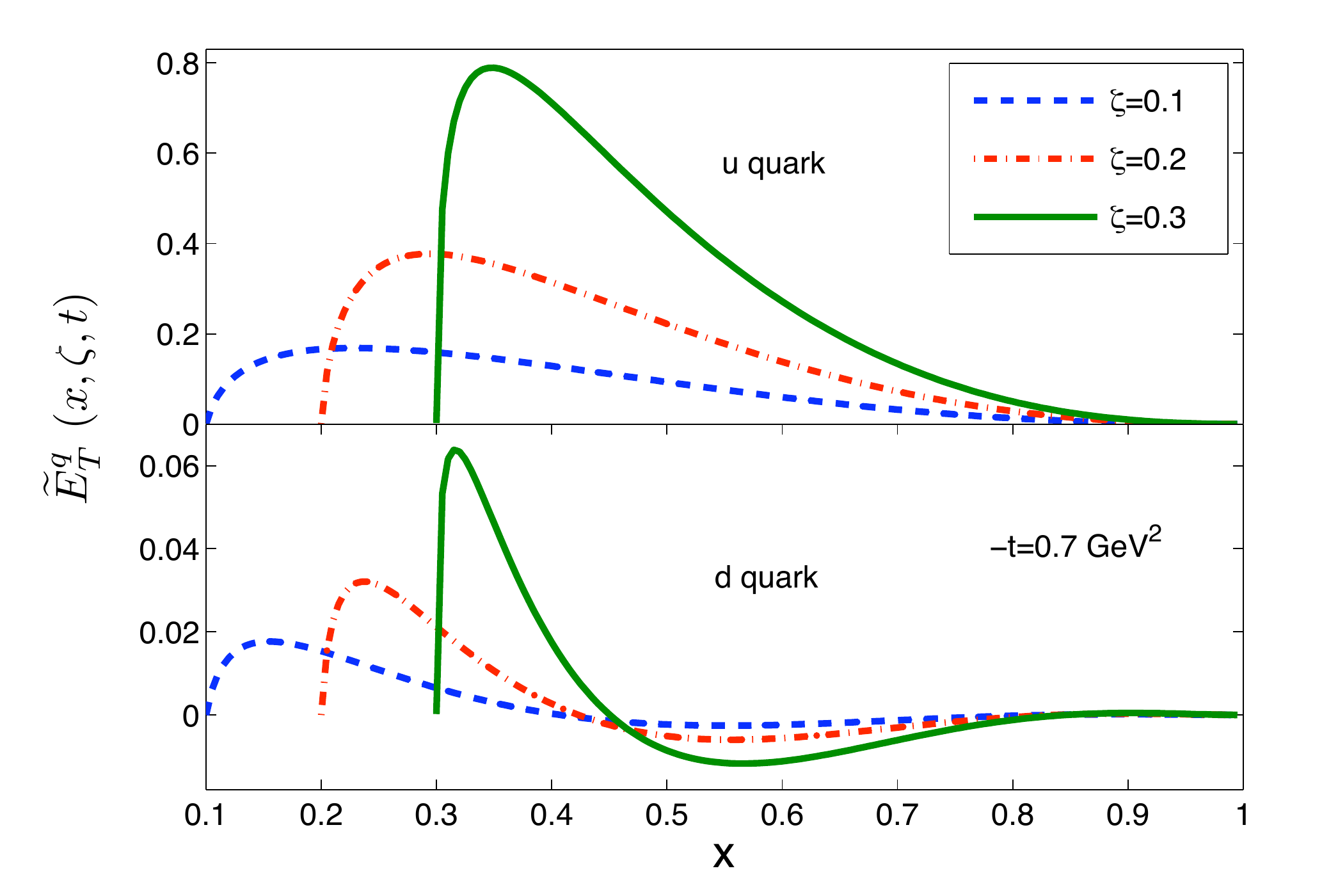}
\end{minipage}
\caption{\label{gt7}(Color online) Plots of chiral odd GPDs for the nonzero skewness vs $x$ and different values of $\zeta$, for fixed value of $t=0.7$ $GeV^2$. (a) $H^q_T$, (b) $\widetilde{H}^q_T$ and (c) $E^q_T$, (d) $\widetilde{E}^q_T$ ; $q$ stands for $u$ and $d$ quark.}
\end{figure}
%%%%%%%%%%%%%%%%%%%%%%%%%%%%%%%%%%%%%%%%%%%%%%%%%%%%%%%%%%%%%%%%%%%%%%%
%%%%%%%%%%%%%%%%%%%...zeta dependent.zt..%%%%%%%%%
\begin{figure}[htbp]
\begin{minipage}[c]{0.98\textwidth}
\small{(a)}
\includegraphics[width=7.5cm,height=5.15cm,clip]{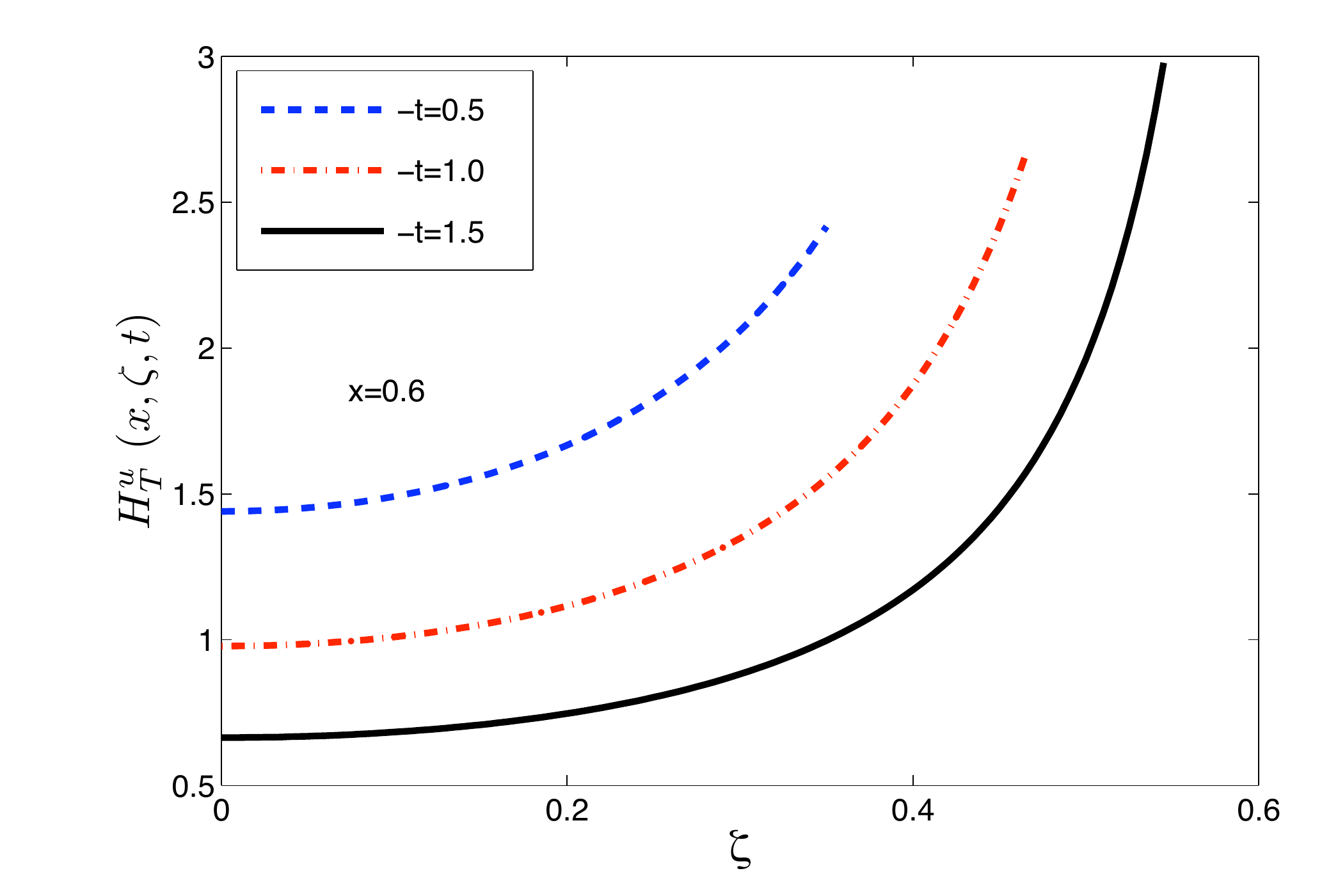}
\hspace{0.1cm}%
\small{(b)}\includegraphics[width=7.5cm,height=5.15cm,clip]{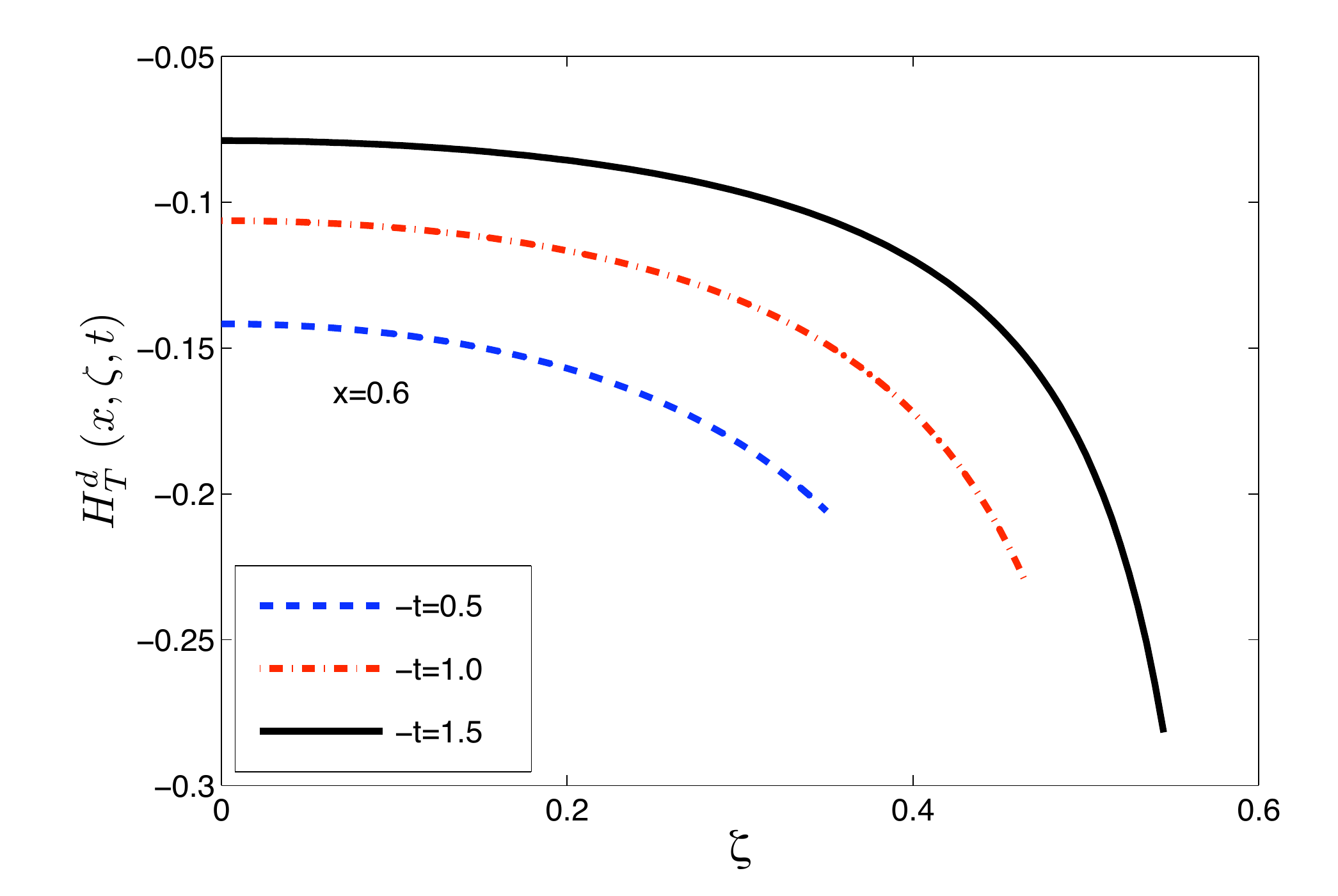}
\end{minipage}
\begin{minipage}[c]{0.98\textwidth}
\small{(c)}\includegraphics[width=7.5cm,height=5.15cm,clip]{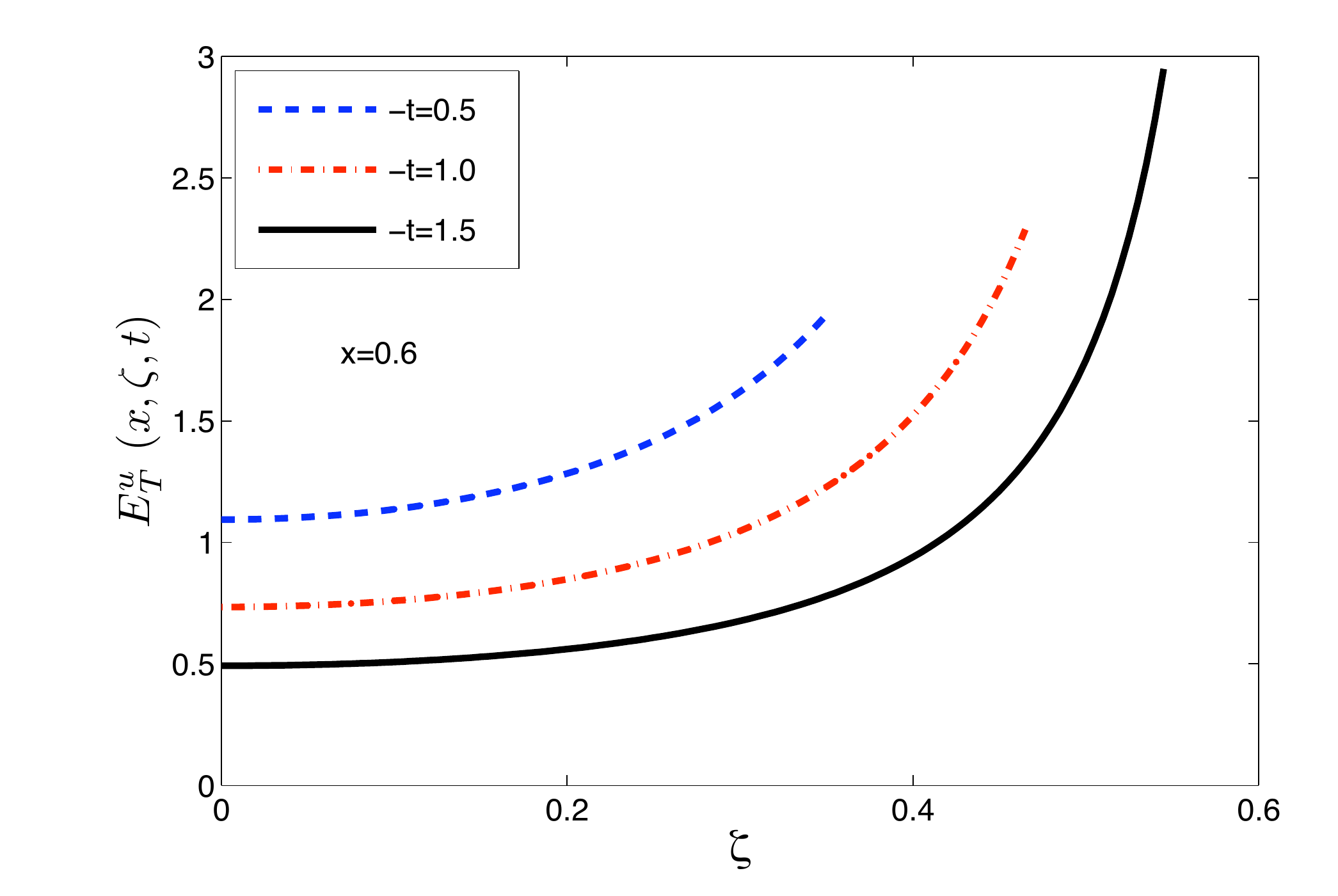}
\hspace{0.1cm}%
\small{(d)}\includegraphics[width=7.5cm,height=5.15cm,clip]{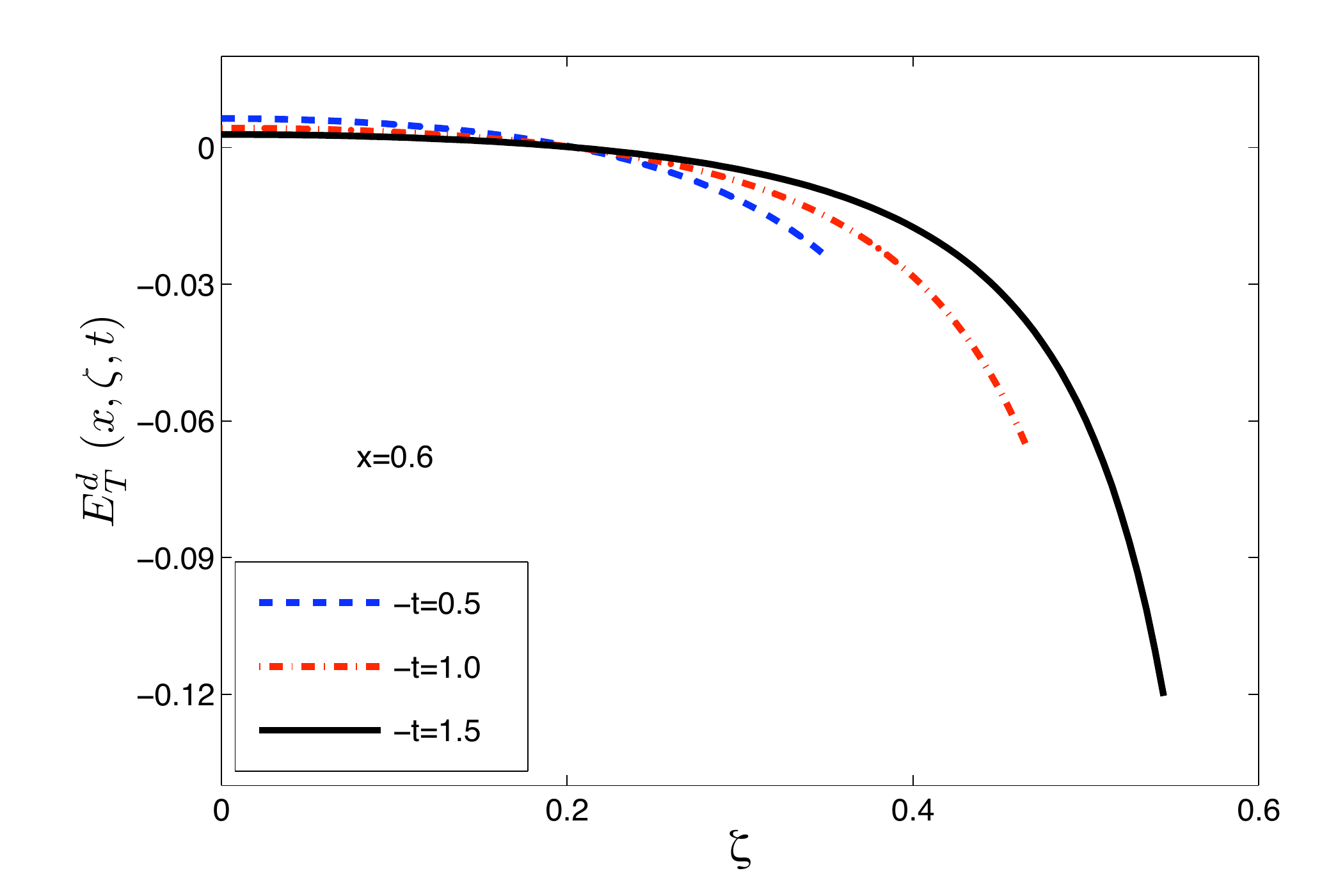}
\end{minipage}
\begin{minipage}[c]{0.98\textwidth}
\small{(e)}\includegraphics[width=7.5cm,height=5.15cm,clip]{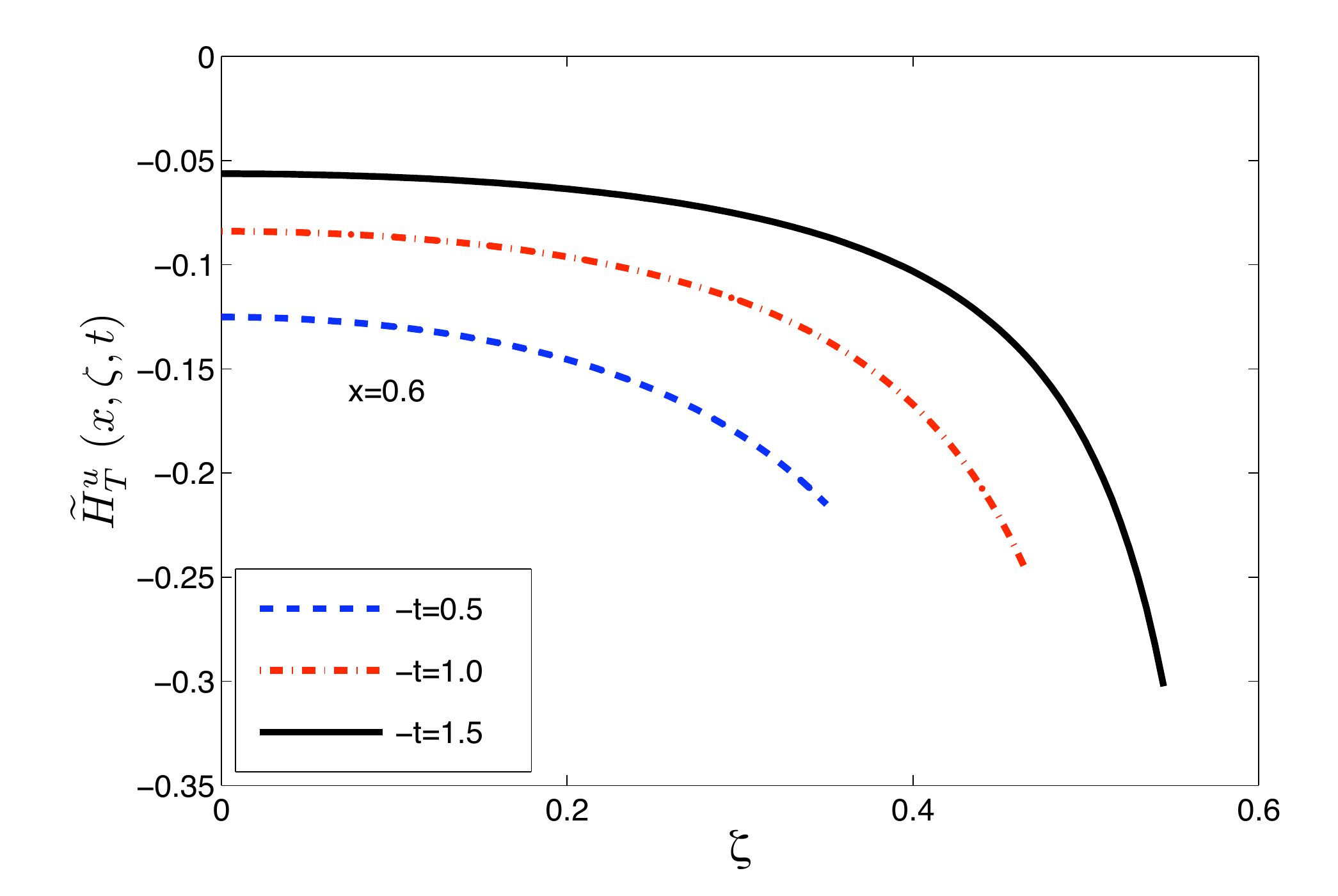}
\hspace{0.1cm}%
\small{(f)}\includegraphics[width=7.5cm,height=5.15cm,clip]{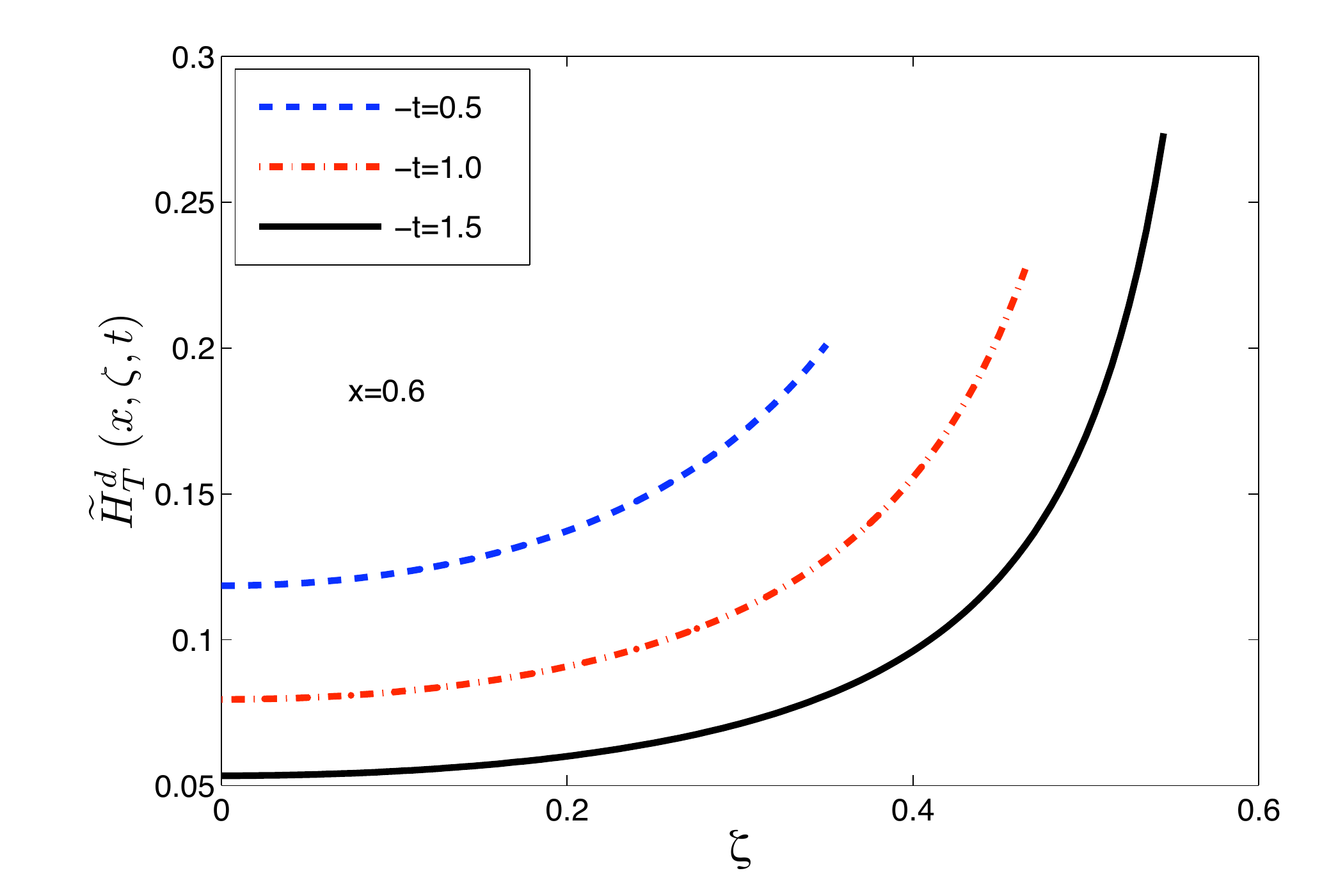}
\end{minipage}
\begin{minipage}[c]{0.98\textwidth}
\small{(g)}\includegraphics[width=7.5cm,height=5.15cm,clip]{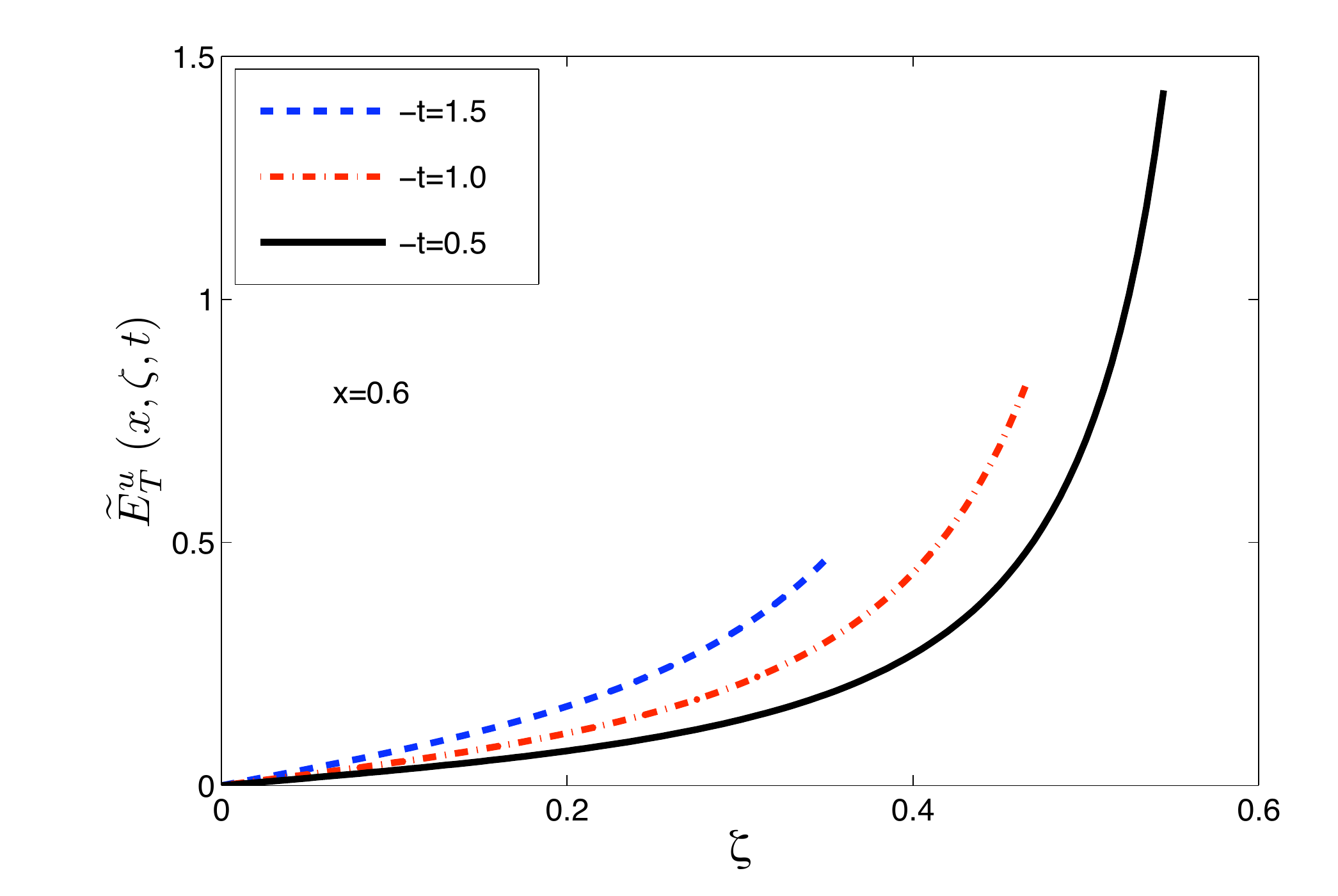}
\hspace{0.1cm}%
\small{(h)}\includegraphics[width=7.5cm,height=5.15cm,clip]{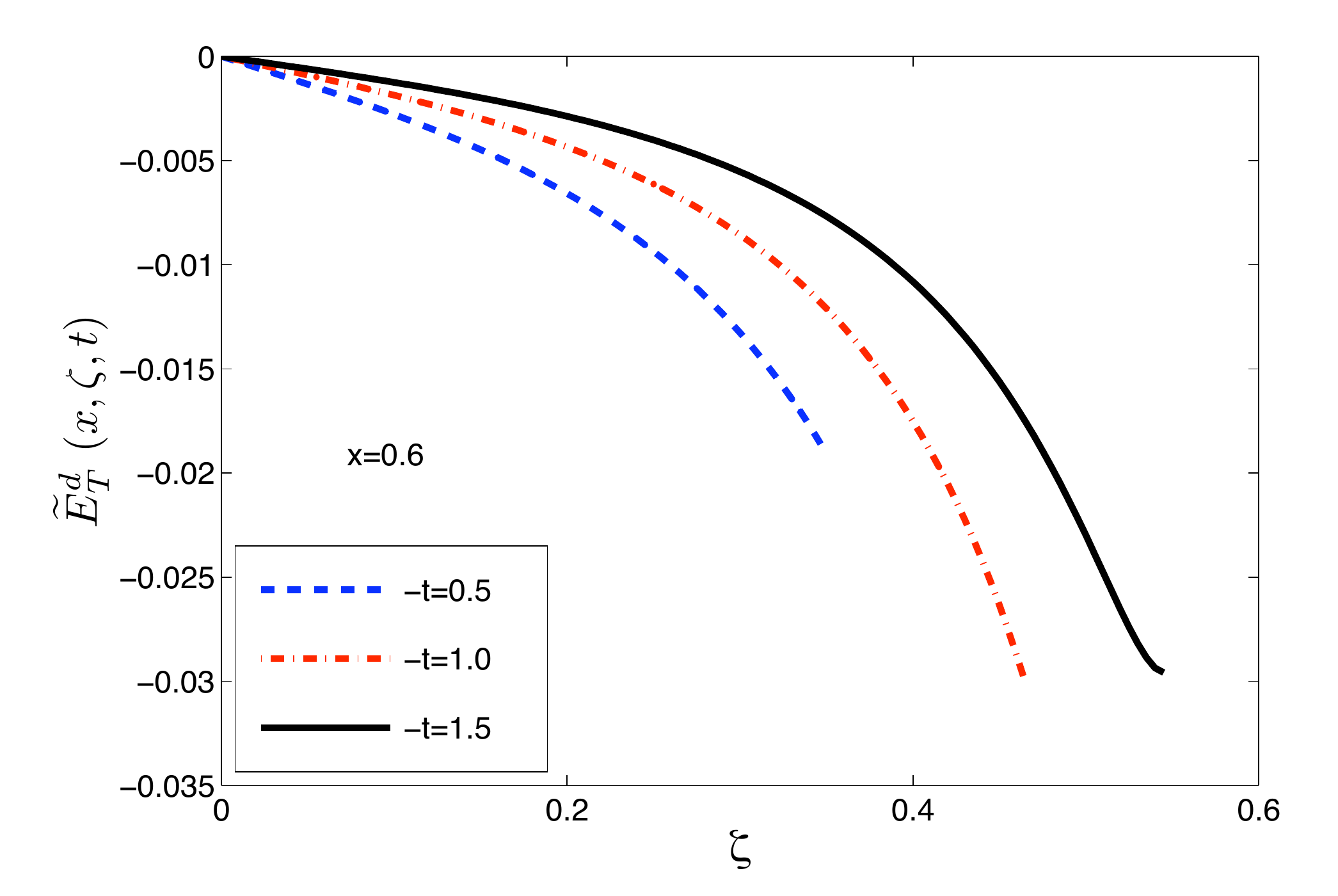}
\end{minipage}
\caption{\label{gx8}(Color online) Plots of chiral odd GPDs for the nonzero skewness vs $\zeta$ and different values of $-t$ in $GeV^2$, for fixed value of $x=0.6$. Left pannel is for $u$ quark and the right pannel is for $d$ quark.}
\end{figure}
%%%%%%%%%%%%%%%%%%%%%%%%%%%%%%%%%%%%%%%%%%%%%%%%%%%%%%%%%%%%%%%%%%%%%%%
%%%%%%%%%%%%%%%%%%%...zeta dependent.zt..%%%%%%%%%
%\begin{figure}[htbp]
%\begin{minipage}[c]{0.98\textwidth}
%\small{(e)}\includegraphics[width=7.5cm,height=5.15cm,clip]{HtTu_vs_zt_x8.pdf}
%\hspace{0.1cm}%
%\small{(f)}\includegraphics[width=7.5cm,height=5.15cm,clip]{EtTu_vs_zt_x8.pdf}
%\end{minipage}
%\begin{minipage}[c]{0.98\textwidth}
%\small{(g)}\includegraphics[width=7.5cm,height=5.15cm,clip]{HtTd_vs_zt_x8.pdf}
%\hspace{0.1cm}%
%\small{(h)}\includegraphics[width=7.5cm,height=5.15cm,clip]{EtTd_vs_zt_x8.pdf}
%\end{minipage}
%\caption{\label{gtx8}(Color online) Plots of chiral odd GPDs for the nonzero skewness vs $\zeta$ and different values of $-t$ in $GeV^2$, for fixed value of $x=0.8$. (a) $\widetilde{H}^u_T$, (b) $\widetilde{E}^u_T$ ; (c) and (d) same as (a) and (b) but for $d$ quark.}
%\end{figure}
%%%%%%%%%%%%%%%%%%%%%%%%%%%%%%%%%%%%%%%%%%%%%%%%%%%%%%%%%%%%%%%%%%%%%%%

Using the overlap representation of light front wave functions, we evaluate the chiral-odd GPDs in light front quark-diquark model. We restrict our discussion to the DGLAP domain, i.e., $\zeta<x<1$ where $\zeta$ is the skewness and $x$ is the light front longitudinal momentum fraction carried by the struck quark. This kinematical domain describes the diagonal $n\rightarrow n$ overlaps where the particle number remain conserved.  This region corresponds to the situation where one removes a quark from the initial proton  with light-front longitudinal momentum $(x+\zeta)P^+$ and re-insert it into the final proton with longitudinal momentum $(x-\zeta)P^+$. The diagonal $2\rightarrow 2$ overlap representation of the matrix elements $T^{q}_{\lambda\lambda'}$ and $\widetilde{T}^{q}_{\lambda\lambda'}$ in terms of light-front wave functions in the quark-diquark model are given by
\be
T^q_{\uparrow\uparrow}&=& \int \frac{d^2\bfk}{16\pi^3}~\bigg[\psi_{+q}^{+*}(x',\bfk')\psi_{-q}^-(x'',\bfk'') 
+\psi_{+q}^{-*}(x',\bfk')\psi_{-q}^+(x'',\bfk'')\bigg],\label{T1}\\
T^q_{\uparrow\downarrow}&=& \int \frac{d^2\bfk}{16\pi^3}~\bigg[\psi_{+q}^{+*}(x',\bfk')\psi_{-q}^+(x'',\bfk'') 
-\psi_{+q}^{-*}(x',\bfk')\psi_{-q}^-(x'',\bfk'')\bigg],\label{T2}\\
%\ee
%\be
\widetilde{T}^q_{\uparrow\uparrow}&=& \int \frac{d^2\bfk}{16\pi^3}~\bigg[\psi_{+q}^{+*}(x',\bfk')\psi_{-q}^+(x'',\bfk'') 
+\psi_{+q}^{-*}(x',\bfk')\psi_{-q}^-(x'',\bfk'')\bigg],\label{T3}\\
\widetilde{T}^q_{\uparrow\downarrow}&=& \int \frac{d^2\bfk}{16\pi^3}~\bigg[\psi_{+q}^{+*}(x',\bfk')\psi_{-q}^-(x'',\bfk'') 
-\psi_{+q}^{-*}(x',\bfk')\psi_{-q}^+(x'',\bfk'')\bigg],\label{T4}
\ee
where, for the final struck quark
\be
x'=\frac{x-\zeta}{1-\zeta}, \quad\quad\quad \bfk'=\bfk+(1-x')\frac{\bf{\Delta}_{\perp}}{2},
\ee
and for the initial struck quark
\be
x''=\frac{x+\zeta}{1+\zeta}, \quad\quad\quad \bfk''=\bfk-(1-x'')\frac{\bf{\Delta}_{\perp}}{2}.
\ee
The explicit calculation of the matrix elements $T^{q}_{\lambda\lambda'}$ and $\widetilde{T}^{q}_{\lambda\lambda'}$ using the light front wave functions of the quark-diquark model given in Eq.(\ref{wf}) gives   
\be
T^q_{\uparrow\uparrow}&=&\frac{1}{\kappa^2}\Big[{\frac{\log x'\log x''}{(1-x')(1-x'')}}\Big]^{1/2}\bigg[(N_q^{(1)})^2(x'x'')^{a_q^{(1)}}\{(1-x')(1-x'')\}^{b_q^{(1)}}\frac{1}{A}\nonumber\\
&-&(N_q^{(2)})^2\frac{1}{M_n^2}(x'x'')^{a_q^{(2)}-1}\{(1-x')(1-x'')\}^{b_q^{(2)}}
\bigg(\frac{B^2}{4A^2}-\frac{1}{4}(1-x')(1-x'')\nonumber\\
&+&\frac{B}{4A}(x''-x')\Big)\frac{Q^2}{A}\Big]
\exp\bigg[Q^2\Big(C-\frac{B^2}{4A}\Big)\bigg] ,\label{T1F}\\
%T^q_{\uparrow\downarrow}&=& ,\label{T2}\
\nonumber\\
T^q_{\uparrow\downarrow}&=&-\frac{N_q^{(1)}N_q^{(2)}}{\kappa^2}\Big[{\frac{\log x'\log x''}{(1-x')(1-x'')}}\Big]^{1/2}\frac{1}{M_n}\bigg[(x')^{a_q^{(1)}}(1-x')^{b_q^{(1)}}(x'')^{a_q^{(2)}-1}(1-x'')^{b_q^{(2)}}\nonumber\\
&\times&\bigg(\frac{BQ}{2A^2}-\frac{Q}{2A}(1-x'')\Big)+(x')^{a_q^{(2)}-1}(1-x')^{b_q^{(2)}}(x'')^{a_q^{(1)}}(1-x'')^{b_q^{(1)}}\nonumber\\
&\times&\bigg(\frac{BQ}{2A^2}+\frac{Q}{2A}(1-x'')\Big)\Big]
\exp\bigg[Q^2\Big(C-\frac{B^2}{4A}\Big)\bigg] \label{T2F},
%T^q_{\uparrow\downarrow}&=& ,\label{T2}
\ee
%%%%%%%%%%%%%%%%%%%%%%%%%%%%%%%%
%%%%%%%%%%%%%%%%%%%%%%%%%%%%%%%%
\be
\widetilde{T}^q_{\uparrow\uparrow}&=&-\frac{N_q^{(1)}N_q^{(2)}}{\kappa^2}\Big[{\frac{\log x'\log x''}{(1-x')(1-x'')}}\Big]^{1/2}\frac{1}{M_n}\bigg[(x')^{a_q^{(1)}}(1-x')^{b_q^{(1)}}(x'')^{a_q^{(2)}-1}(1-x'')^{b_q^{(2)}}\nonumber\\
&\times&\bigg(\frac{BQ}{2A^2}-\frac{Q}{2A}(1-x'')\Big)-(x')^{a_q^{(2)}-1}(1-x')^{b_q^{(2)}}(x'')^{a_q^{(1)}}(1-x'')^{b_q^{(1)}}\nonumber\\
&\times&\bigg(\frac{BQ}{2A^2}+\frac{Q}{2A}(1-x'')\Big)\Big]
\exp\bigg[Q^2\Big(C-\frac{B^2}{4A}\Big)\bigg] ,\label{T3F}\\
%T^q_{\uparrow\downarrow}&=& ,\label{T2}\
\nonumber\\
\widetilde{T}^q_{\uparrow\downarrow}&=&\frac{1}{\kappa^2}\Big[{\frac{\log x'\log x''}{(1-x')(1-x'')}}\Big]^{1/2}\bigg[(N_q^{(1)})^2(x'x'')^{a_q^{(1)}}\{(1-x')(1-x'')\}^{b_q^{(1)}}\frac{1}{A}\nonumber\\
&+&(N_q^{(2)})^2\frac{1}{M_n^2}(x'x'')^{a_q^{(2)}-1}\{(1-x')(1-x'')\}^{b_q^{(2)}}
\bigg(\frac{B^2}{4A^2}-\frac{1}{4}(1-x')(1-x'')\nonumber\\
&+&\frac{B}{4A}(x''-x')\Big)\frac{Q^2}{A}\Big]
\exp\bigg[Q^2\Big(C-\frac{B^2}{4A}\Big)\bigg] ,
%T^q_{\uparrow\downarrow}&=& ,\label{T2}
\label{T4F}\ee
where ${{\Delta}}_{\perp}^2=Q^2=-t(1-\zeta^2)-4M_n^2\zeta^2$. $A$, $B$ and $C$ are  functions of $x'$ and $x''$,
\be
A&=&A(x',x'')=-\frac{\log x'}{2\kappa^2(1-x')^2}-\frac{\log x''}{2\kappa^2(1-x'')^2},\nonumber\\
B&=&B(x',x'')=\frac{\log x'}{2\kappa^2(1-x')}-\frac{\log x''}{2\kappa^2(1-x'')},\\
C&=&C(x',x'')=\frac{1}{4}\Big[\frac{\log x'}{2\kappa^2}+\frac{\log x''}{2\kappa^2}\Big].\nonumber
\ee
Using the matrix elements calculated in Eqs.(\ref{T1F}-\ref{T4F}) we evaluate the chiral-odd GPDs in Eqs.(\ref{HT}-\ref{EtT}). All the GPDs are suitably scaled by the flavor factors $P_q$ where $P_u=\frac{4}{3}$, and $P_d=-\frac{1}{3}$ are dictated by SU(6) spin-flavor symmetry \cite{Karl}.

In Fig.\ref{gz0} we show the $t$ dependence chiral-odd GPDs $H^q_T$, $\widetilde{H}^q_T$ and $E^q_T$ for up and down quarks in the quark-diquark model when the skewness $\zeta=0$ . Being an odd function of $\zeta$, the GPD $\tilde E^q_T$ vanishes at $\zeta=0$ in this model. The similar behavior of $\tilde E^q_T$ has been reported in~\cite{diehl01, Pasquini1}. One can notice that the sign of all three GPDs for $u$ quark are opposite with respect to $d$ quark and $\widetilde{H}^q_T$ shows opposite sign of ${H}^q_T(x,0,t)$ as expected from $SU(6)$ symmetry. The peaks of all the distributions move to higher values of $x$ as $-t$ increases. For $\zeta\ne 0$, all the four GPDs are shown in Fig.\ref{gz15} and \ref{gt7}. In Fig.\ref{gz15} the GPDs are shown for fixed value of $\zeta=0.15$ but for different values of $-t$. In Fig.\ref{gt7}, we plot the GPDs for fixed value of $-t=0.7~ \rm{GeV^2}$ and different values of $\zeta$. One can notice that the height of the peaks of the distributions increase  and shift to 
higher $x$ 
with increasing $\zeta$ for fixed $-t$. In all cases, the GPDs vanish at $x=\zeta$. The reason is that in our approach we consider only the contribution from the valence quarks. In this model we can not evaluate the total (sea+valence) GPDs as the model itself depends only on the valence quarks.  %Physically the GPDs should not vanish at $\bar x=\zeta$ 
The similar behavior of the chiral-odd GPDs has been found in the relativistic constituent quark model calculated in~\cite{Pasquini1}. Also, the region for $x<\zeta$, the so called ERBL region, where quark-antiquark pair creation/annihilation are involved are not included in this model.
In Fig.\ref{gx8}, we have shown the GPDs as functions of $\zeta$ for different values of $-t$ and fixed $x$. It can be noticed that only $\widetilde{E}^q_T(x,\zeta,t)$ ( Fig.\ref{gx8}(g) and  Fig.\ref{gx8}(h)) shows markedly different behavior from the other GPDs. $\widetilde{E}^q_T(x,\zeta,t)$ rises smoothly from zero as $\zeta$ increases for all $t$ values whereas the other GPDs have   different values at $\zeta=0$ for different values of $-t$. The similar behaviors  of the chiral-odd GPDs  have been observed  \cite{CMM1} in a QED model.

%%%%%%%%%%%%%%%%%%%%%%%%%%%%%%%%%%%%%%%%%%%%
\subsection{Mellin moments of chiral-odd GPDs}
%%%%%%%%%%%%%%%%%%%%%%%%%%%%%%%%%%%%%%%%%%%%
\begin{figure}[htbp]
\begin{minipage}[c]{0.98\textwidth}
\small{(a)}
\includegraphics[width=7.5cm,height=5.15cm,clip]{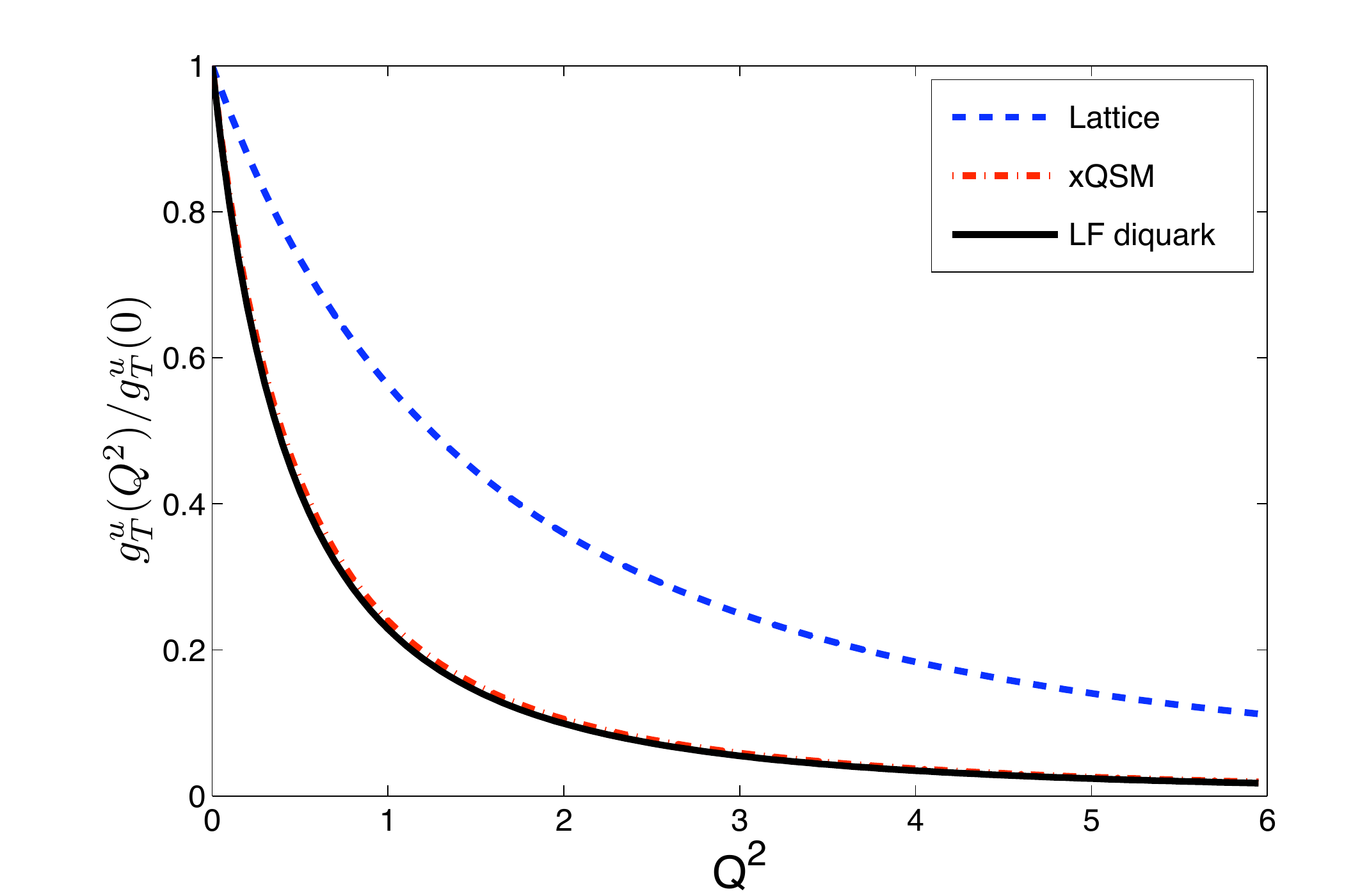}
\hspace{0.1cm}%
\small{(b)}\includegraphics[width=7.5cm,height=5.15cm,clip]{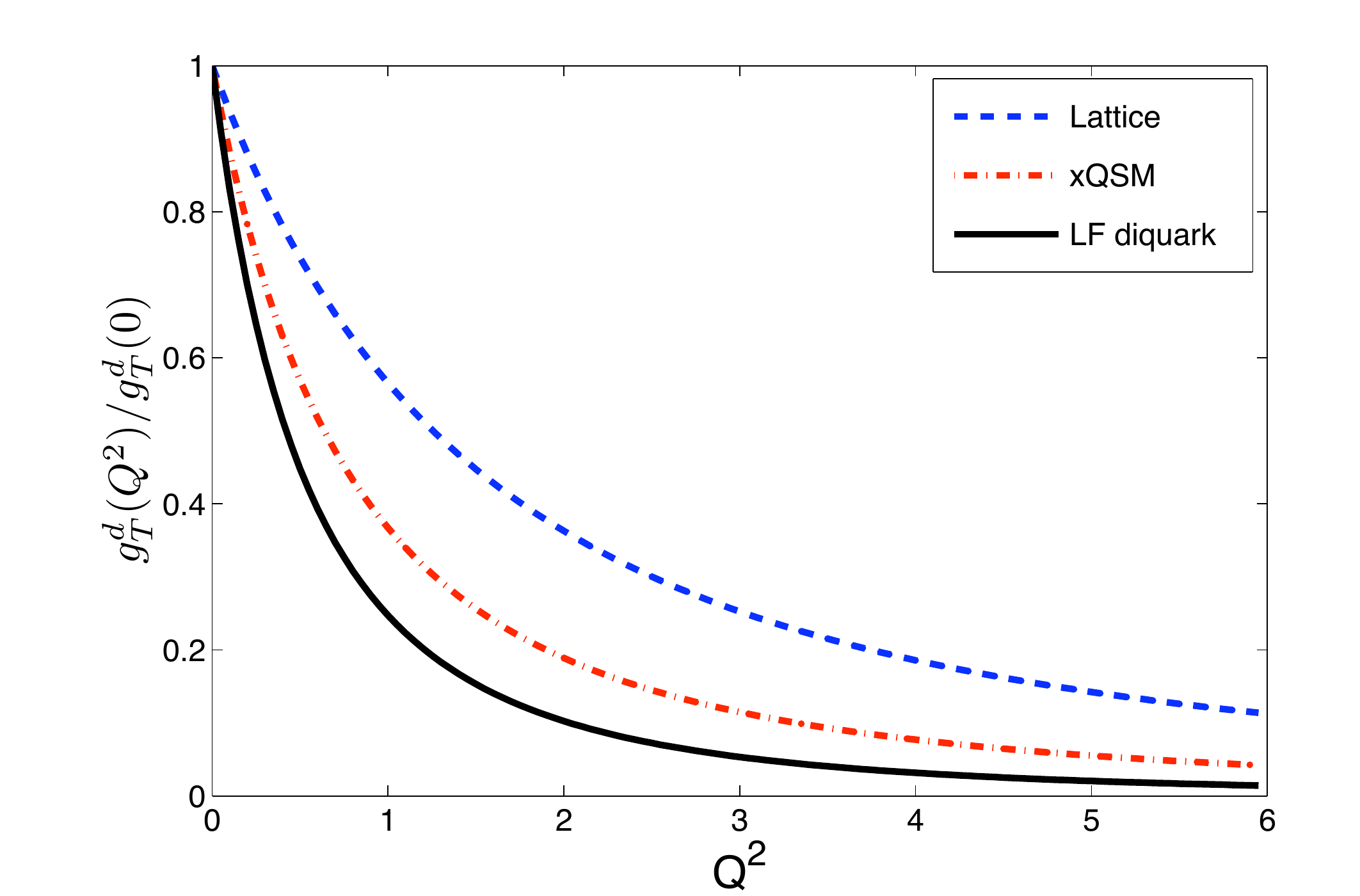}
\end{minipage}
\begin{minipage}[c]{0.98\textwidth}
\small{(c)}\includegraphics[width=7.5cm,height=5.15cm,clip]{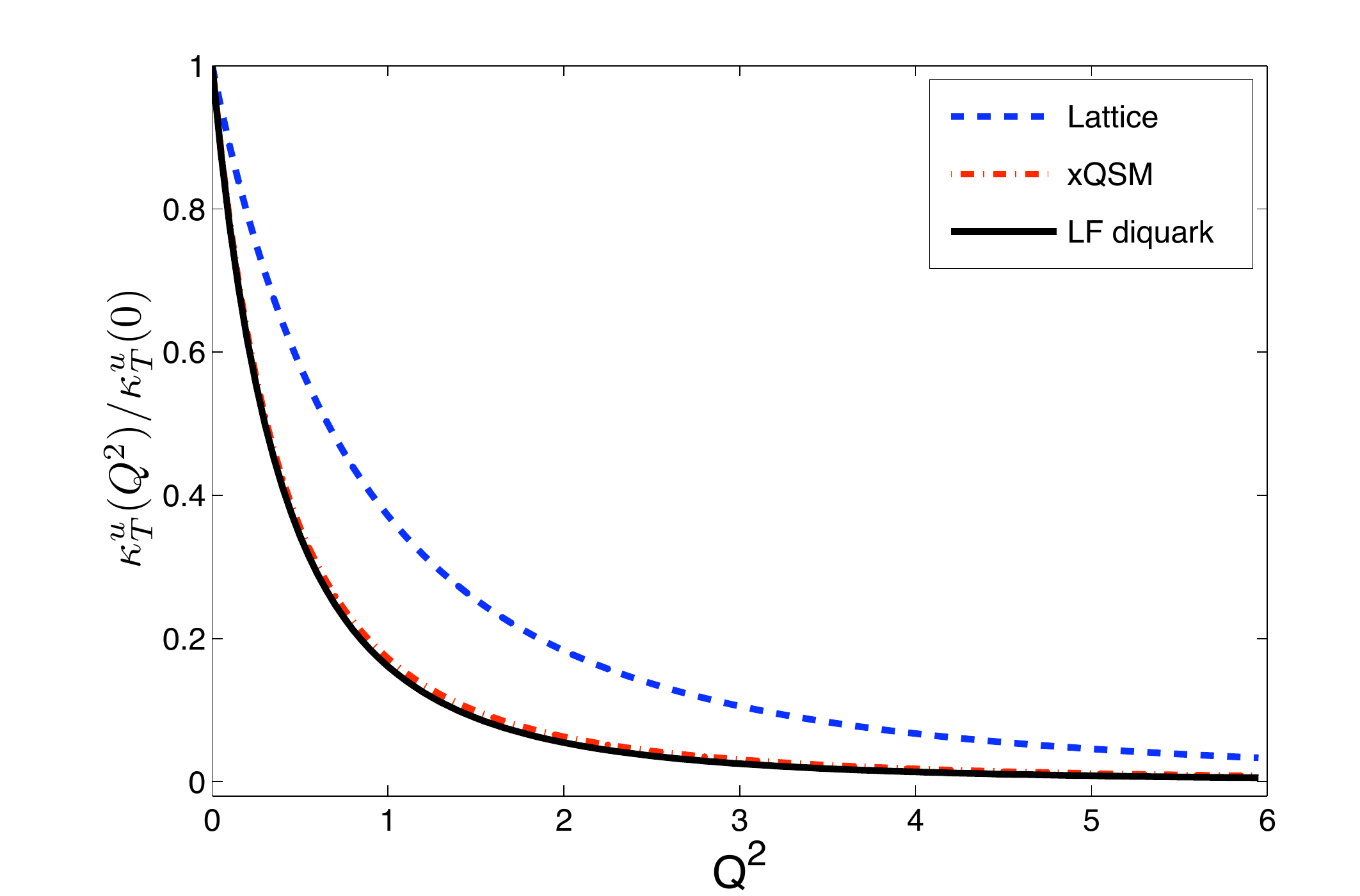}
\hspace{0.1cm}%
\small{(d)}\includegraphics[width=7.5cm,height=5.15cm,clip]{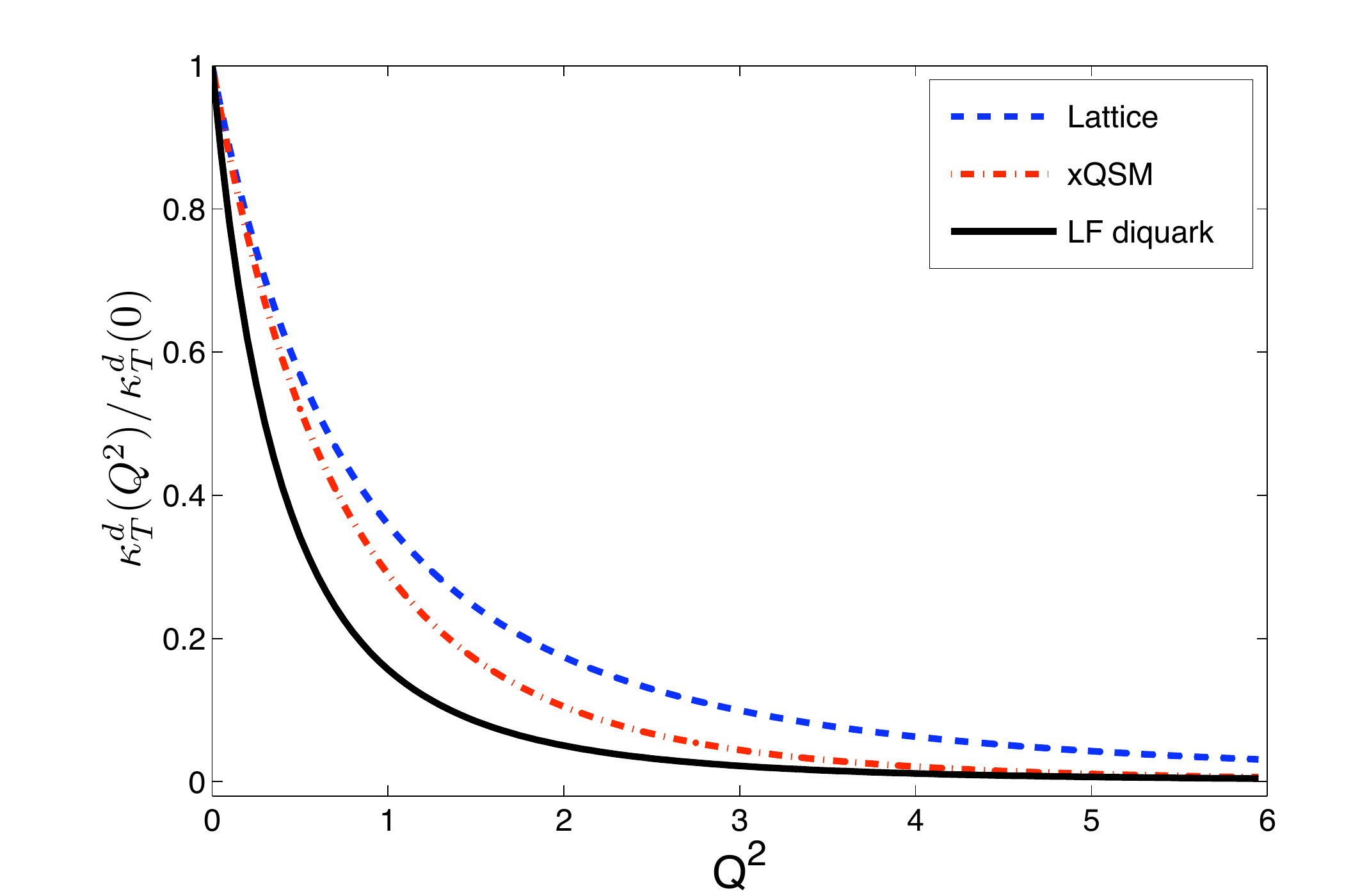}
\end{minipage}
\caption{\label{tensor}(Color online) Tensor form factors for $u$ and $d$ quarks are compared with the lattice~\cite{gock} and chiral quark-soliton model (xQSM)~\cite{xQSM1,xQSM2} results.}
\end{figure}
%%%%%%%%%%%%%%%%%%%
\begin{figure}[htbp]
\begin{minipage}[c]{0.98\textwidth}
\small{(a)}
\includegraphics[width=7.5cm,height=5.15cm,clip]{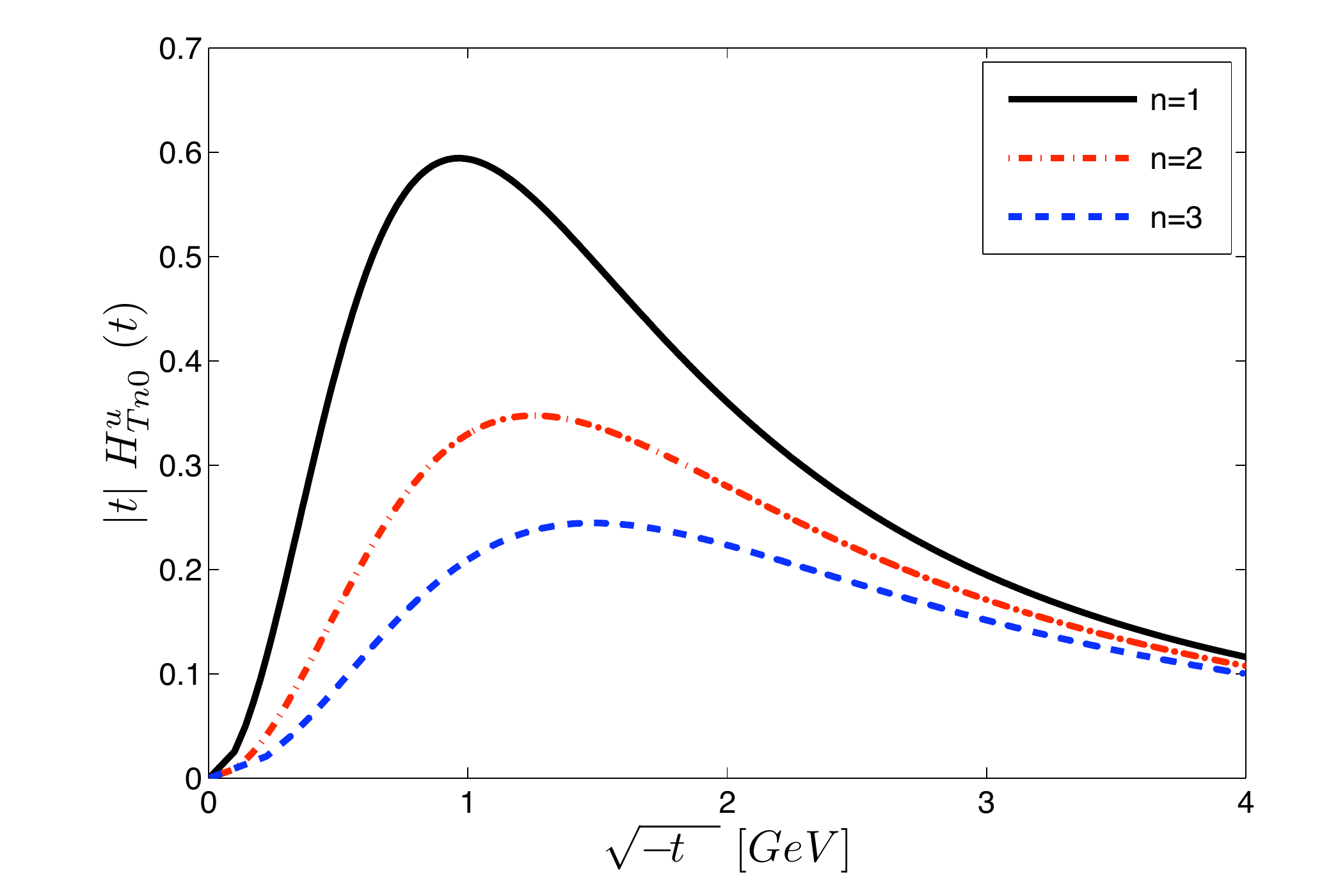}
\hspace{0.1cm}%
\small{(b)}\includegraphics[width=7.5cm,height=5.15cm,clip]{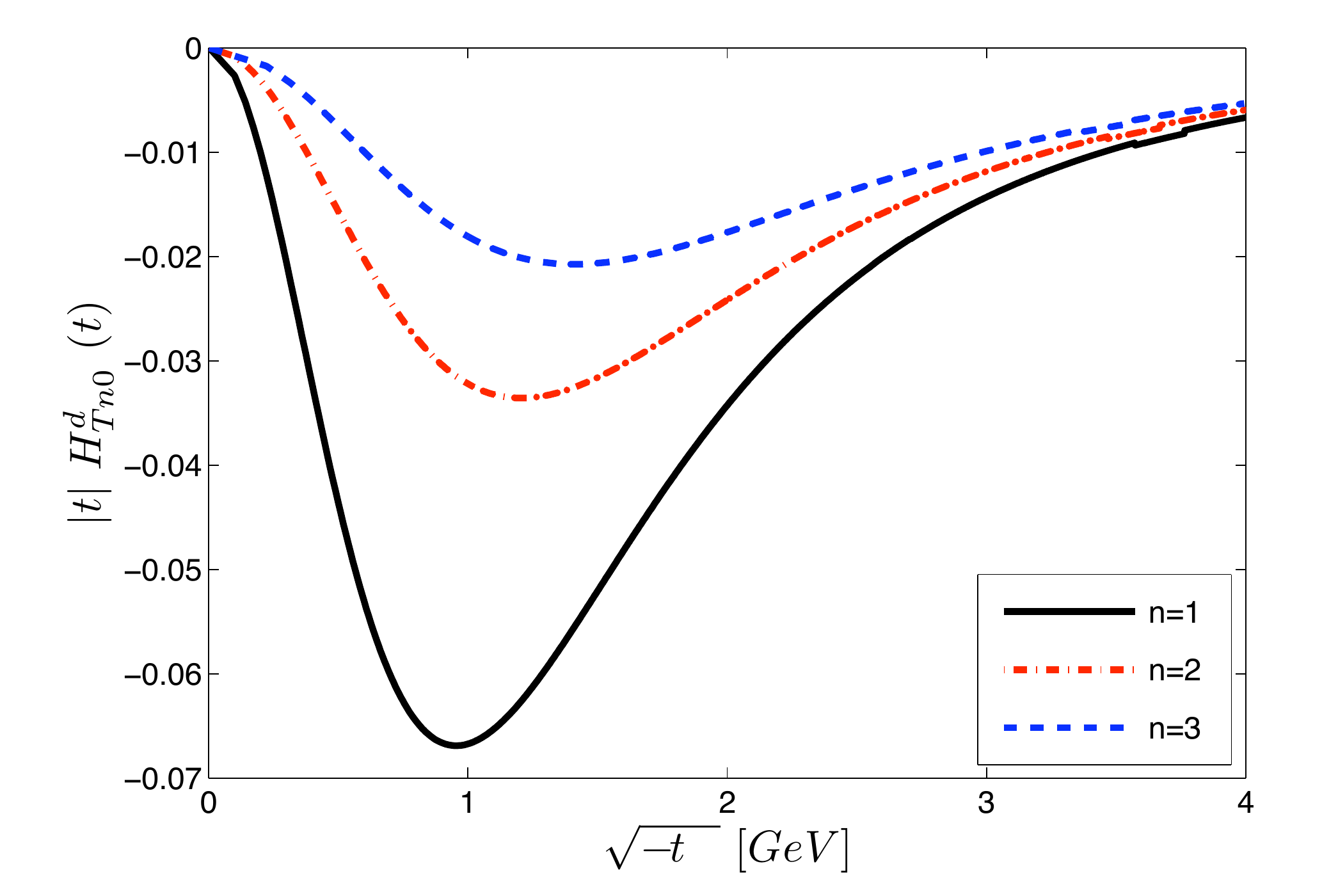}
\end{minipage}
\begin{minipage}[c]{0.98\textwidth}
\small{(c)}\includegraphics[width=7.5cm,height=5.15cm,clip]{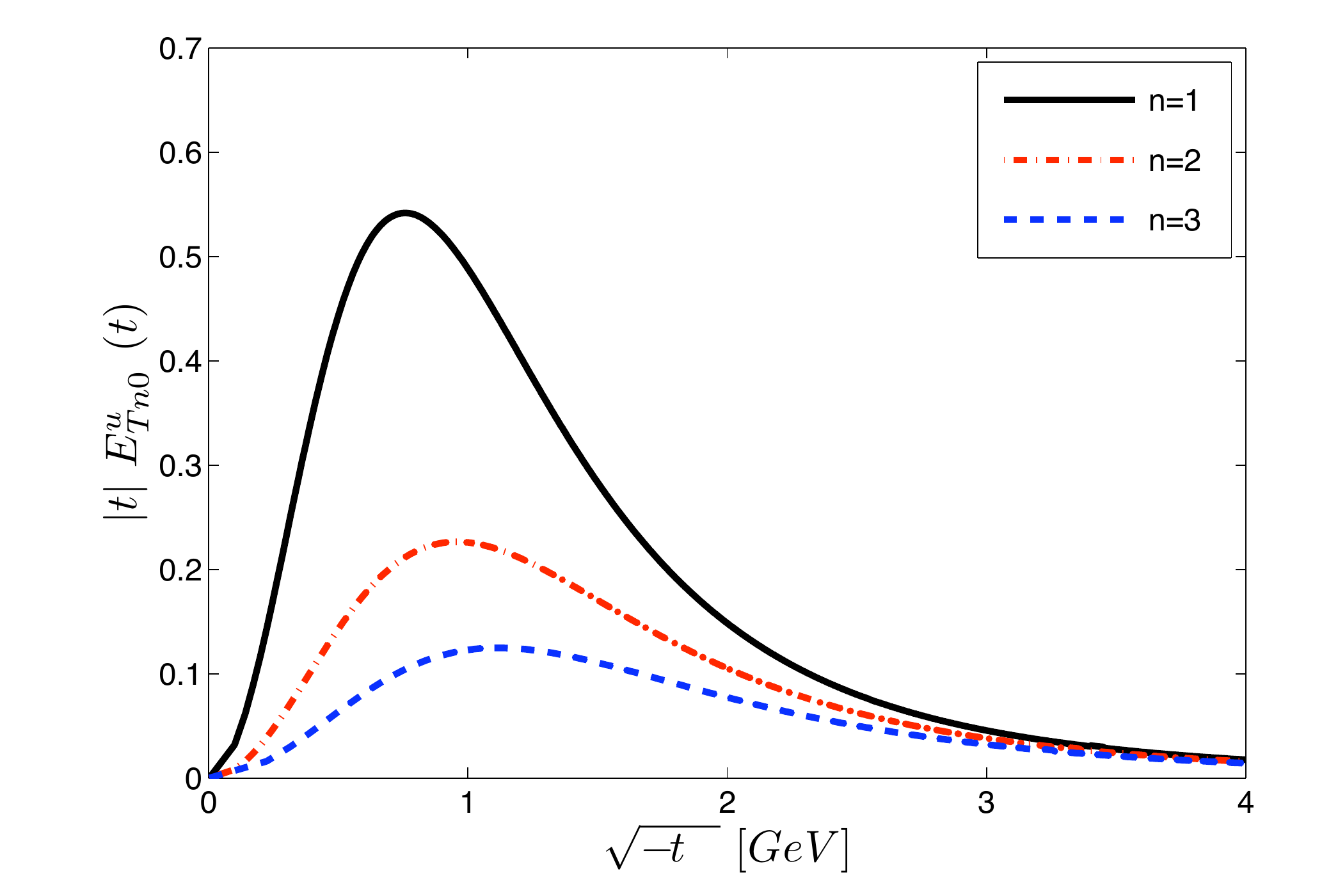}
\hspace{0.1cm}%
\small{(d)}\includegraphics[width=7.5cm,height=5.15cm,clip]{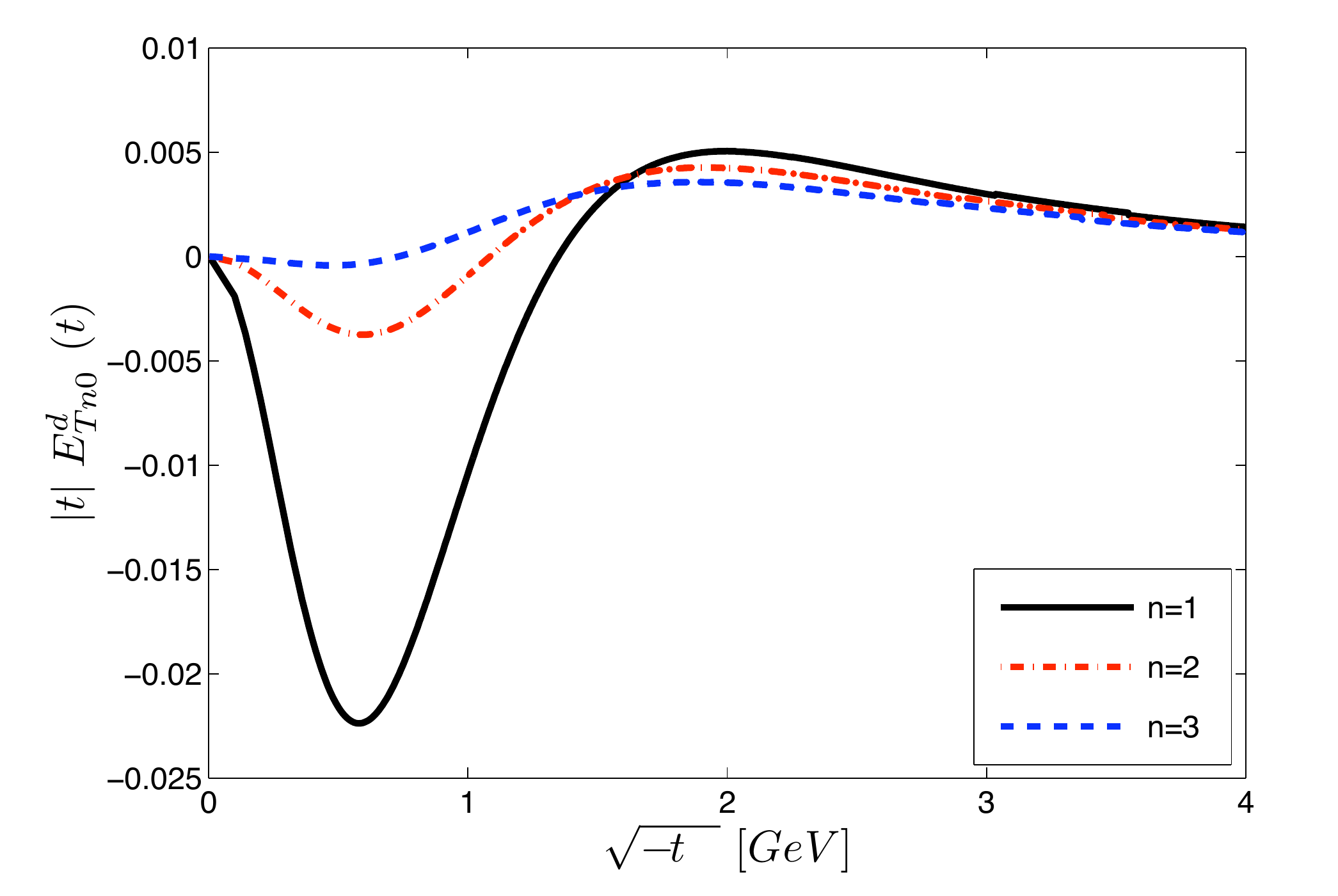}
\end{minipage}
\begin{minipage}[c]{0.98\textwidth}
\small{(e)}\includegraphics[width=7.5cm,height=5.15cm,clip]{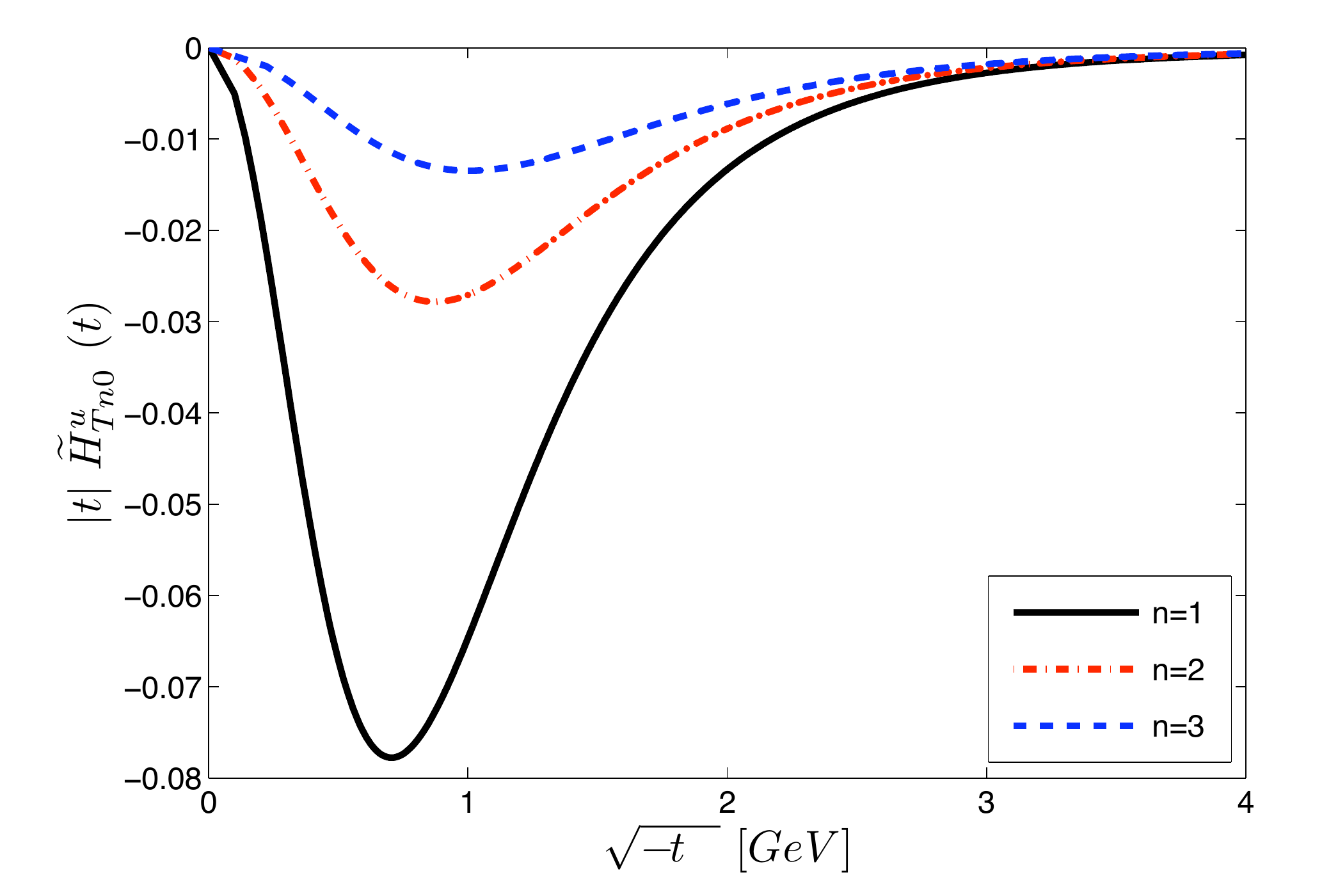}
\hspace{0.1cm}%
\small{(f)}\includegraphics[width=7.5cm,height=5.15cm,clip]{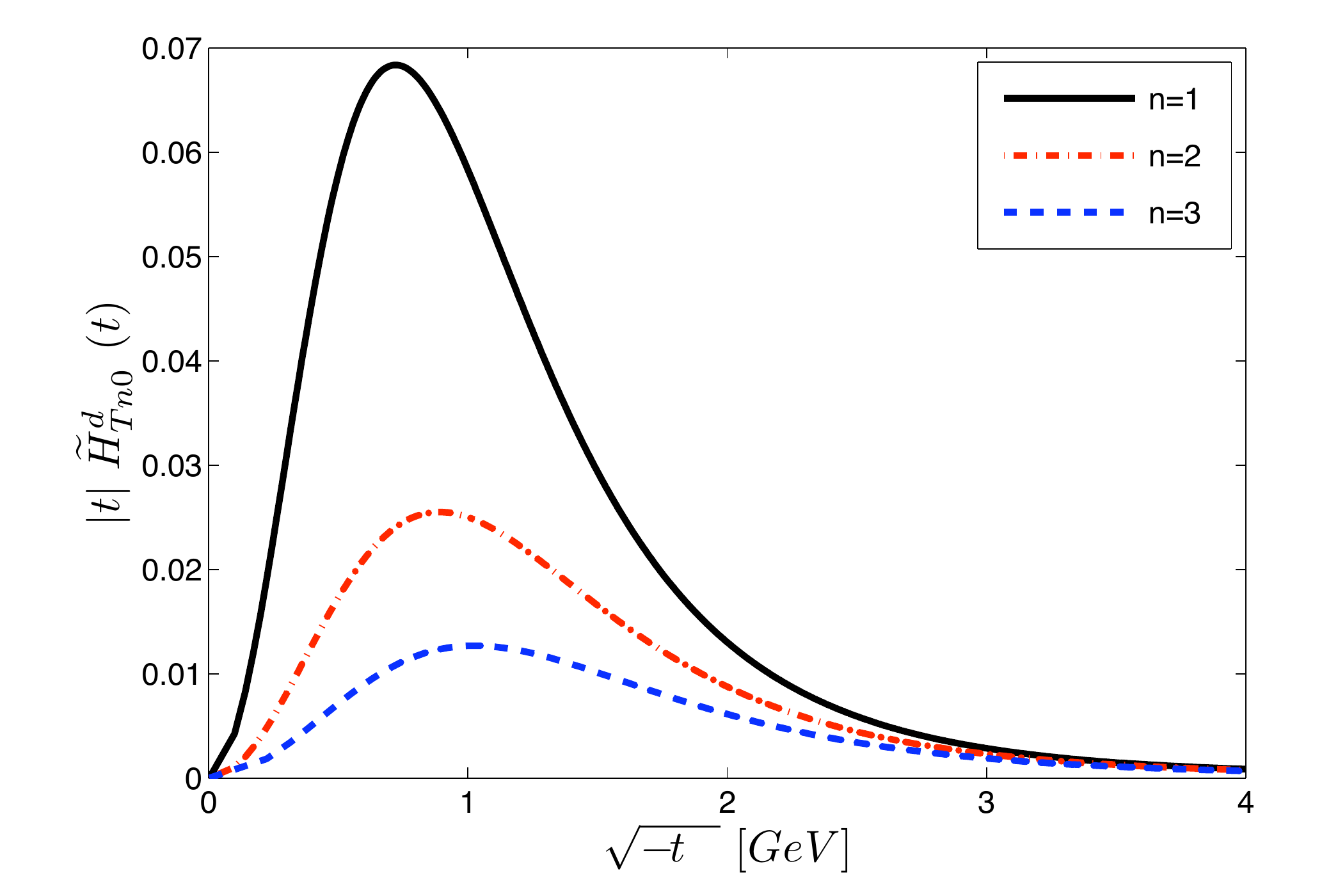}
\end{minipage}
\caption{\label{moments}(Color online) Plots of first three moments of the chiral odd GPDs for zero skewness vs $\sqrt{-t}$ in $GeV$. Left pannel is for $u$ quark and the right pannel is for $d$ quark.}
\end{figure}
%%%%%%%%%%%%%%%%%%%%%%%%%%%%%%%%%%%%%%%%%%%%%%%%%%

%We now use these chiral-odd GPDs to compute higher order moments in $x$. 
The Mellin moments of the valence GPDs are defined as
\be
H^q_{Tn0}(t)=\int_0^1~dx x^{n-1}H^q_{T}(x,0,t),\label{moment_formula}
\ee
where the index $n=1,2,3$ etc., and the second subscript indicates that the moments are evaluated at zero skewness. The moments of  the other GPDs, $E^q_{Tn0}(t)$ and $\widetilde{H}^q_{Tn0}(t)$ can also be defined in the same way as (\ref{moment_formula}).
The first moments of chiral-odd GPDs give the tensor form factors. The forward values, $t = 0$, of the form factor $g_T=H_{T10}(t=0)$ can be identified as the tensor charge~\cite{Hagler:2009}. The combination of tensor form factors $\bar{E}^q_{T10}=(E^q_{T10}+2\widetilde{H}^q_{T10})$ in the forward limit plays a role very similar to that of anomalous magnetic moment $\kappa^q$ and therefore may be identified with a tensor magnetic moment, $\kappa^q_T=\bar{E}^q_{T10}(t=0)$~\cite{Burk3}. In Fig.(\ref{tensor}), we have compared our result for the tensor form factors with the corresponding results from lattice~\cite{gock} and chiral quark soliton model \cite{xQSM1,xQSM2}. The tensor form factors for $u$ quark in this model  agrees well with the chiral quark-soliton model ($\chi$QSM) but both(this model and $\chi$QSM model) deviate from lattice results for both $u$ and $d$ quarks.  
The second moments of these GPDs correspond to the gravitational form factors of quarks with transverse spin in an unpolarized nucleon.  A linear combination of $H^q_{T20}(t)$, $E^q_{T20}(t)$ and $\widetilde{H}^q_{T20}(t)$ gives the angular momentum carried by quarks with transverse spin in an unpolarized nucleon\cite{Burk3}, in analogy to Ji's angular momentum sum rule. The third moments of the GPDs generate form factors of a twist-two operator having
two covariant derivatives~\cite{rev} and the higher order moments give the form factors of higher-twist operators. 

In Fig.\ref{moments} we show the first three moments of the chiral-odd GPDs $|t|H^q_{Tn0}(t)$, $|t|E^q_{Tn0}(t)$ and $|t|\widetilde{H}^q_{Tn0}(t)$ as functions of $\sqrt{-t}$ for $u$ and $d$ quarks. We find a strong decrease in the  magnitudes of the moments with increasing  $n$. 
This can be understood from the behavior of the GPDs with $x$ as shown in Fig.\ref{gz0}. Higher moments involve higher power of $x$ and hence the dominant contributions come from the large $x$ region($x\to1$), but  the GPDs decrease rapidly as $x$ increases, and hence the higher moments become smaller.
%The overall nature of the moments with $-t$ is the same as the behavior of GPDs with $x$.
We also observe that as the index $n$ increases, the decrease of the moments becomes slower with increasing $-t$. This again can be  explained in terms of the decrease of the GPDs with momentum fraction $x$, which results in a weaker $t$ slope for the higher moments. The similar behavior has been observed in lattice QCD calculations of the moments of chiral-odd GPD~\cite{gock}.  

%%%%%%%%%%%%%%%%%%%%%%%%%%%%%%%%%%%%%%%%%%%%%%%%%%%%%%%%%%%%%%%%%%%%%%
\section{Impact parameter representation of chiral-odd GPDs}\label{chiral_impact}
%%%%%%%%%%%%%%%%%%%%%%%%%%%%%%%%%%%%%%%%%%%%%%%%%%%%%%%%%%%%%%%%%%%%%%%
%%%%%%%%%%%%%%%%%%%..zeta dependent..xb..%%%%%%%%
\begin{figure}[htbp]
\begin{minipage}[c]{0.98\textwidth}
\small{(a)}
\includegraphics[width=7.5cm,height=5.15cm,clip]{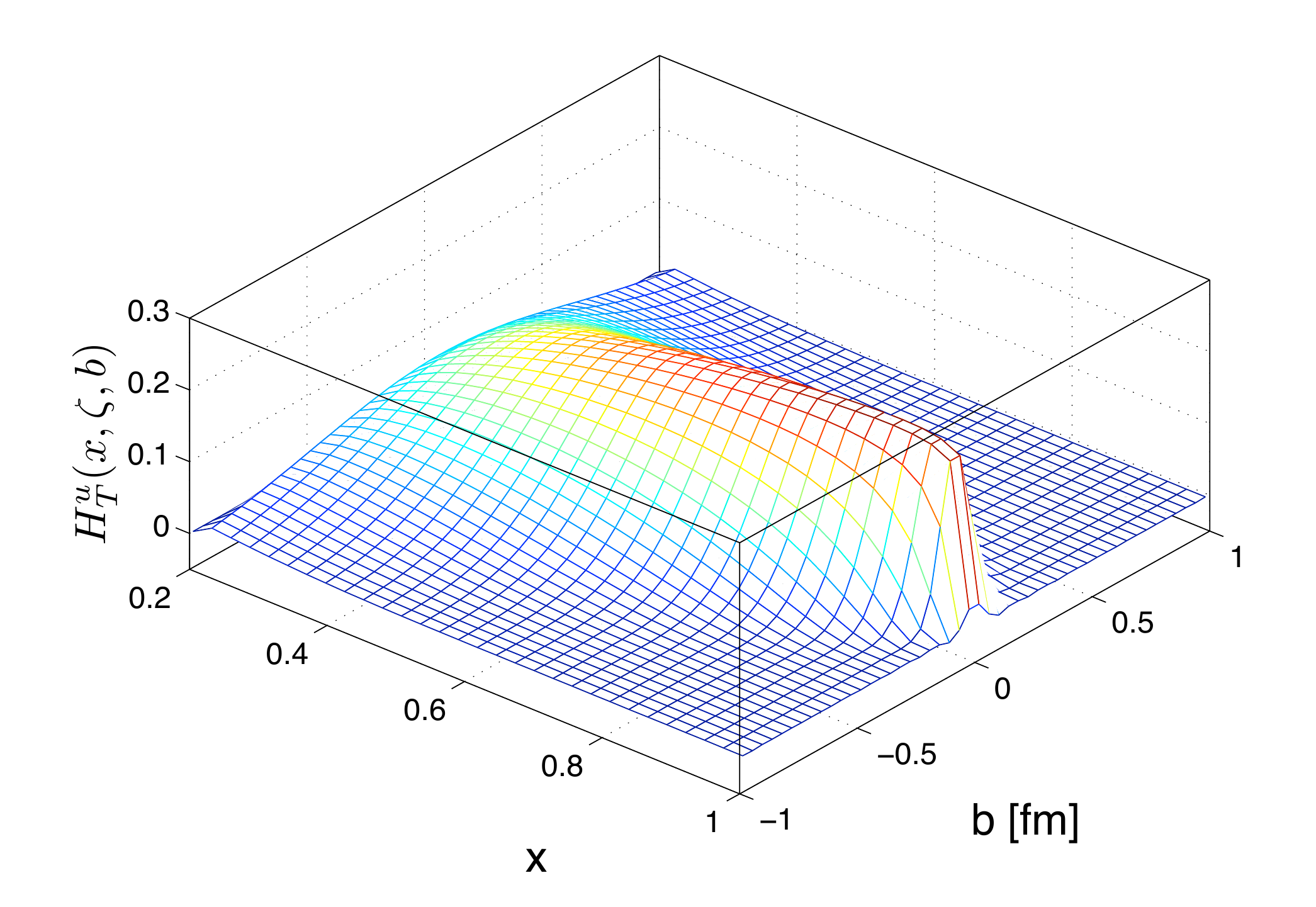}
\hspace{0.1cm}%
\small{(b)}\includegraphics[width=7.5cm,height=5.15cm,clip]{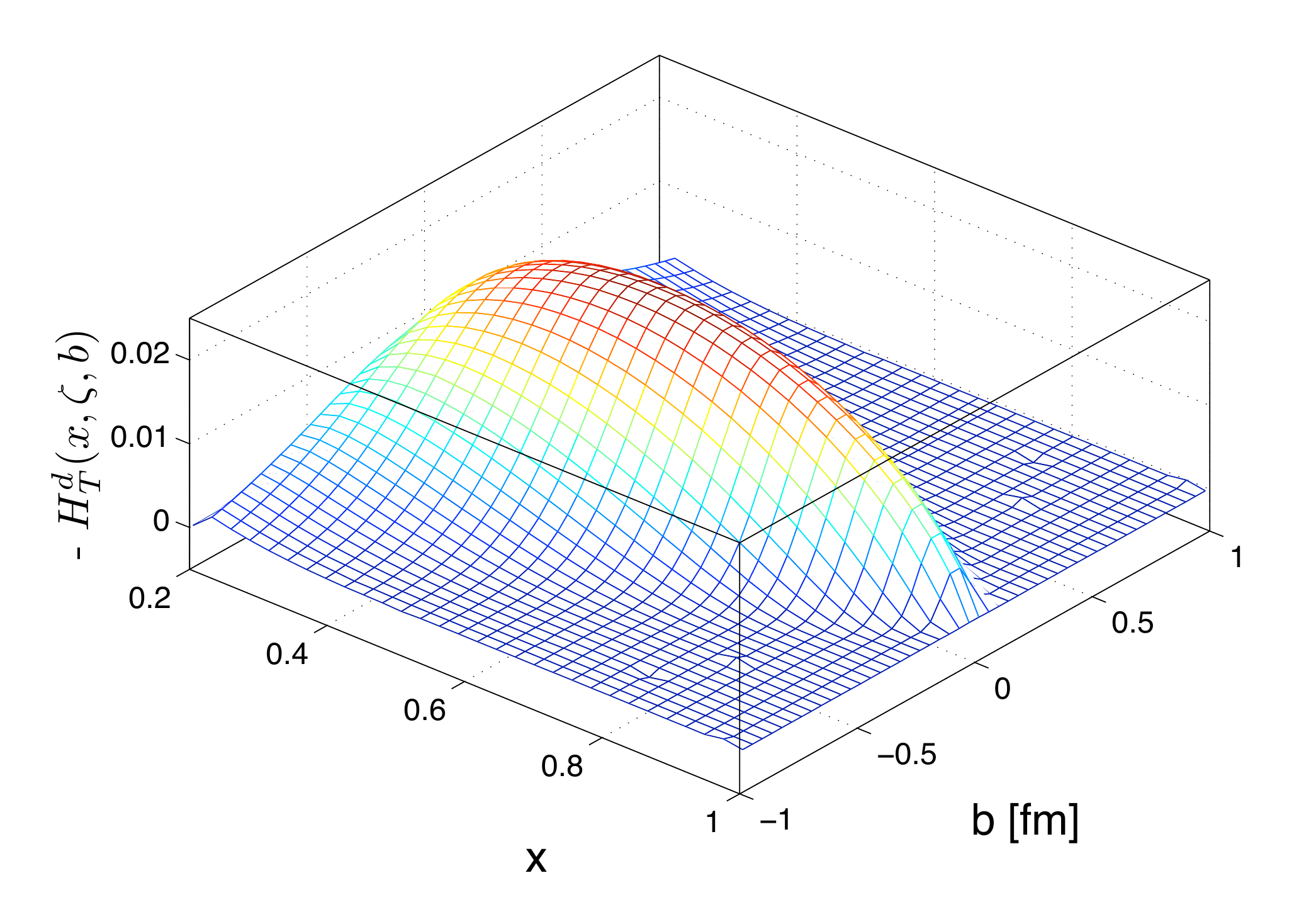}
\end{minipage}
\begin{minipage}[c]{0.98\textwidth}
\small{(c)}\includegraphics[width=7.5cm,height=5.15cm,clip]{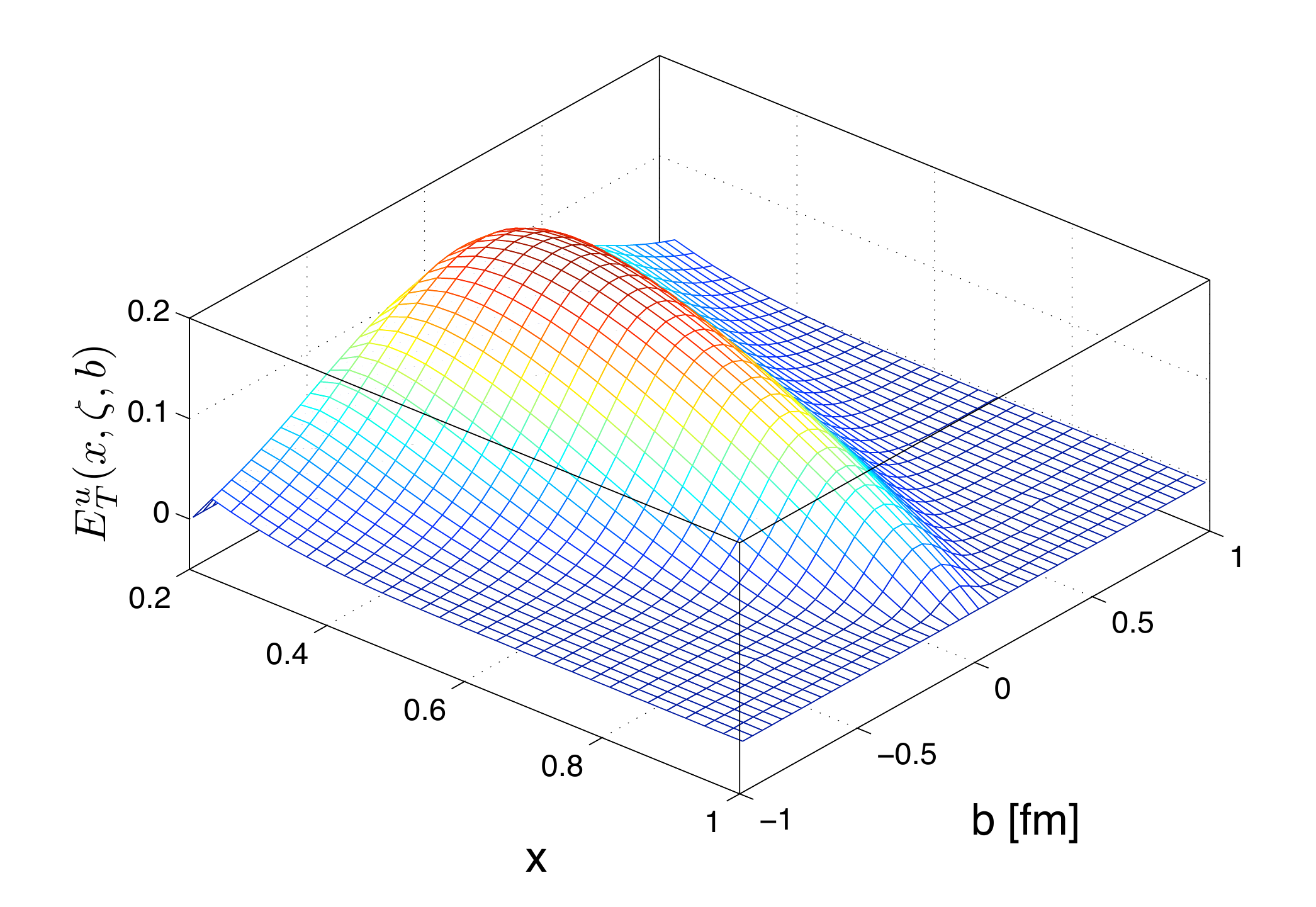}
\hspace{0.1cm}%
\small{(d)}\includegraphics[width=7.5cm,height=5.15cm,clip]{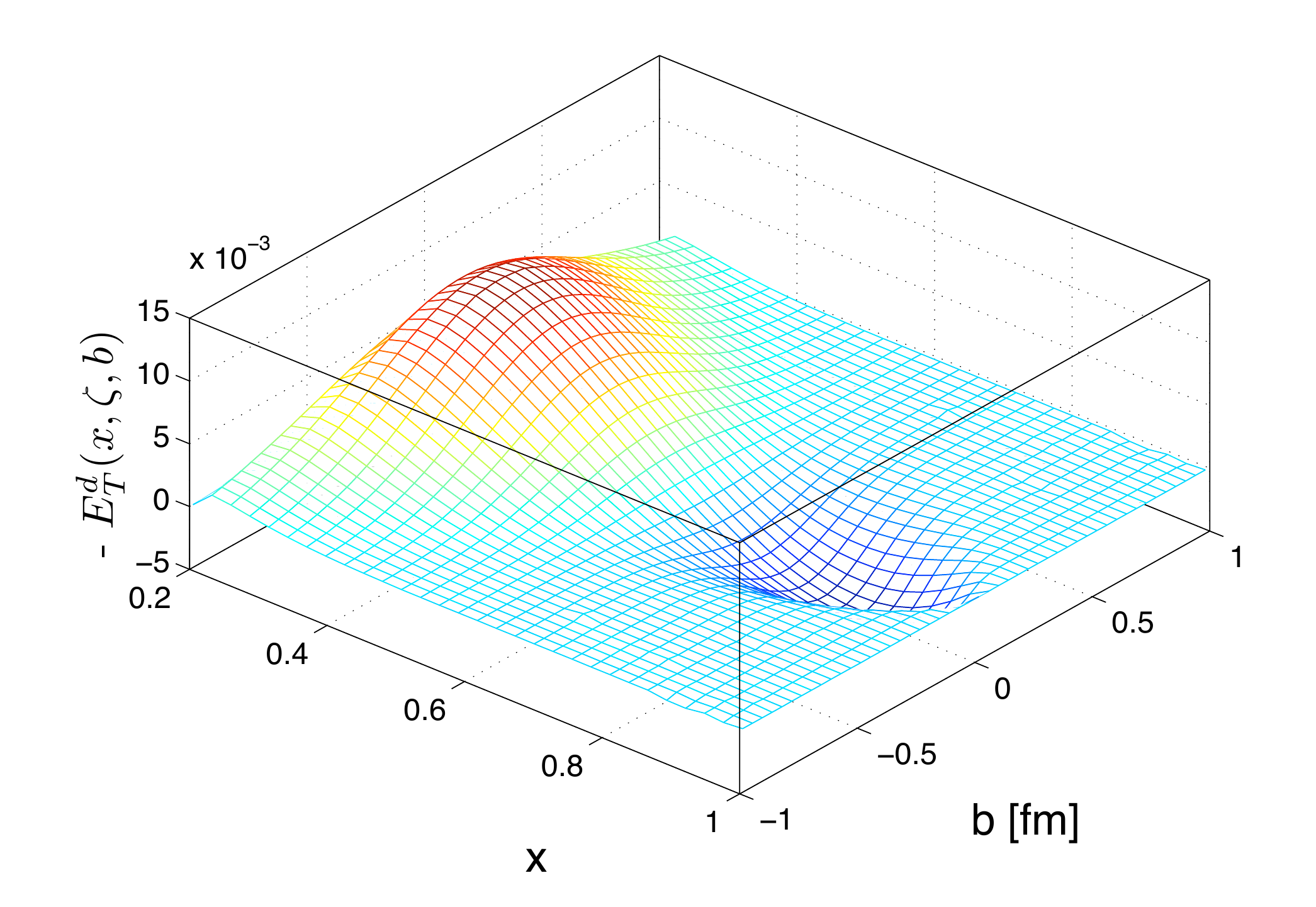}
\end{minipage}
\begin{minipage}[c]{0.98\textwidth}
\small{(e)}\includegraphics[width=7.5cm,height=5.15cm,clip]{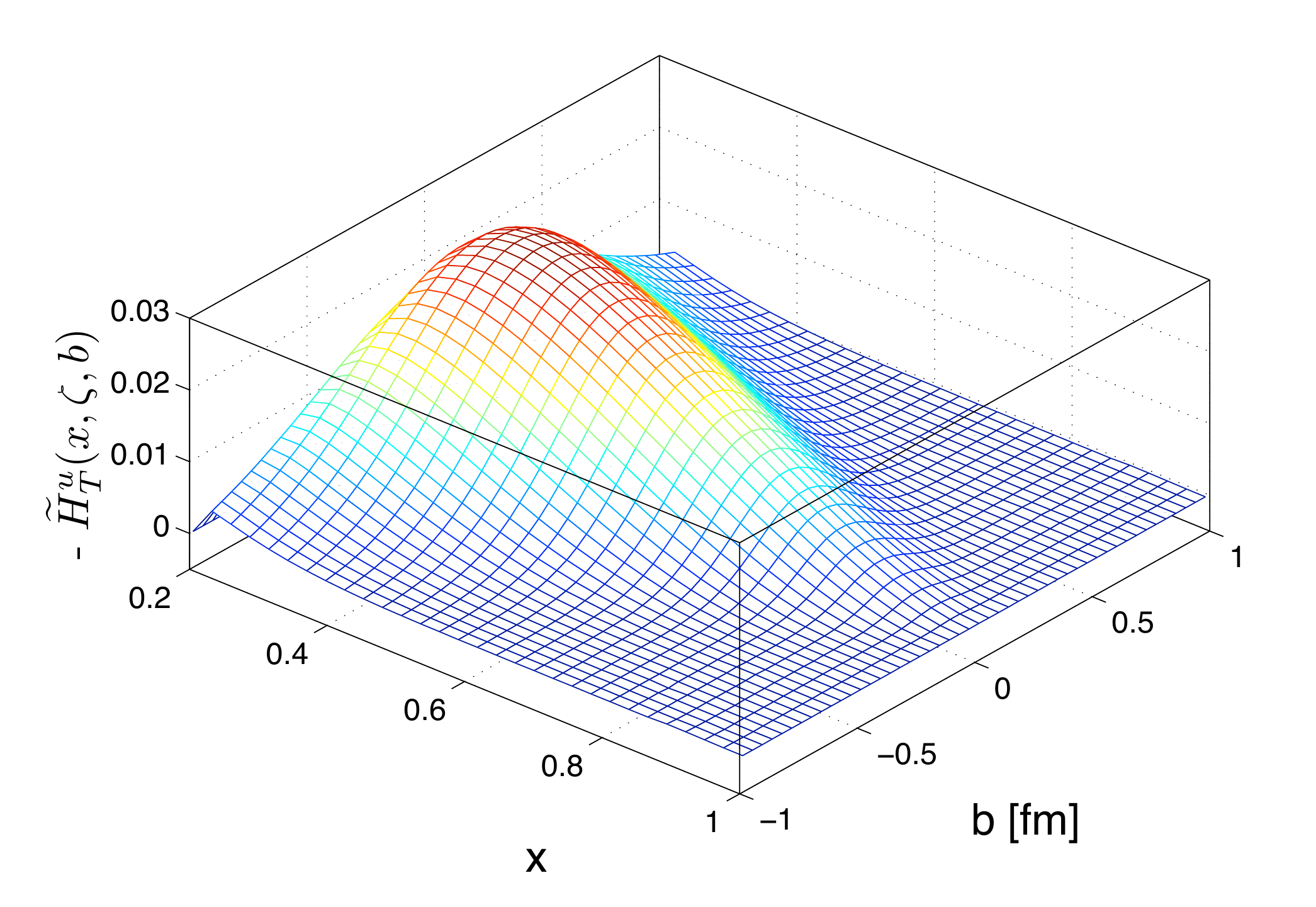}
\hspace{0.1cm}%
\small{(f)}\includegraphics[width=7.5cm,height=5.15cm,clip]{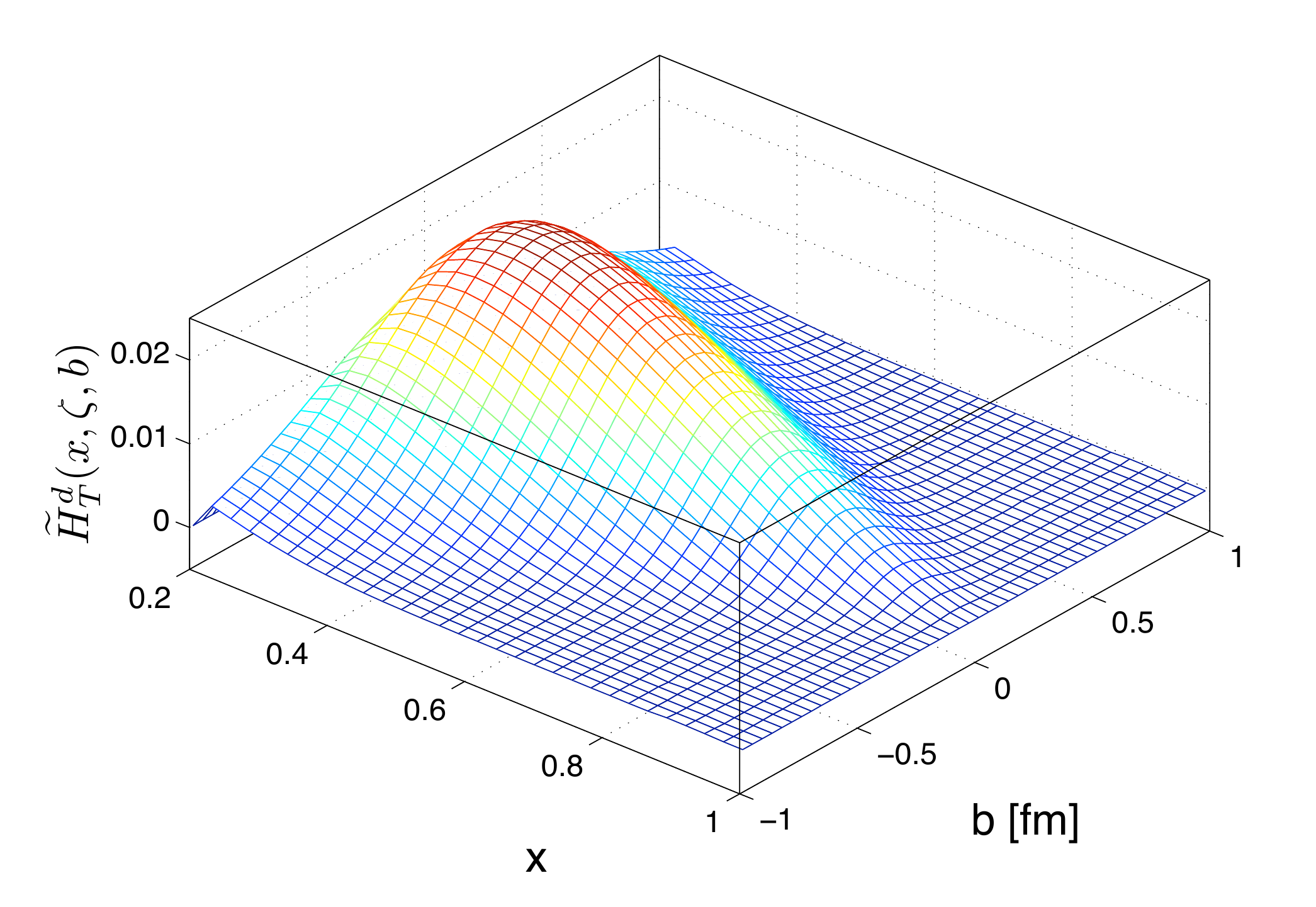}
\end{minipage}
\begin{minipage}[c]{0.98\textwidth}
\small{(g)}\includegraphics[width=7.5cm,height=5.15cm,clip]{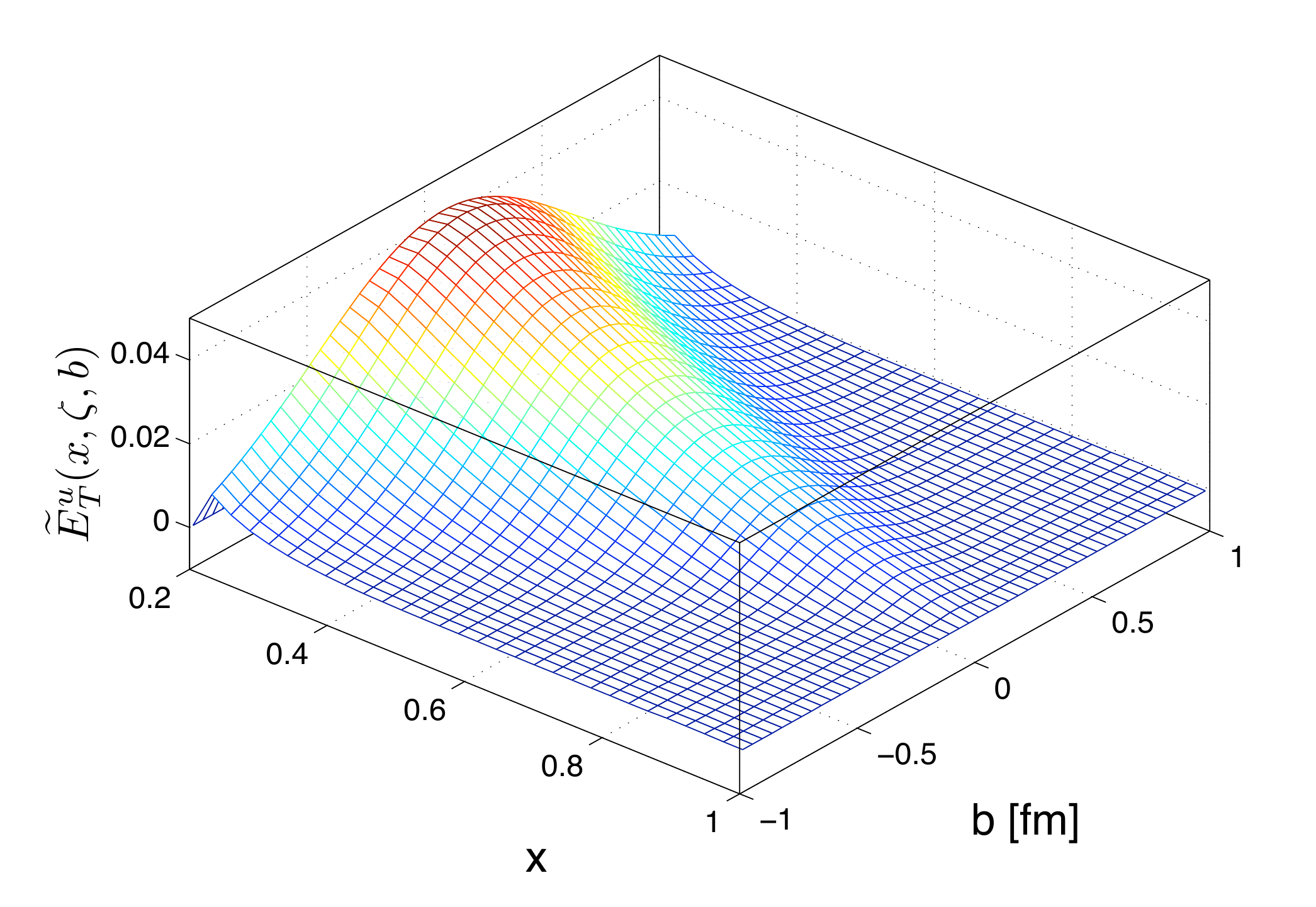}
\hspace{0.1cm}%
\small{(h)}\includegraphics[width=7.5cm,height=5.15cm,clip]{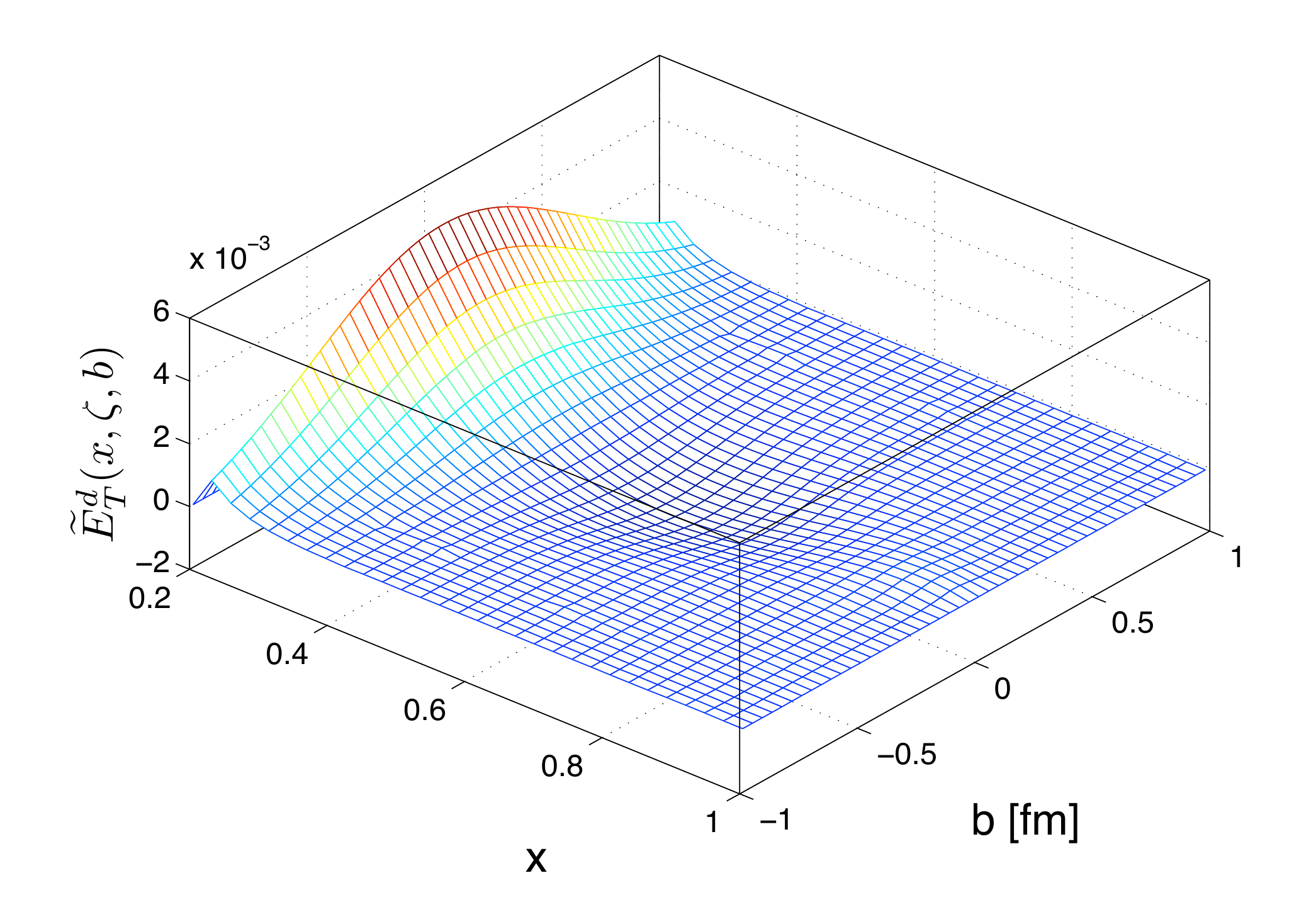}
\end{minipage}
\caption{\label{giz2}(Color online) Plots of chiral odd GPDs for the nonzero skewness in impact space vs $x$ and $b=|\bf b|$ for fixed value of $\zeta=0.2$. Left pannel is for $u$ quark and the right pannel is for $d$ quark.}
\end{figure}
%%%%%%%%%%%%%%%%%%%..zeta dependent..zb..%%%%%%%%
\begin{figure}[htbp]
\begin{minipage}[c]{0.98\textwidth}
\small{(a)}
\includegraphics[width=7.5cm,height=5.15cm,clip]{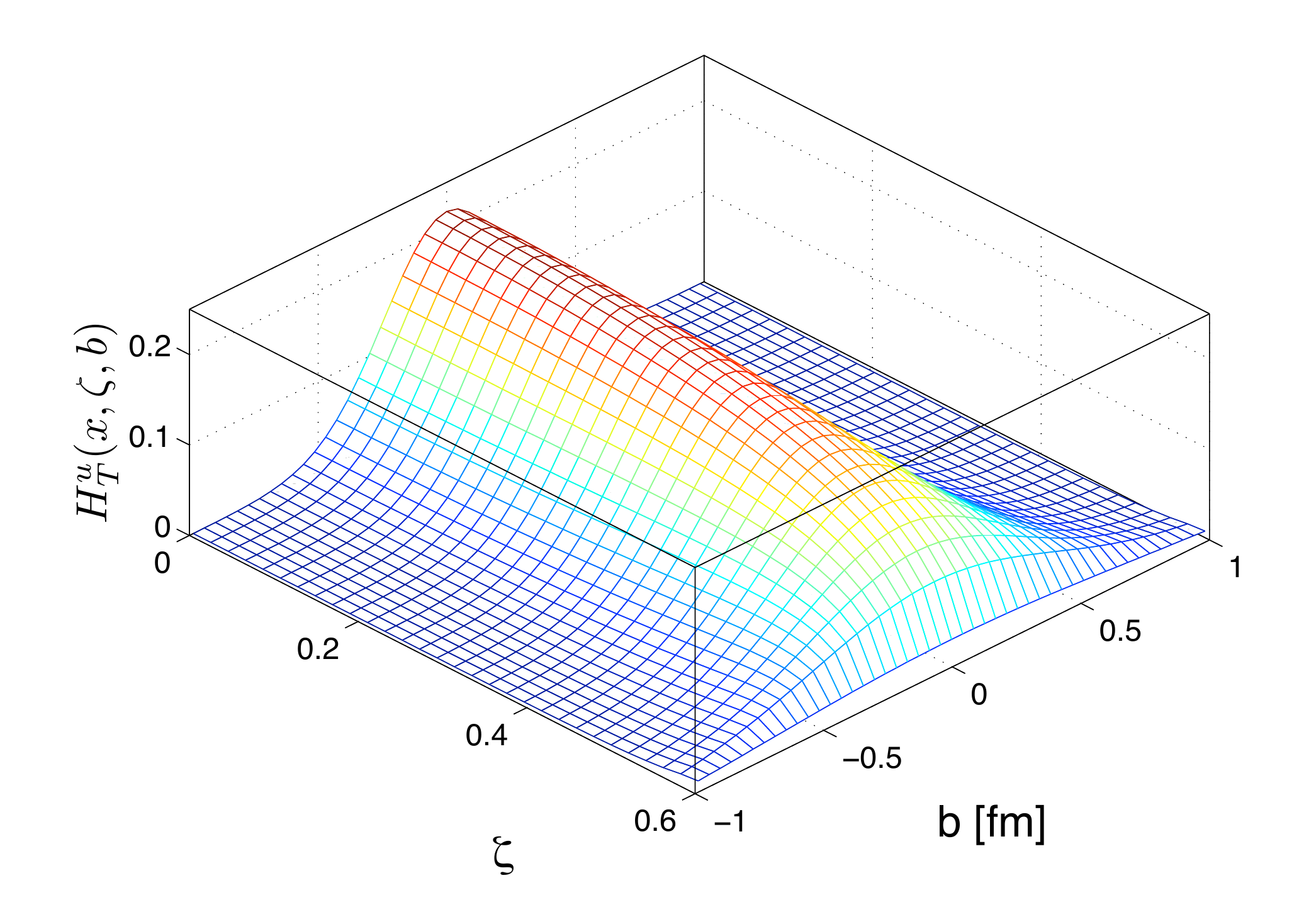}
\hspace{0.1cm}%
\small{(b)}\includegraphics[width=7.5cm,height=5.15cm,clip]{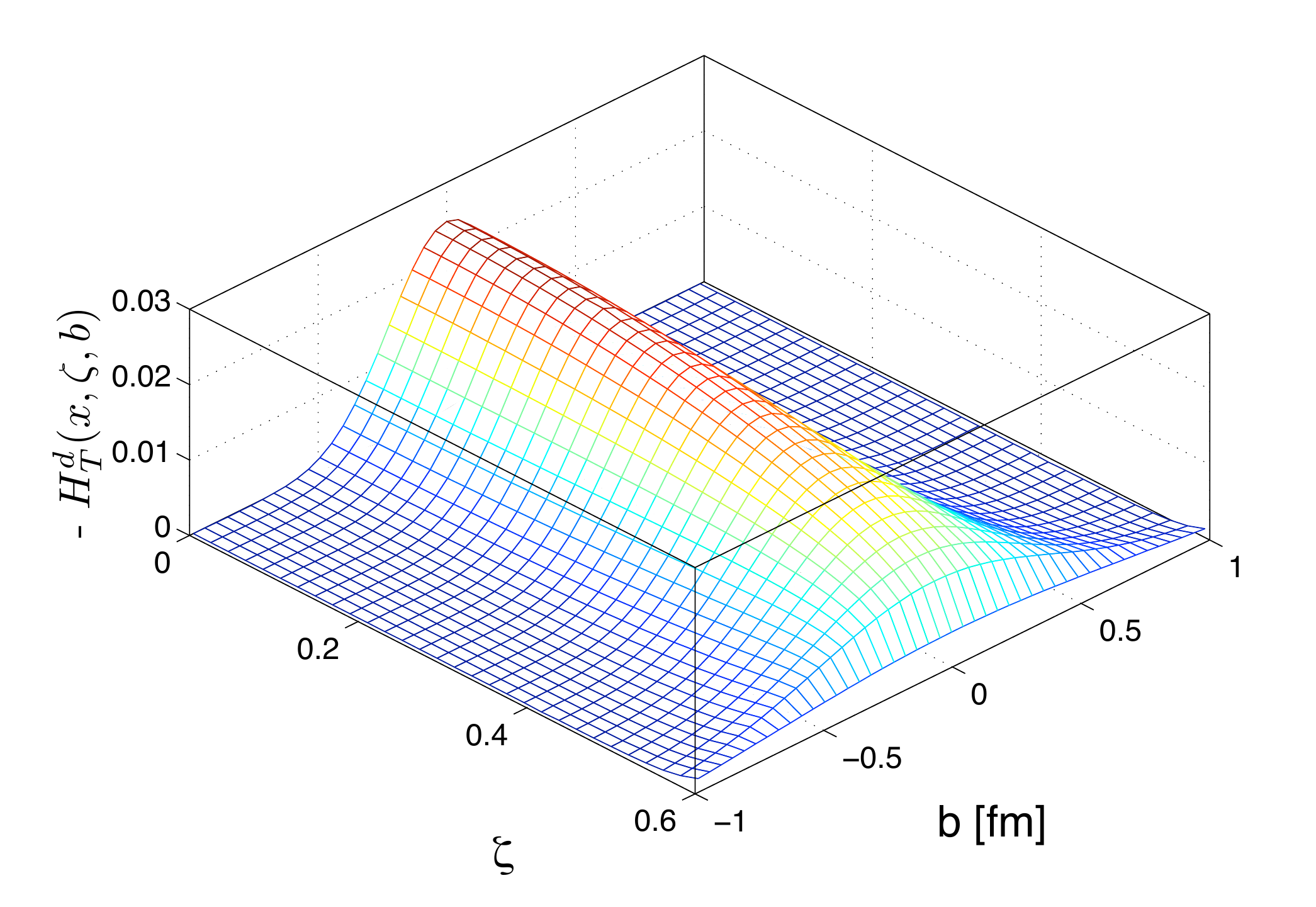}
\end{minipage}
\begin{minipage}[c]{0.98\textwidth}
\small{(c)}\includegraphics[width=7.5cm,height=5.15cm,clip]{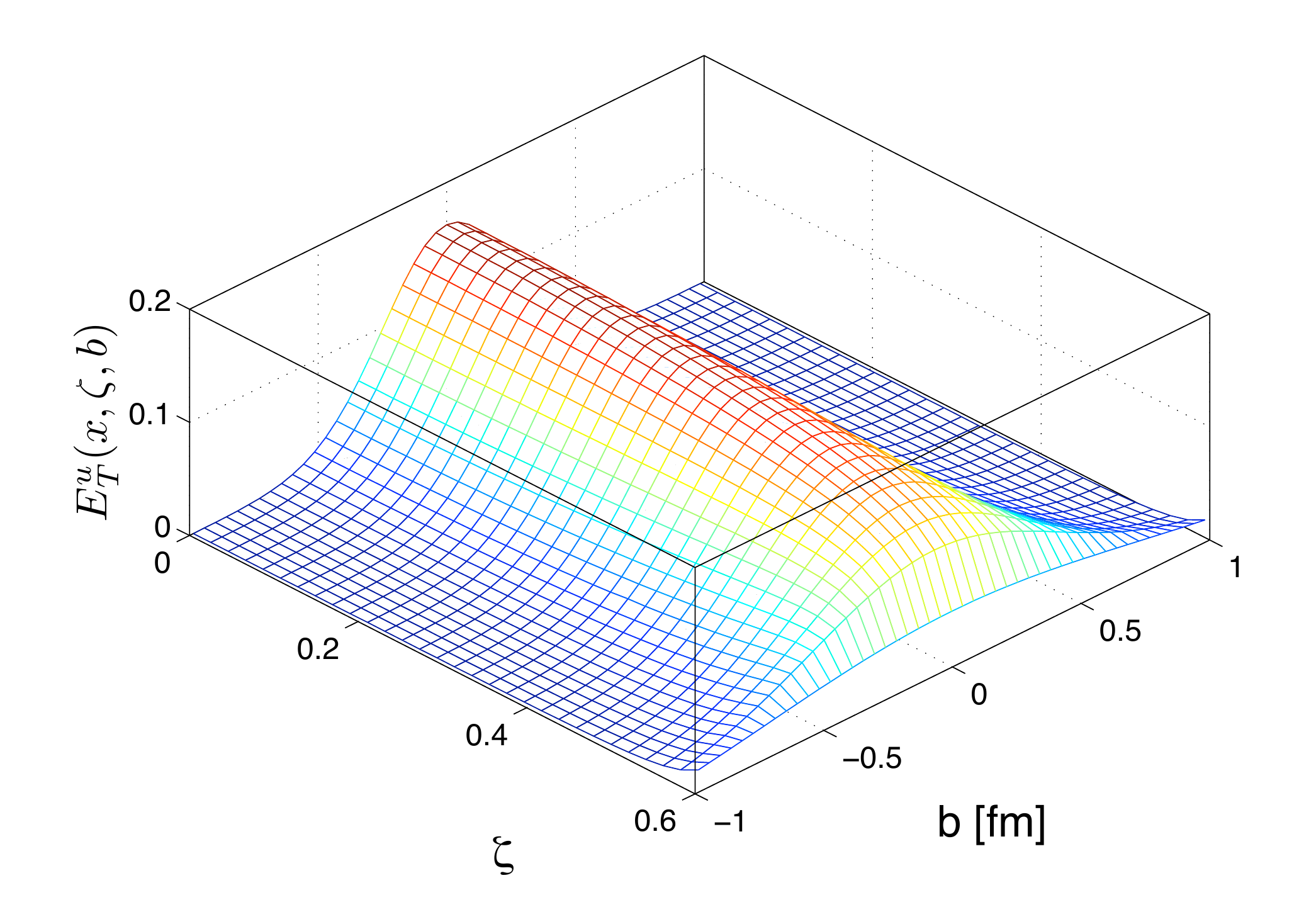}
\hspace{0.1cm}%
\small{(d)}\includegraphics[width=7.5cm,height=5.15cm,clip]{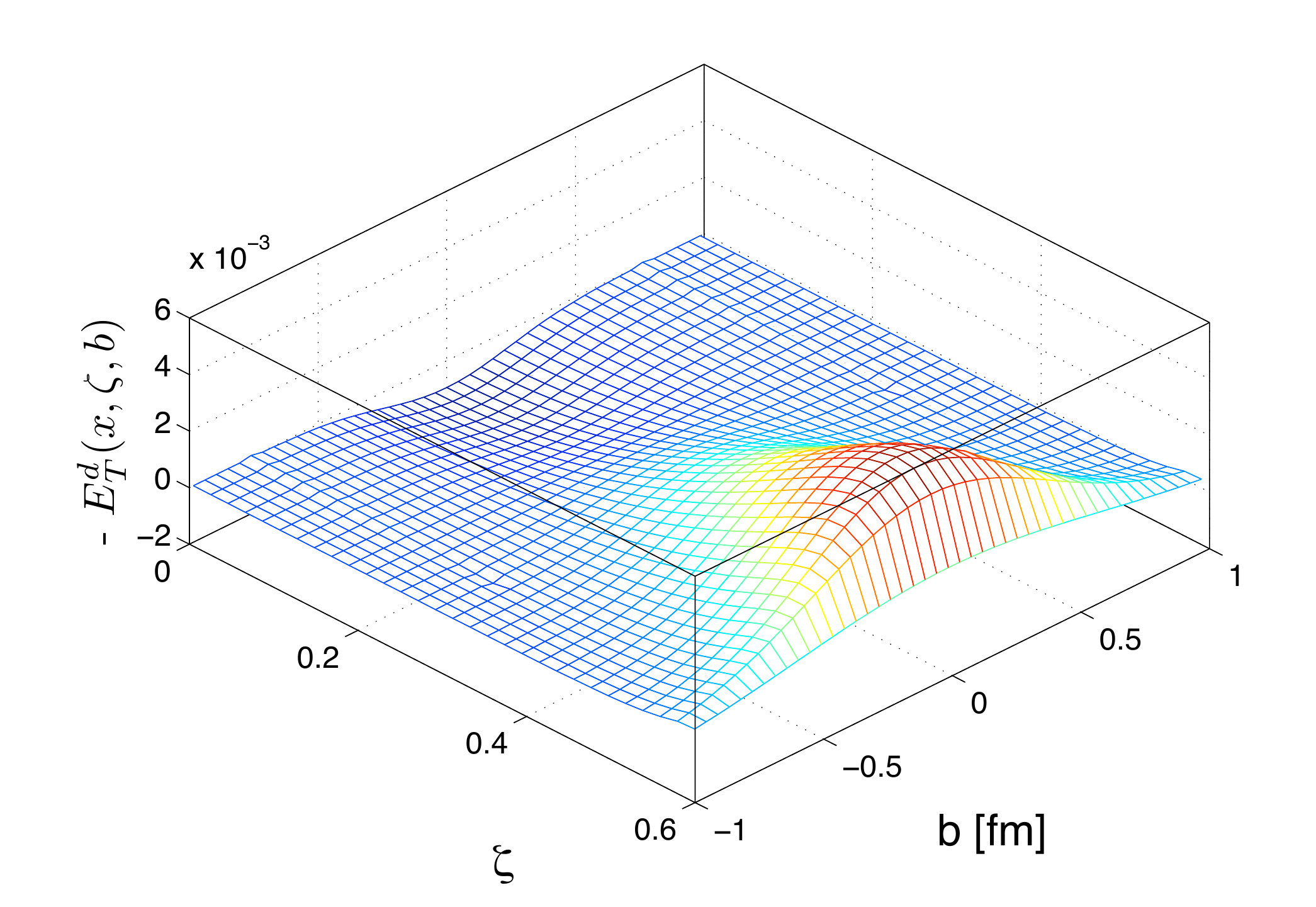}
\end{minipage}
\begin{minipage}[c]{0.98\textwidth}
\small{(e)}\includegraphics[width=7.5cm,height=5.15cm,clip]{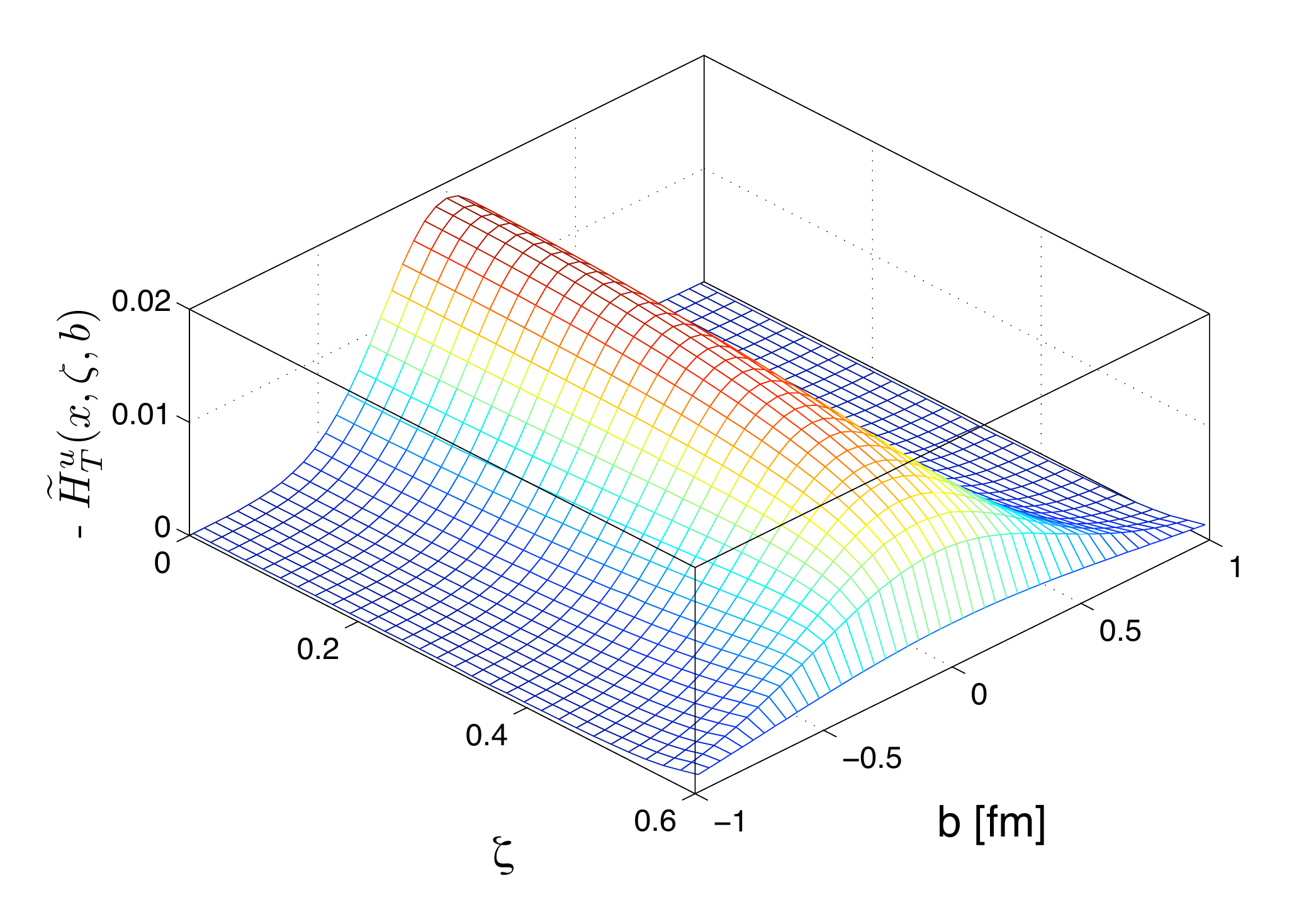}
\hspace{0.1cm}%
\small{(f)}\includegraphics[width=7.5cm,height=5.15cm,clip]{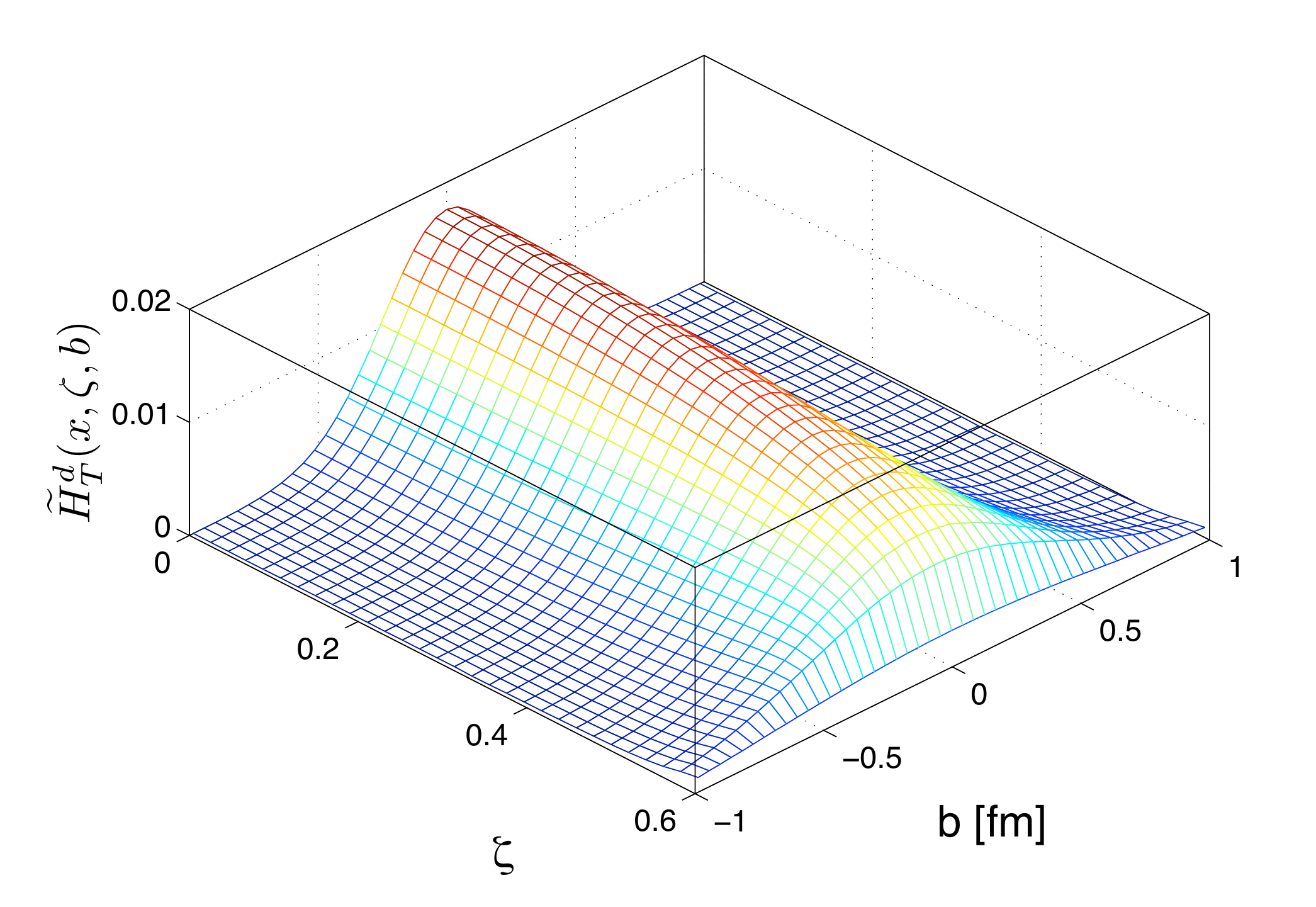}
\end{minipage}
\begin{minipage}[c]{0.98\textwidth}
\small{(g)}\includegraphics[width=7.5cm,height=5.15cm,clip]{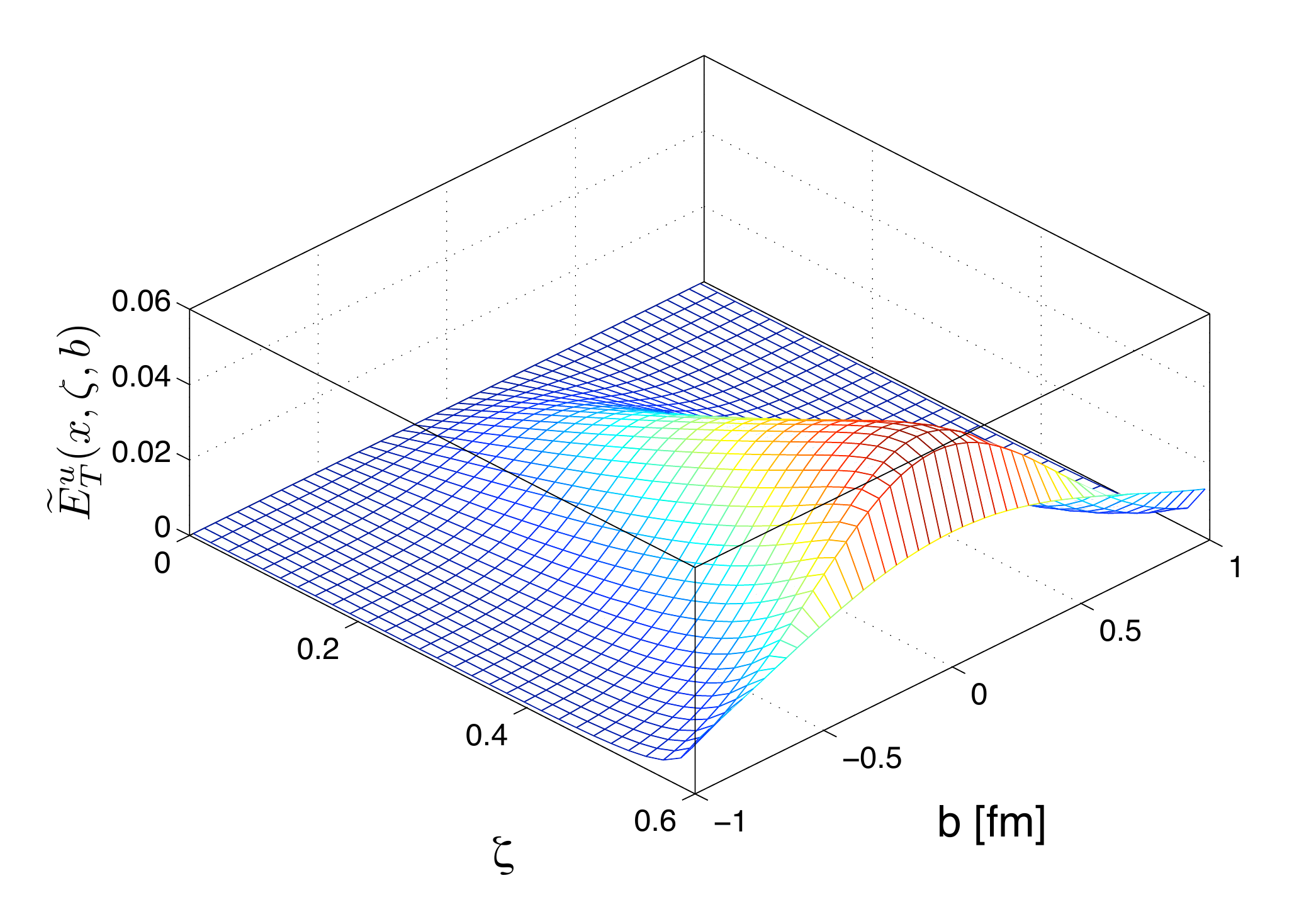}
\hspace{0.1cm}%
\small{(h)}\includegraphics[width=7.5cm,height=5.15cm,clip]{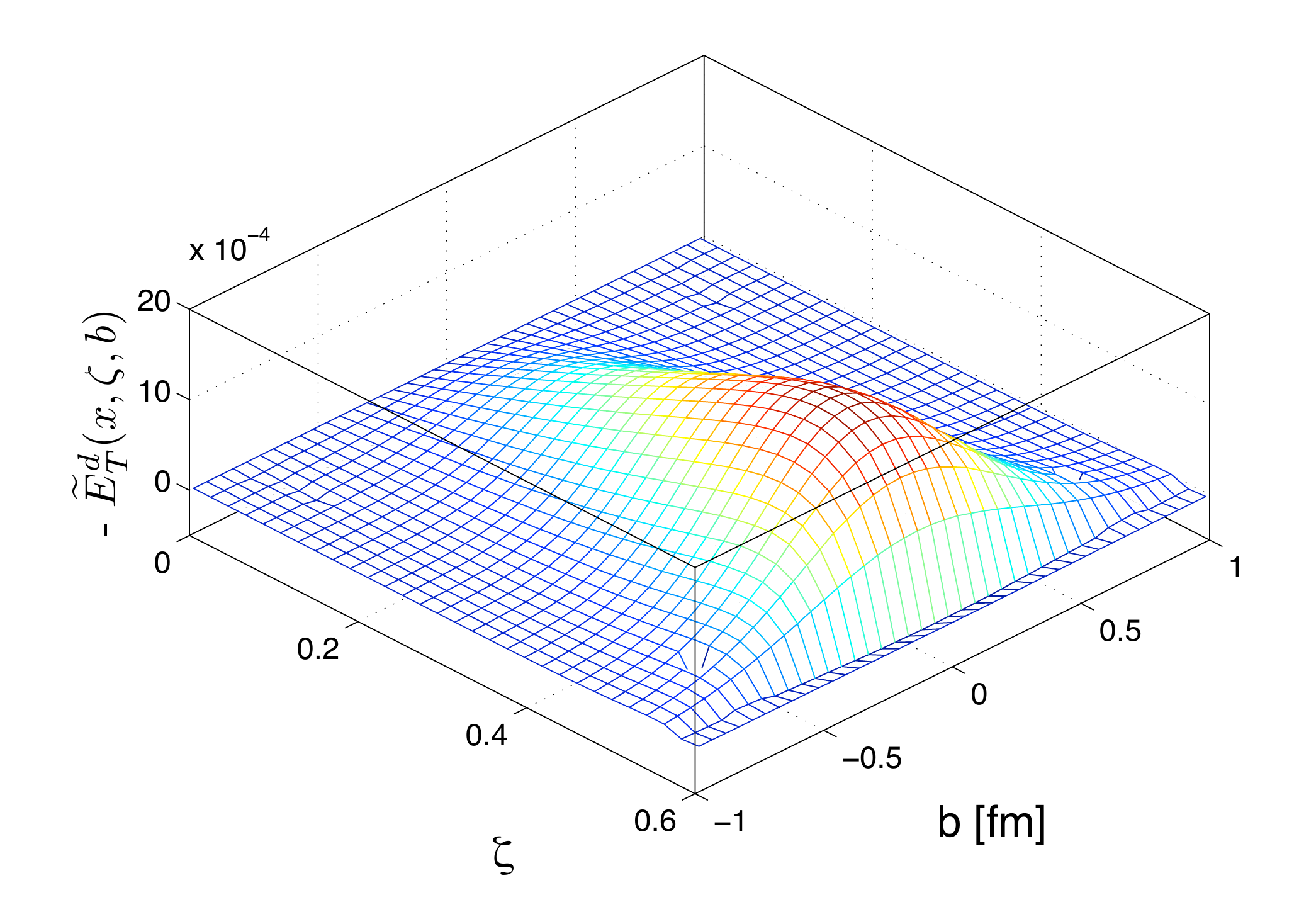}
\end{minipage}
\caption{\label{gix6}(Color online) Plots of chiral odd GPDs for the nonzero skewness in impact space vs $\zeta$ and $b=|\bf b|$ for fixed value of $x=0.6$. Left pannel is for $u$ quark and the right pannel is for $d$ quark.}
\end{figure}
%%%%%%%%%%%%%%%%%%%%%%%%%%%%%%%%%%%%%%%%%%%%%%%%%%%%%%%%%%%%%%%%%%%%%%%%%%%
GPDs in transverse impact parameter space are defined by a two-dimensional Fourier transform in $\Delta_{\perp}$ as follows~\cite{burk1,burk2}:
\be
{H}_T(x,\zeta,b_{\perp})&=&{1\over (2 \pi)^2} \int d^2 \Delta_{\perp} e^{-i \Delta_{\perp} \cdot b_{\perp}}
H_T(x,\zeta,t),\\
{E}_T(x,\zeta,b_{\perp})&=&{1\over (2 \pi)^2} \int d^2 \Delta_{\perp} e^{-i \Delta_{\perp} \cdot b_{\perp}}
E_T(x,\zeta,t),
\\
\widetilde{{H}}_T(x,\zeta,b_{\perp})&=&{1\over (2 \pi)^2} \int d^2 \Delta_{\perp} e^{-i \Delta_{\perp} \cdot b_{\perp}}
\widetilde{H}_T(x,\zeta,t),\\
\widetilde{{E}}_T(x,\zeta,b_{\perp})&=&{1\over (2 \pi)^2} \int d^2 \Delta_{\perp} e^{-i \Delta_{\perp} \cdot b_{\perp}}
\widetilde{E}_T(x,\zeta,t).
\ee
Here, $b_{\perp}$ is the transverse impact parameter conjugate to $\Delta_{\perp}$. 
For zero skewness, $\bfb$ gives a measure of  the transverse distance between the struck parton and the center of momentum of the hadron. $b_{\perp}$ satisfies the condition $\sum_i x_i b_{\perp i}=0$, where the sum is over the number of partons. The relative distance between the struck parton and the center of momentum of the spectator system is given by  ${\mid \bfb \mid\over 1-x}$, which provides us an estimate of the size of the bound state \cite{diehl}. 
%However, the exact estimation of the nuclear size is not possible as the spatial extension of the spectator system is not available from the GPDs. 
In the DGLAP region $x>\zeta$, the impact parameter $b_{\perp}$ implies the location where the quark is pulled out and re-insert to the nucleon. In the ERBL domain $x<\zeta$, $b_{\perp}$ provides the transverse distance of the quark-antiquark pair inside the nucleon. For zero skewness, the chiral-odd GPDs also have a density interpretation in transverse impact parameter space like chiral-even GPDs depending on the polarization of both the active quark and the nucleon. A combination of ${E}_T(x,b_{\perp})$ and $\widetilde{{H}}_T(x,b_{\perp})$ i.e. $({E}_T+2\widetilde{{H}}_T)$ is responsible for a deformation in the transversity asymmetry quarks in an unpolarized target~\cite{Burk3,Diehl05,Dahiya07}. This is similar to the role played by ${E}(x,b_{\perp})$ for both unpolarized active quark and the nucleon. On the other hand a combination of ${H}_T(x,b_{\perp})$ and $\widetilde{{H}}_T(x,b_{\perp})$ provides a distortion in the transverse spin density when the active quark and the nucleon are transversely polarized~\cite{Diehl05,Pasquini2}. Note 
that the density interpretation is possible only in the limit $\zeta=0$, but in most experiment $\zeta$ is nonzero. So, it is interesting to investigate the chiral-odd GPDs in the impact parameter space with nonzero $\zeta$. 

We show the skewness dependent chiral-odd GPDs in transverse impact parameter space for fixed $\zeta=0.2$ as functions of $b=|b_{\perp}|$ and $x$ for $u$ and $d$ quark in Fig.\ref{giz2}. Similarly, all the chiral-odd GPDs as functions of $\zeta$ and $b$ for a fixed value of $x=0.6$ are shown in Fig.\ref{gix6}. 
The peak of the distribution $H_T(x,\zeta,b_{\perp})$ for fixed $\zeta$ appears at lower $x$ for $d$ quark and shifts to higher $x$ for $u$ quark while for $\widetilde{H}_T(x,\zeta,b_{\perp})$ we get the peak at lower $x$ for both $u$ and $d$ quarks. For both $E_T(x,\zeta,b_{\perp})$ and $\widetilde{E}_T(x,\zeta,b_{\perp})$, the peaks arise at lower $x$ for both $u$ and $d$ quarks but we also get an oscillatory behavior  for both the GPDs of $d$ quark. This is due to the fact that  $E_T^d(x,\zeta,t)$ and $\widetilde{E}_T^d(x,\zeta,t)$  have slight oscillatory behavior as can be seen in  Fig. \ref{gz15}(c)-(d). Except $\widetilde{H}_T(x,\zeta,b_{\perp})$, the peak of $u$ quark in all other distributions are sufficiently large compare to $d$ quark. For $\widetilde{H}_T(x,\zeta,b_{\perp})$, the peak of $u$ quark is slightly large compare to $d$ quark. For small $b$, $\widetilde{E}_T(x,\zeta,b_{\perp})$ falls off slowly at large $x$ for $u$ quark compare to $d$ quark. 
With increasing $x$, the width of all the distributions in transverse impact parameter space  decreases, which implies that the distributions are more localized near the center of momentum for higher values of $x$. Substantial differences is observed in $\widetilde{E}_T(x,\zeta,b_{\perp})$ from other GPDs when the GPDs are plotted against $\zeta$ and $b$ for fixed values of $x$ in Fig.\ref{gix6}. $\widetilde{E}_T(x,\zeta,b_{\perp})$ 
increases with increasing $\zeta$.
Another interesting behavior of all the GPDs is that the peaks of all the distributions become broader as $\zeta$ increases for a fixed value of $x$.  This  means that as the  momentum transfer in the longitudinal direction increases the probability of hitting the transversely polarized active quark  at a larger transverse impact parameter $b$  increases.

%%%%%%%%%%%%%%%%%%%%%%%%%%
\subsection{GPDs in longitudinal impact parameter space}\label{chiral_longi_impact}
%%%%%%%%%%%%%%%%%%%..zeta dependent..zb..%%%%%%%%
\begin{figure}[htbp]
\begin{minipage}[c]{0.98\textwidth}
\small{(a)}
\includegraphics[width=7.5cm,height=5.15cm,clip]{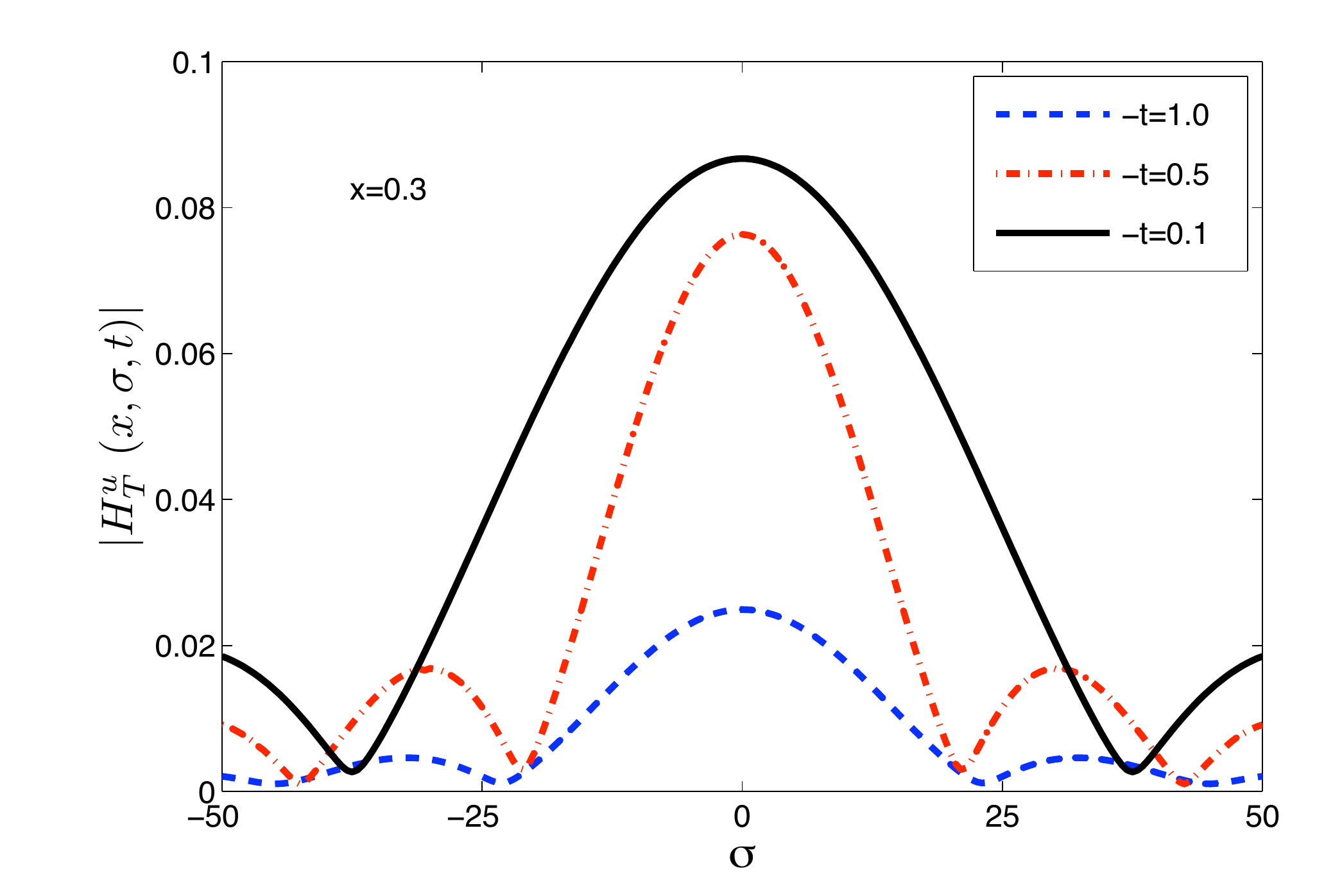}
\hspace{0.1cm}%
\small{(b)}\includegraphics[width=7.5cm,height=5.15cm,clip]{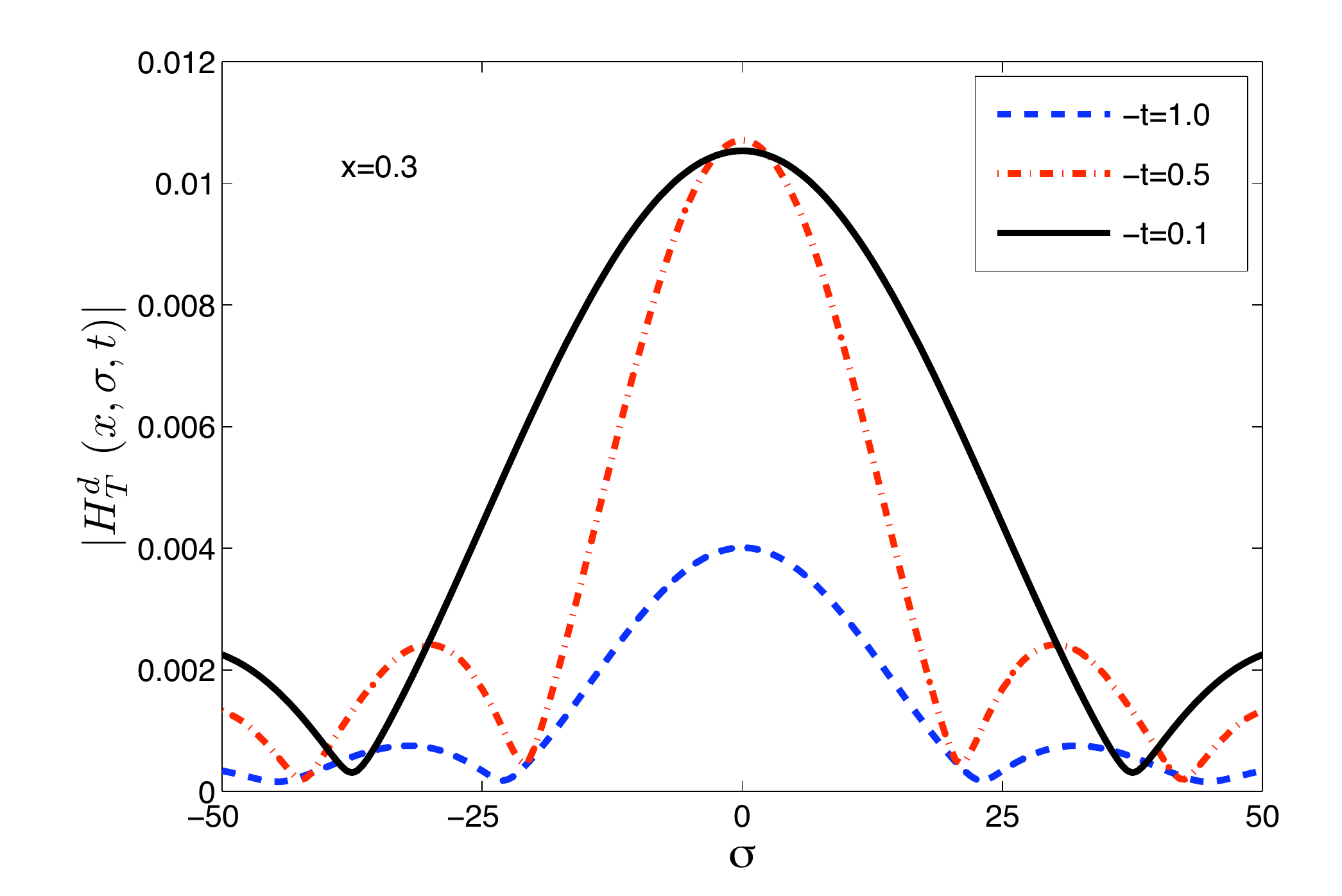}
\end{minipage}
\begin{minipage}[c]{0.98\textwidth}
\small{(c)}\includegraphics[width=7.5cm,height=5.15cm,clip]{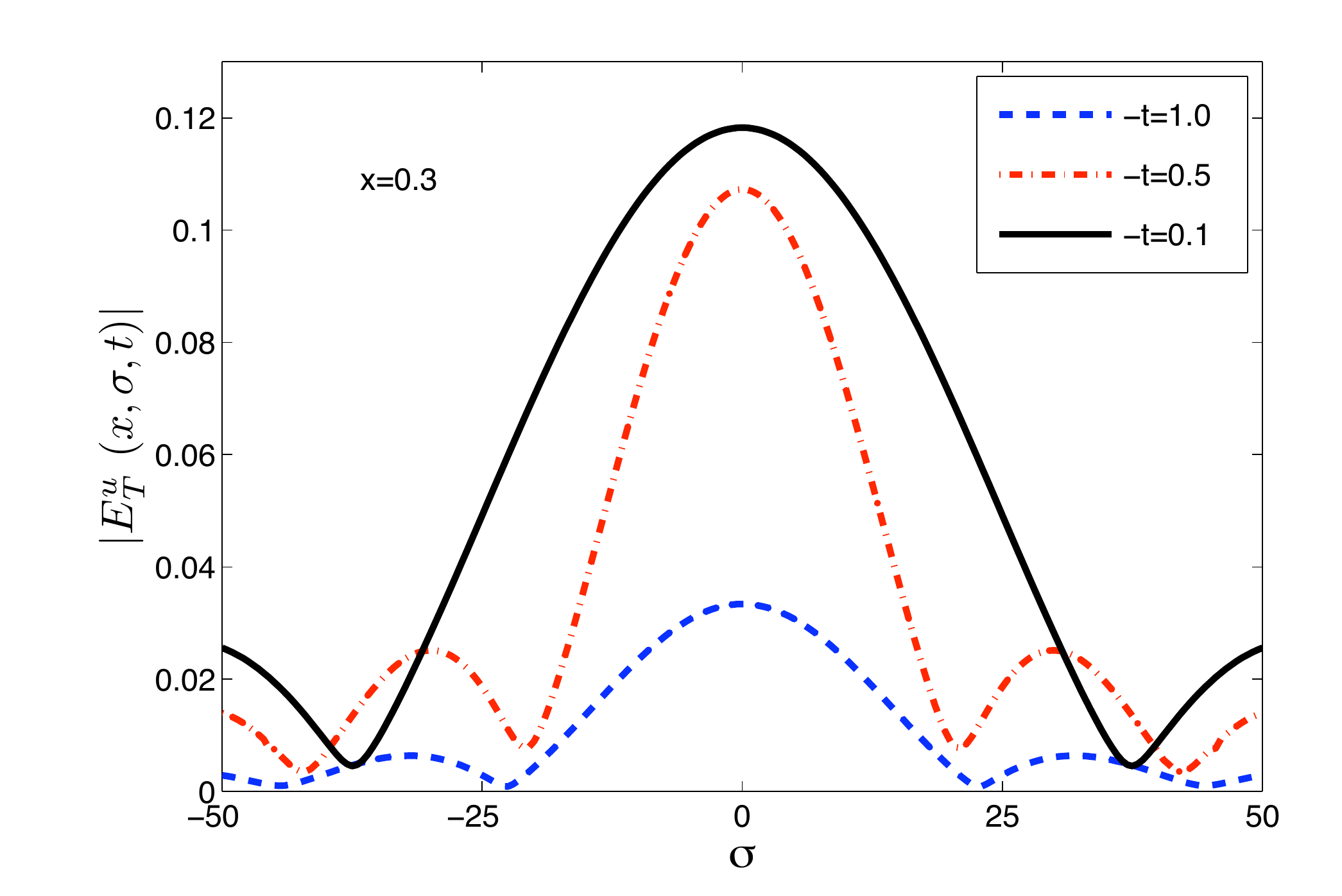}
\hspace{0.1cm}%
\small{(d)}\includegraphics[width=7.5cm,height=5.15cm,clip]{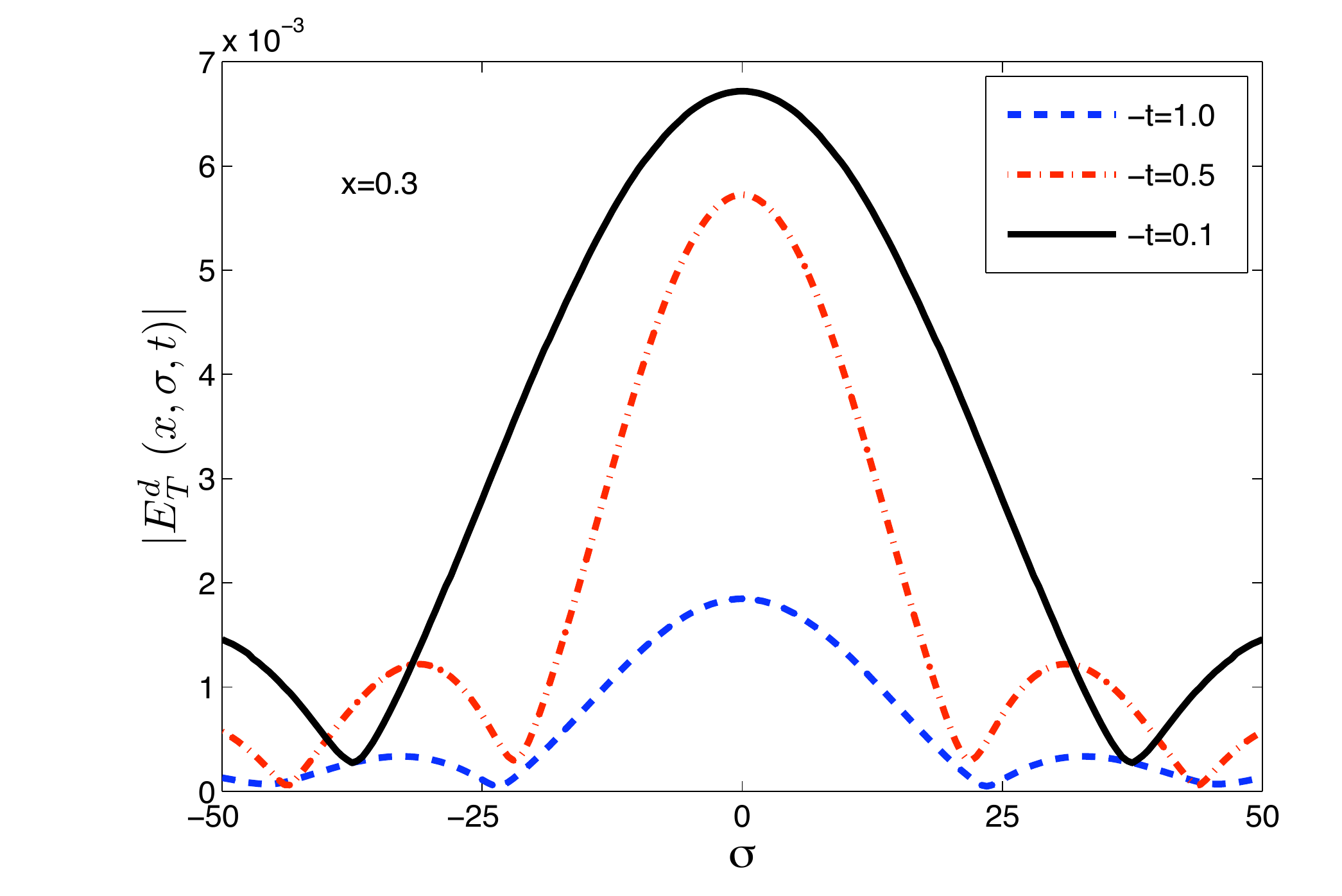}
\end{minipage}
\begin{minipage}[c]{0.98\textwidth}
\small{(a)}\includegraphics[width=7.5cm,height=5.15cm,clip]{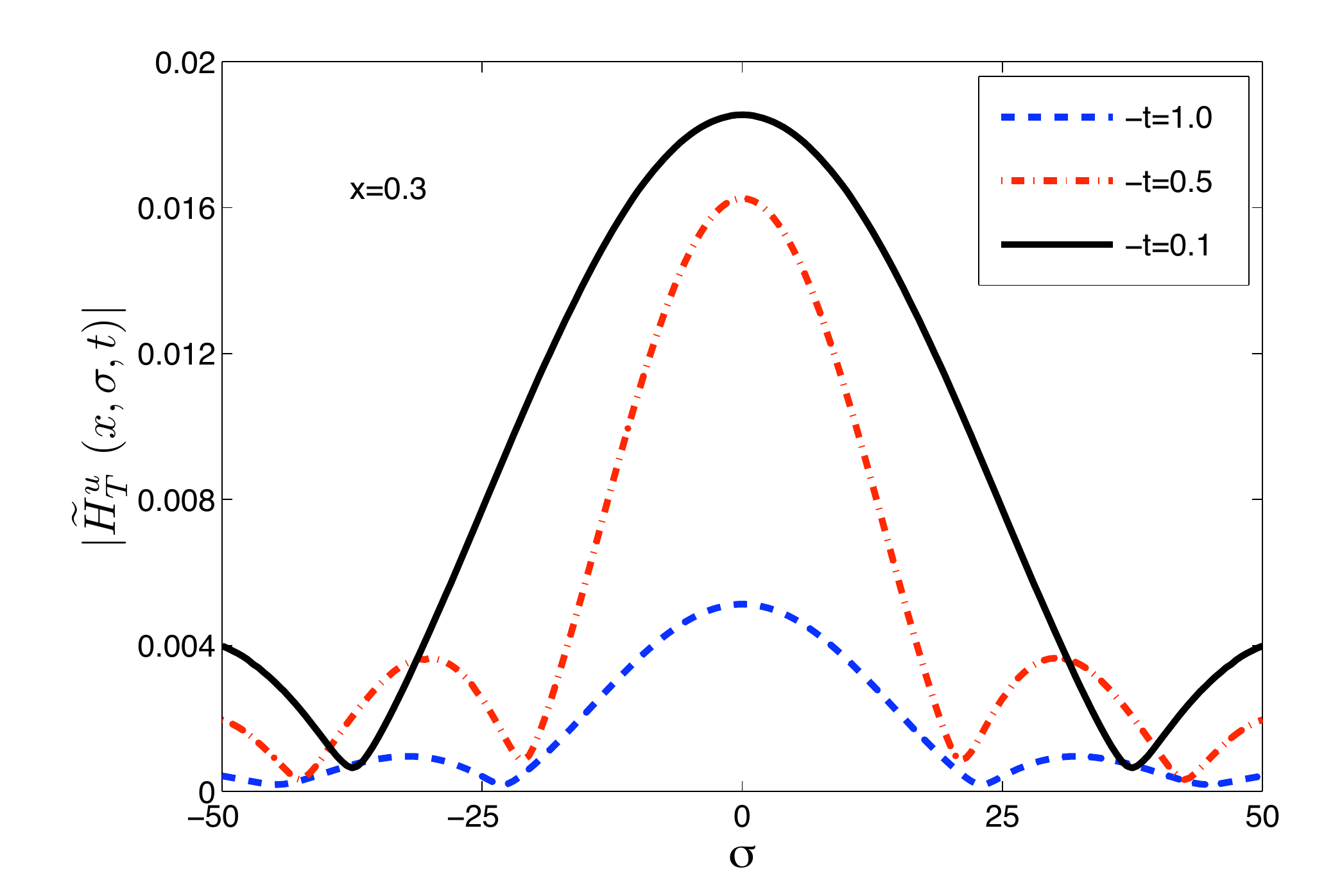}
\hspace{0.1cm}%
\small{(b)}\includegraphics[width=7.5cm,height=5.15cm,clip]{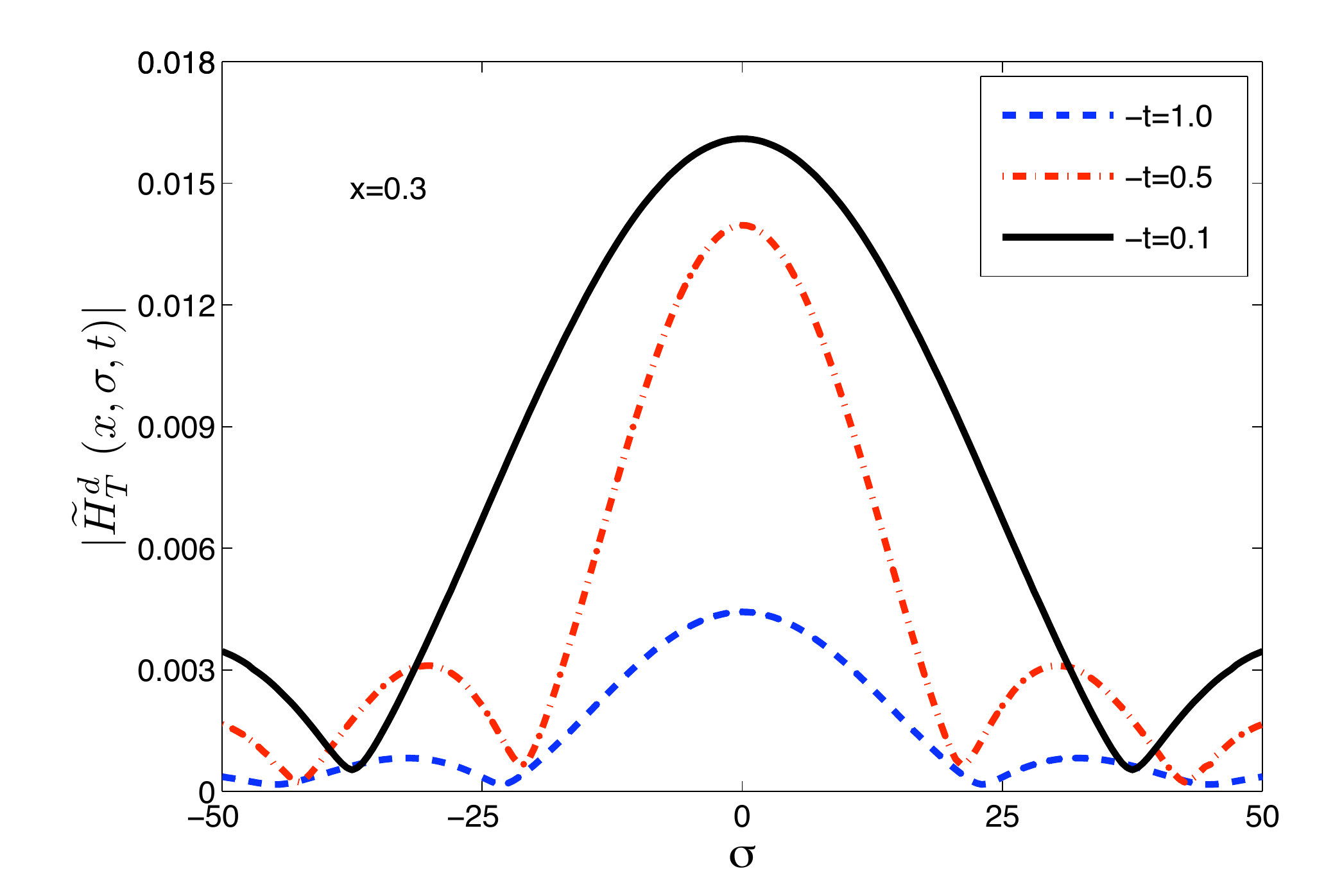}
\end{minipage}
\begin{minipage}[c]{0.98\textwidth}
\small{(c)}\includegraphics[width=7.5cm,height=5.15cm,clip]{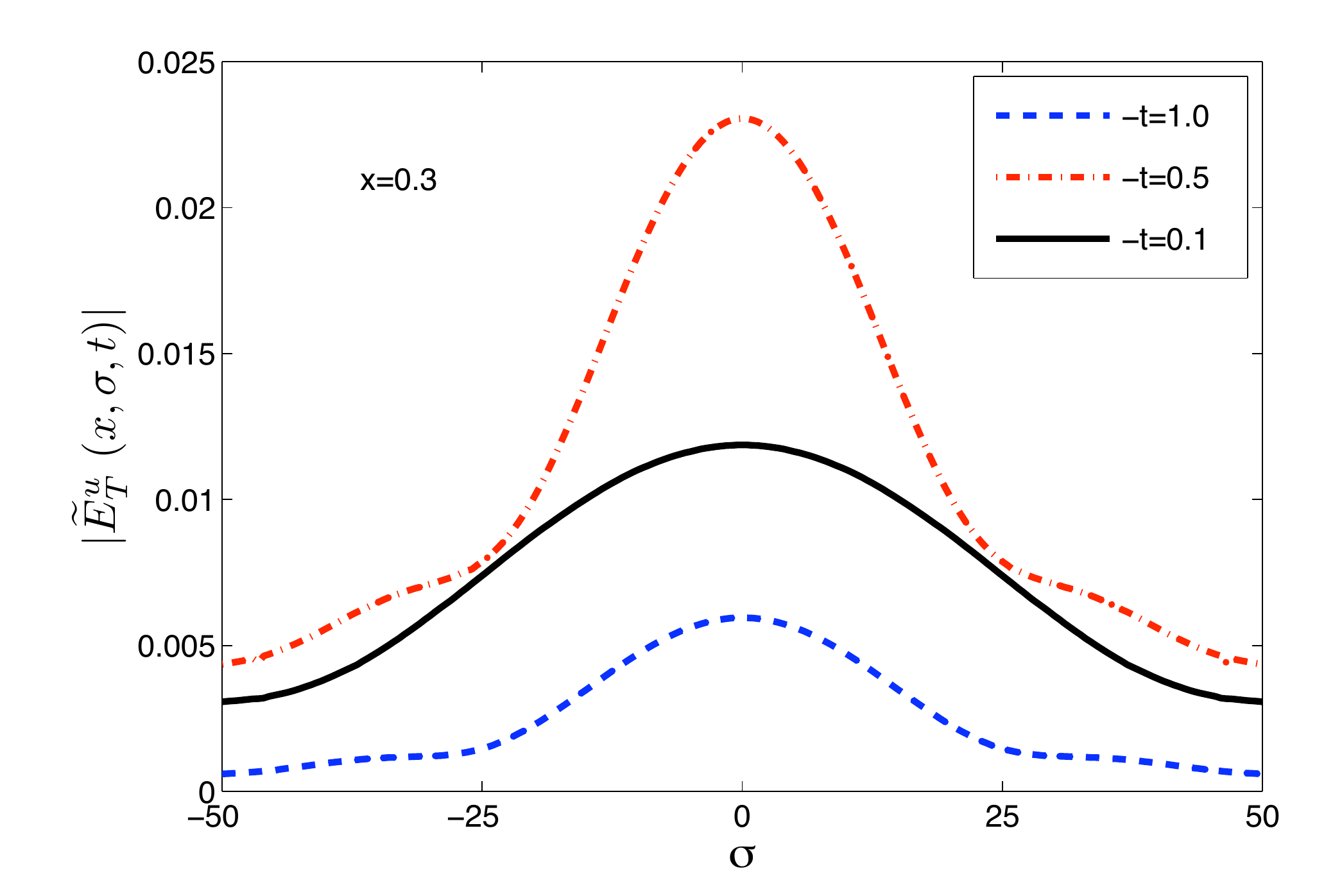}
\hspace{0.1cm}%
\small{(d)}\includegraphics[width=7.5cm,height=5.15cm,clip]{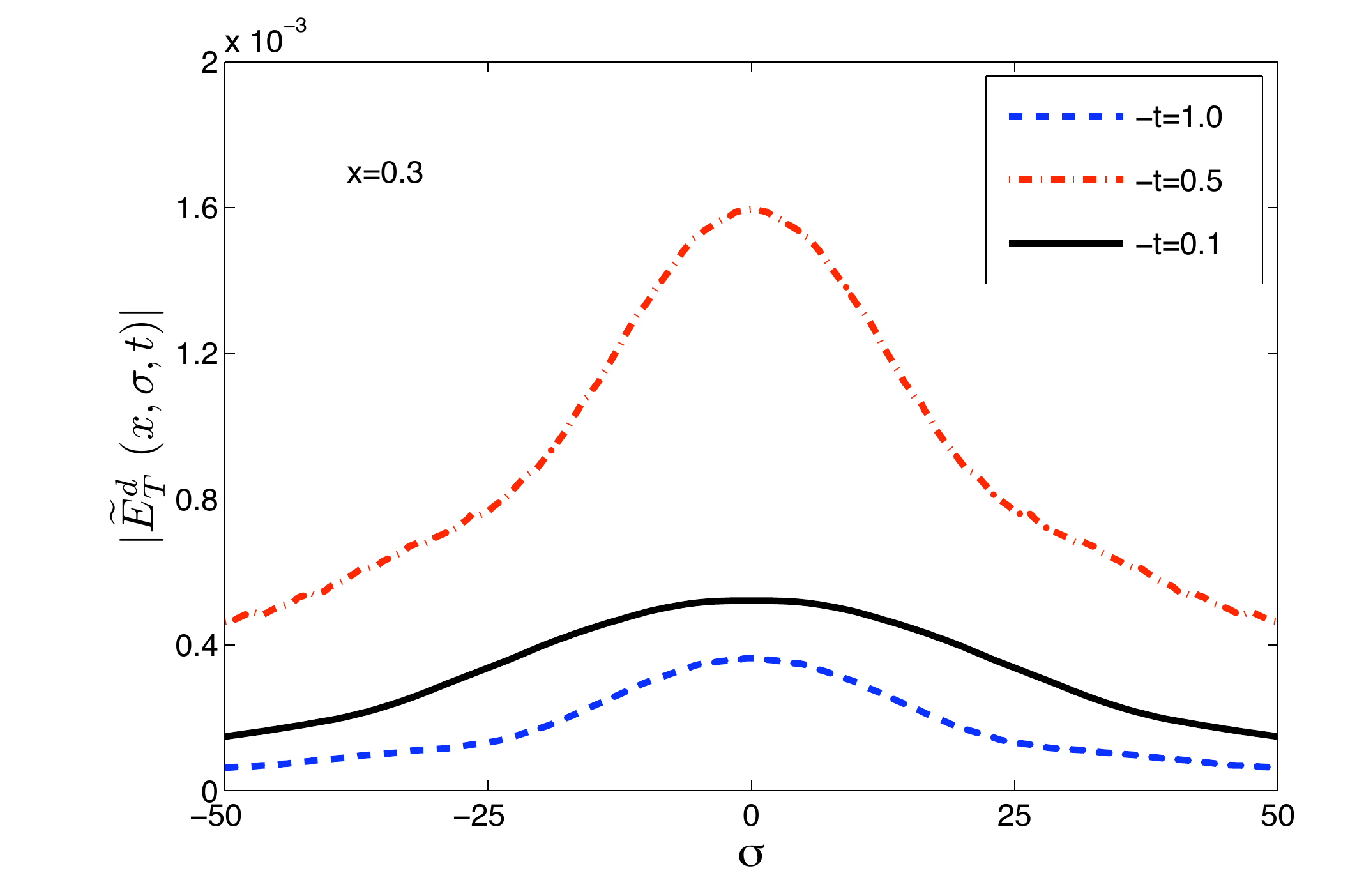}
\end{minipage}
\caption{\label{glx6}(Color online) Plots of the chiral-odd GPDs in longitudinal impact space vs $\sigma$ and different values of $-t$ in $\rm{GeV}^2$, for fixed value of $x=0.3$. Left pannel is for $u$ quark and the right pannel is for $d$ quark.}
\end{figure}
\begin{figure}[htbp]
\begin{minipage}[c]{0.98\textwidth}
\small{(a)}
\includegraphics[width=7.5cm,height=5.15cm,clip]{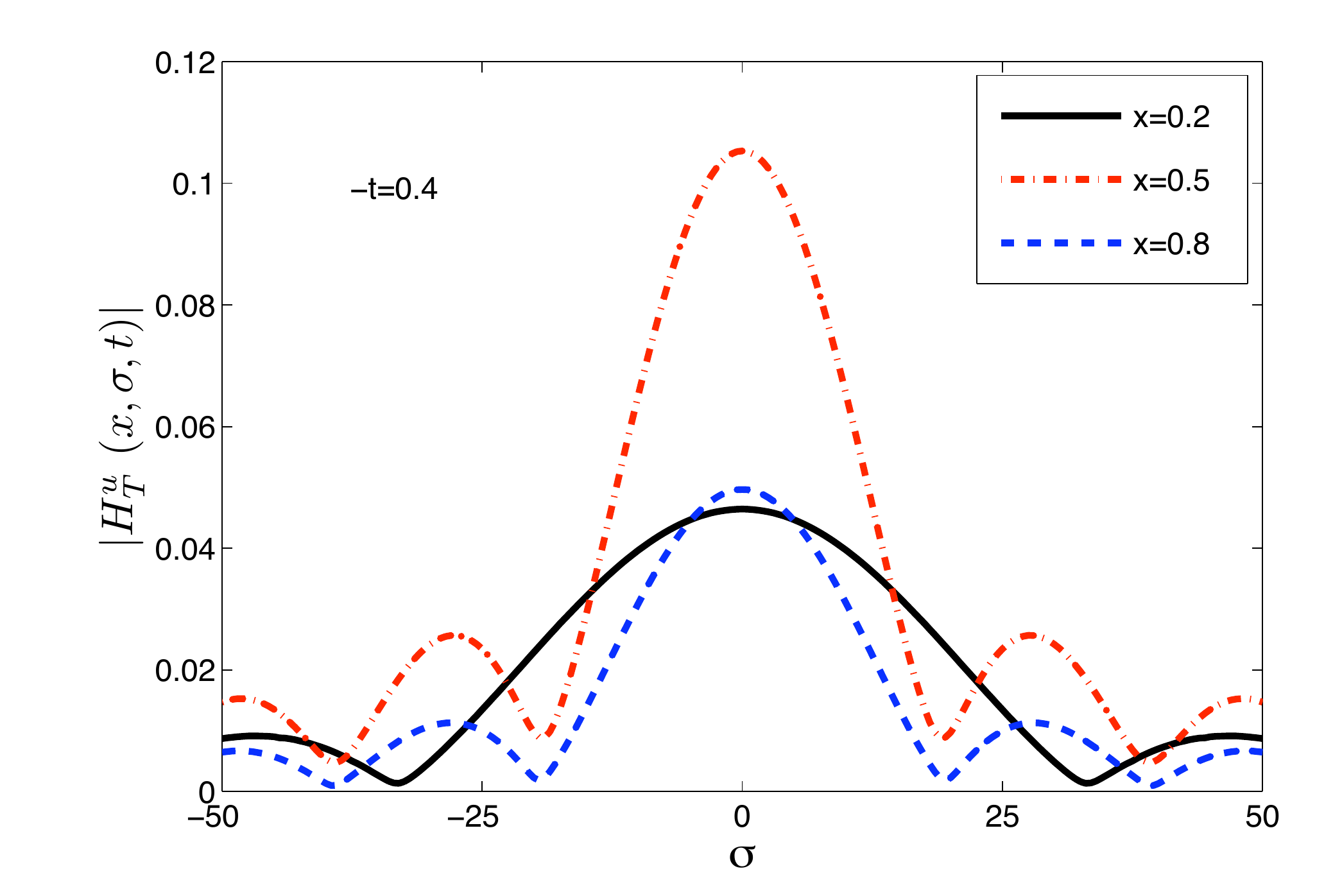}
\hspace{0.1cm}%
\small{(b)}\includegraphics[width=7.5cm,height=5.15cm,clip]{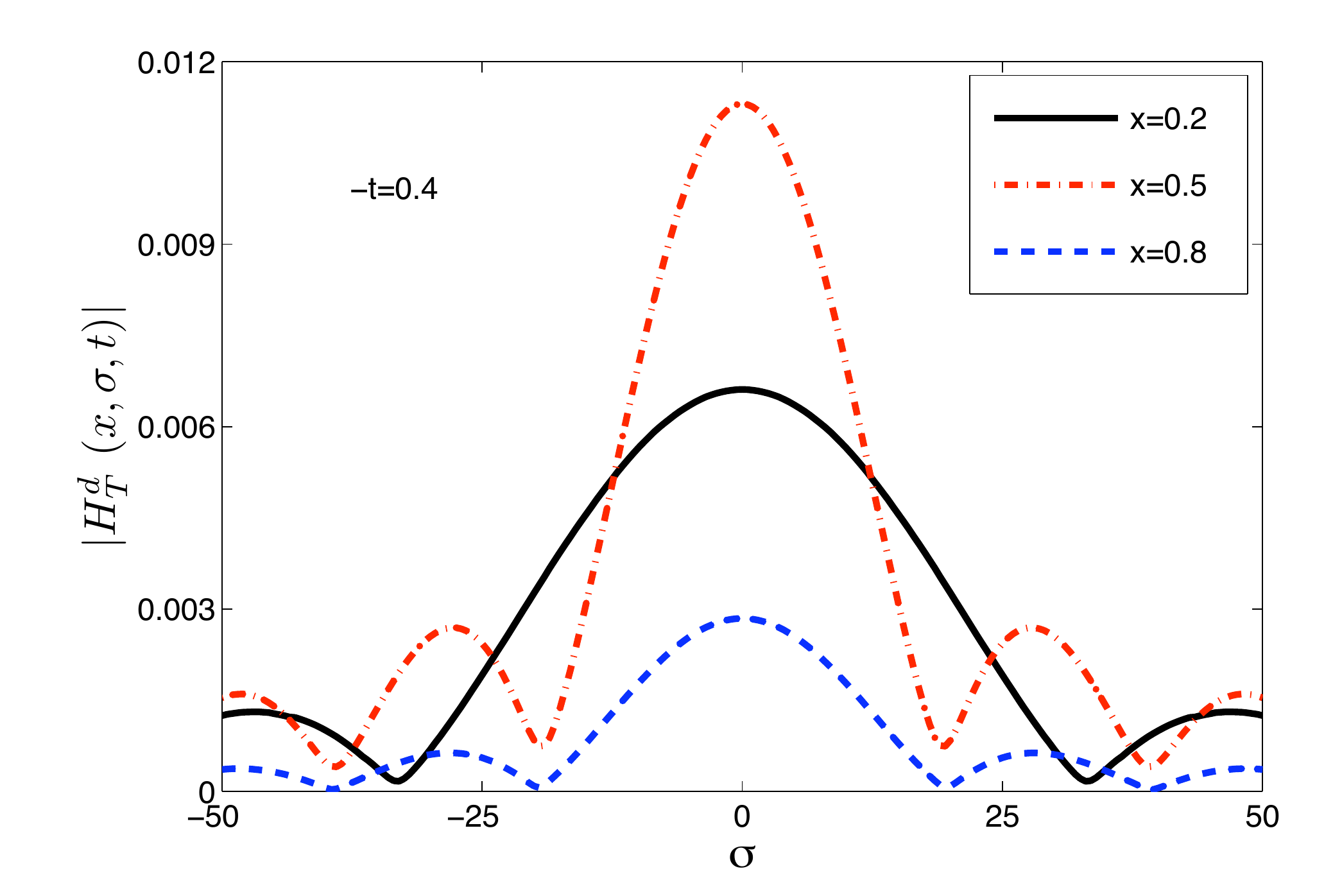}
\end{minipage}
\begin{minipage}[c]{0.98\textwidth}
\small{(c)}\includegraphics[width=7.5cm,height=5.15cm,clip]{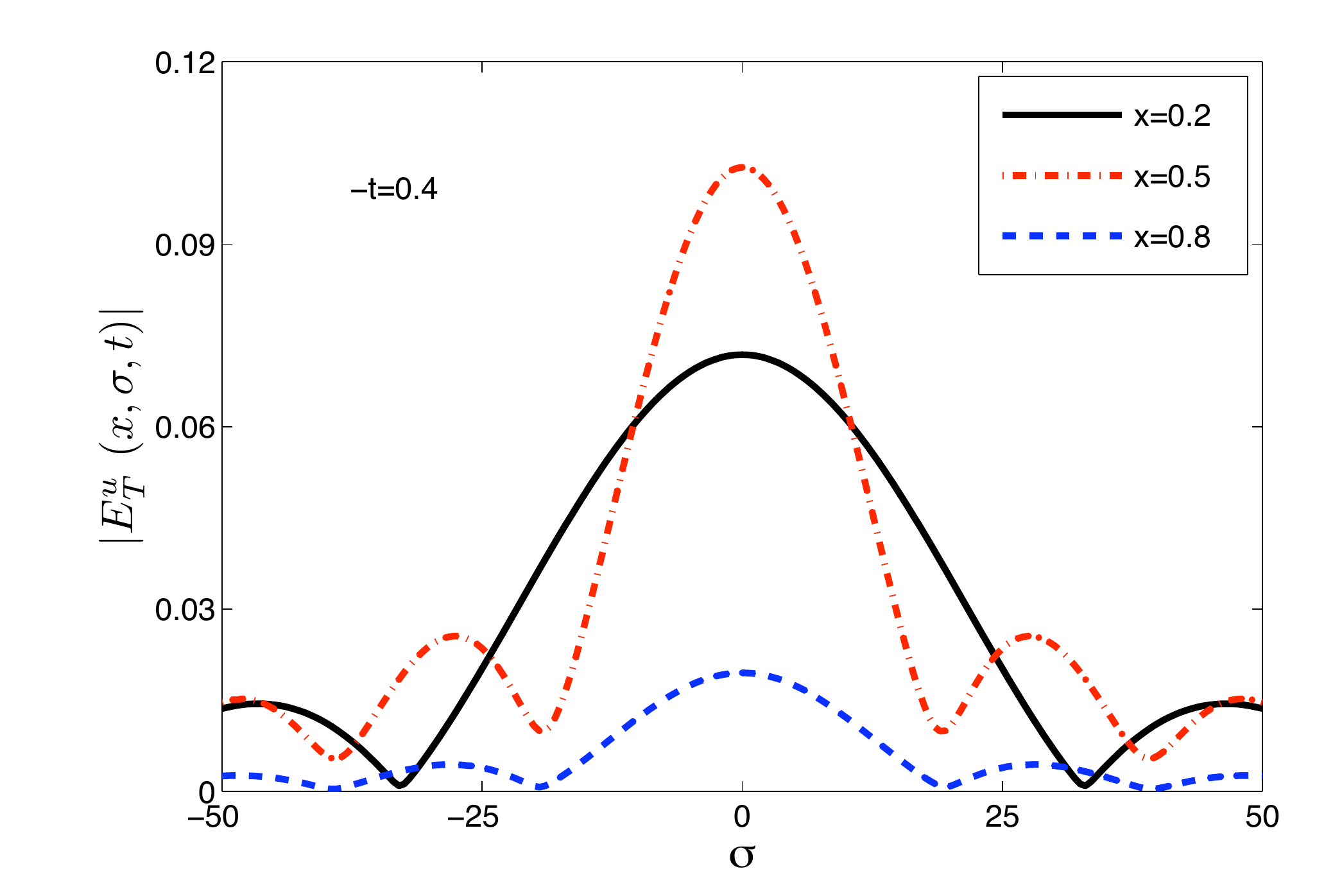}
\hspace{0.1cm}%
\small{(d)}\includegraphics[width=7.5cm,height=5.15cm,clip]{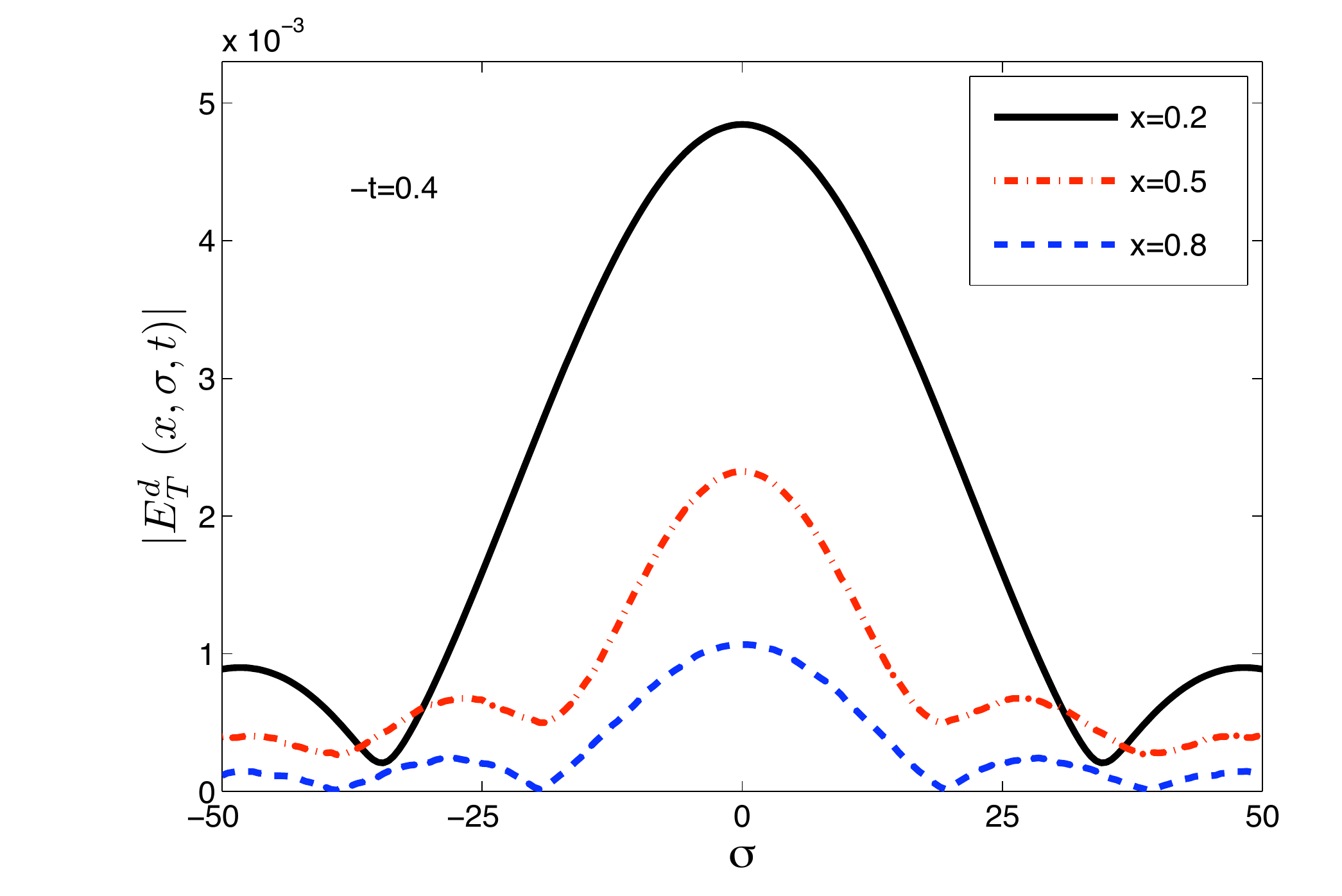}
\end{minipage}
\begin{minipage}[c]{0.98\textwidth}
\small{(a)}\includegraphics[width=7.5cm,height=5.15cm,clip]{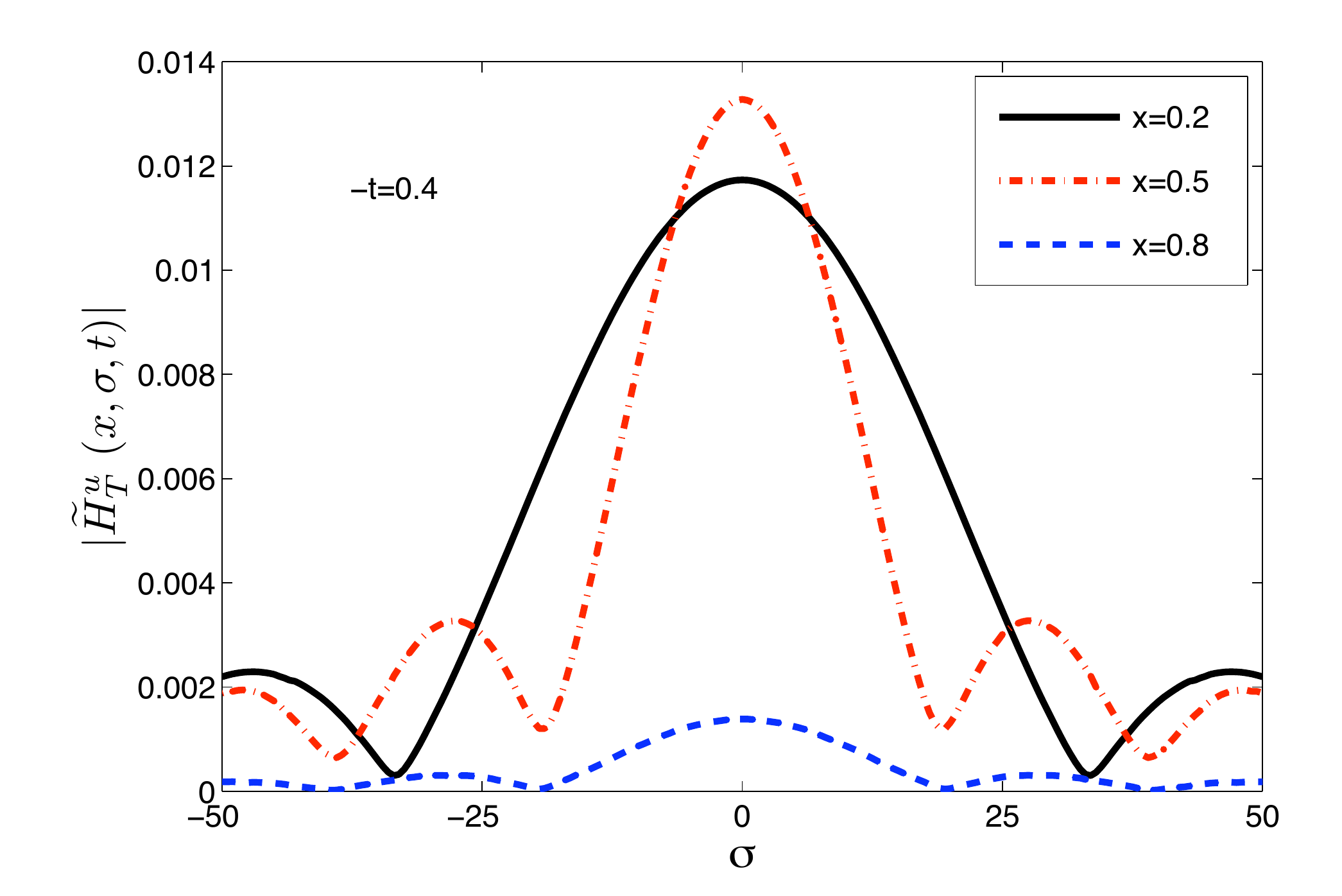}
\hspace{0.1cm}%
\small{(b)}\includegraphics[width=7.5cm,height=5.15cm,clip]{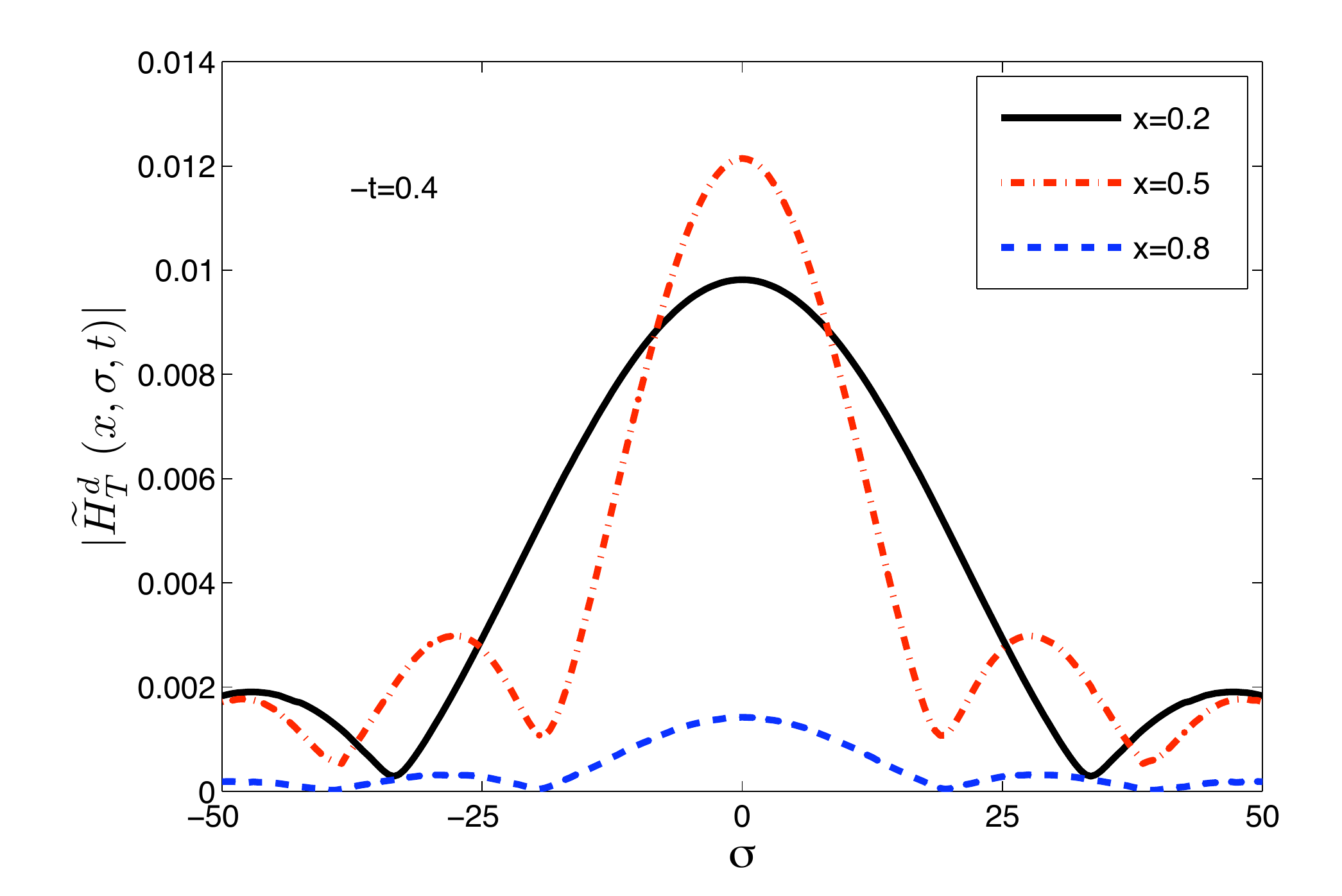}
\end{minipage}
\begin{minipage}[c]{0.98\textwidth}
\small{(c)}\includegraphics[width=7.5cm,height=5.15cm,clip]{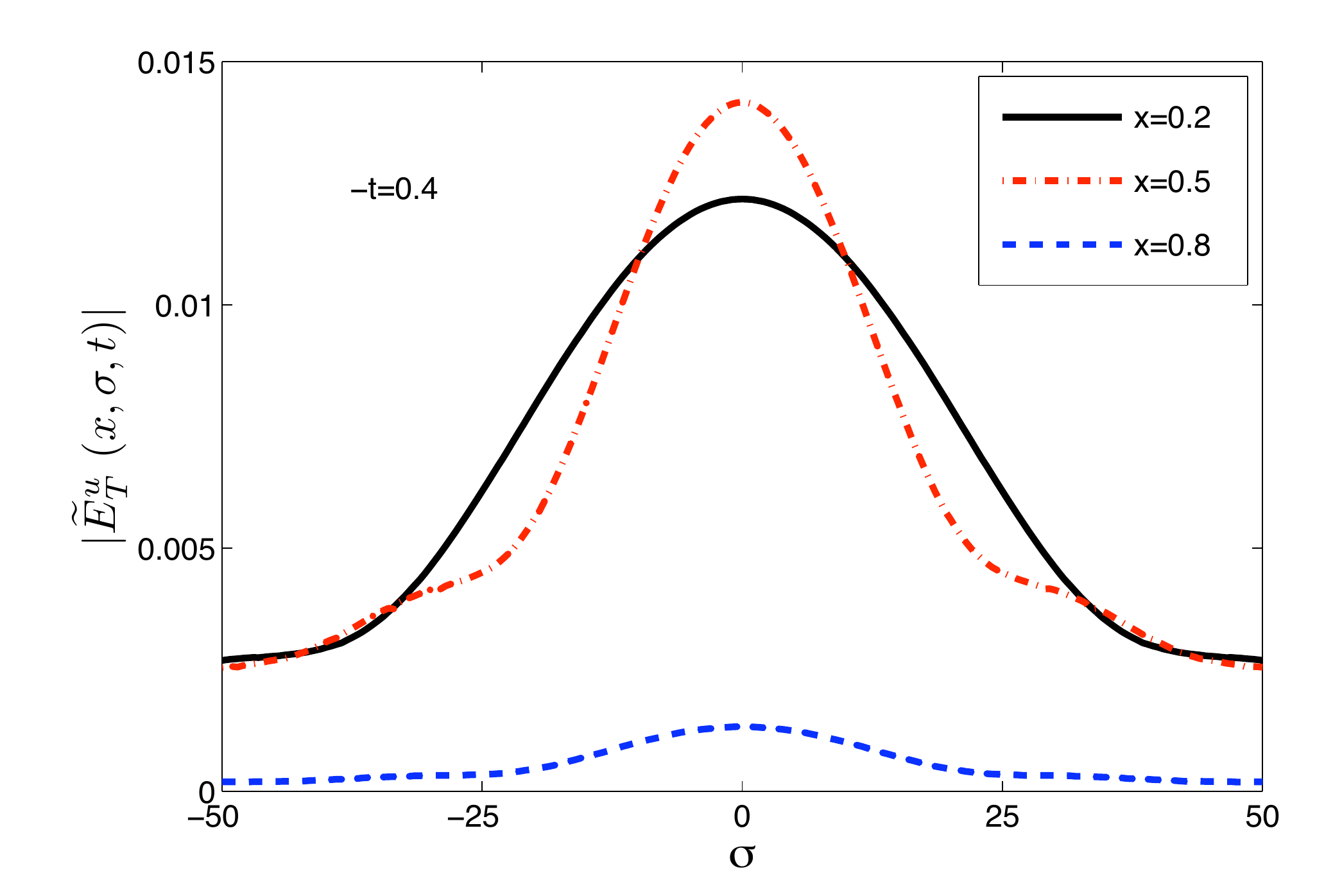}
\hspace{0.1cm}%
\small{(d)}\includegraphics[width=7.5cm,height=5.15cm,clip]{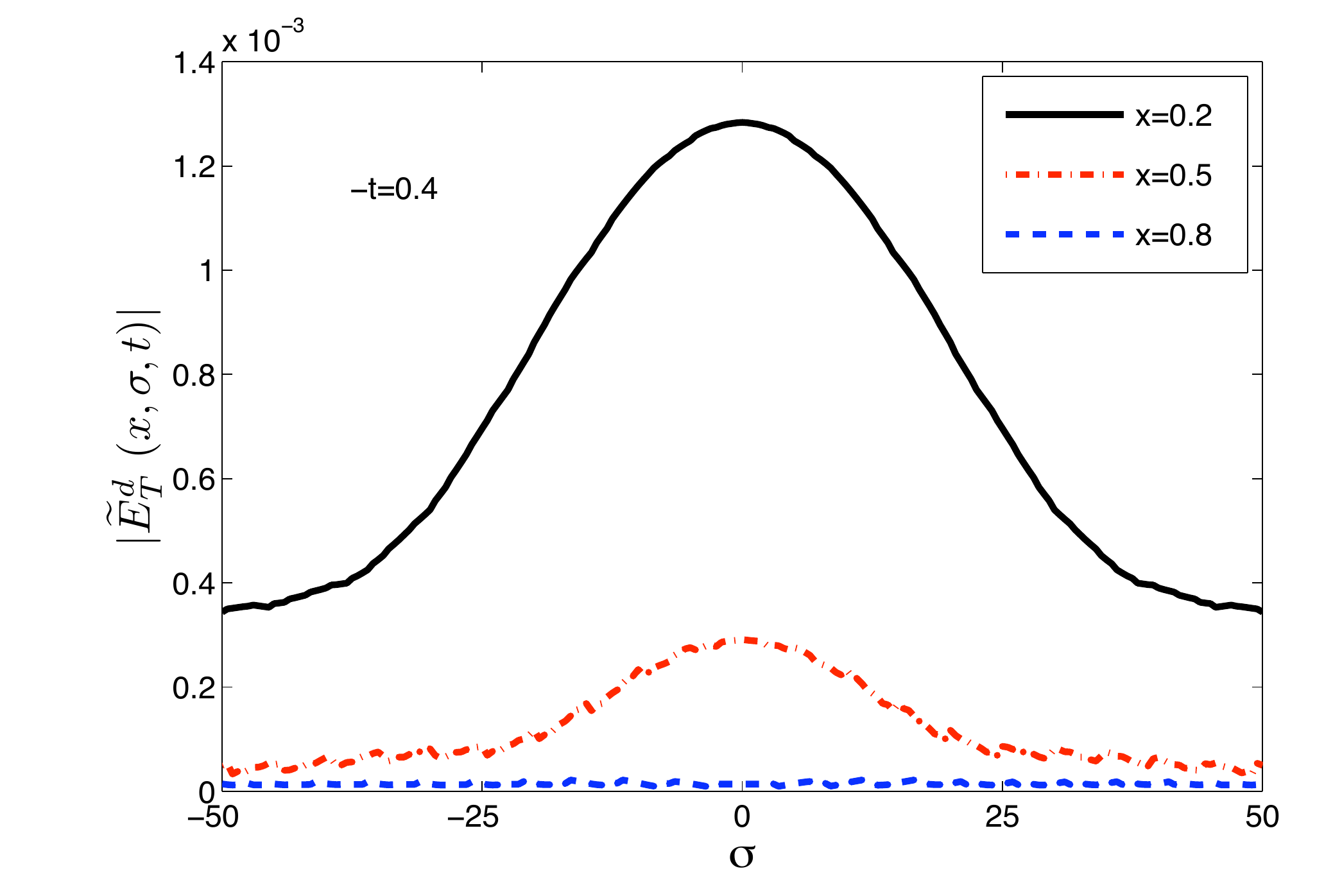}
\end{minipage}
\caption{\label{glt5}(Color online) Plots of the chiral-odd GPDs in longitudinal impact space vs $\sigma$ and different values of $x$, for fixed value of $-t=0.4$ $GeV^2$. Left pannel is for $u$ quark and the right pannel is for $d$ quark.}
\end{figure}

%%%%%%%%%%%%%%%%%%%%%%%%%%%%%%%%%%%%%%%%%%%%%%%%%%%%%%%%%%%%%%%%%%%%%%%
The boost invariant longitudinal impact parameter is defined as $\sigma=\frac{1}{2}b^-P^+$ which is conjugate to $\zeta$, the measure of longitudinal momentum transfer. The parameter $\sigma$ was first introduced in \cite{BDHAV}. The DVCS amplitude in a QED model of a dressed electron  shows an interesting diffraction pattern in the longitudinal impact parameter space analogous to diffractive scattering of a wave in optics~\cite{BDHAV}. 
%The correlation defined in the three dimensional position space $b_\perp$ and $\sigma$ is frame independent as the Lorentz boosts are kinematical in the light front.  
In analogy with optics, the finite size of the $\zeta$ can be interpreted  as a slit of finite width and produces the diffraction pattern.  It should be mentioned here that the FT with a finite range of $\zeta$ of any function does not show the diffraction pattern~\cite{CMM2}. The pattern depends on the behavior of the function. The chiral-odd GPDs calculated in \cite{CMM1} for
a simple relativistic spin half system of an electron dressed  with a photon exhibit the similar diffraction pattern in the longitudinal impact parameter space. A phenomenological model for proton GPDs show the similar diffraction pattern \cite{CMM2}. Similar diffraction patterns are also observed for the chiral-even GPDs  in this light front quark-diquark model~\cite{chandan} as well as  in QED model~\cite{Kumar1}.   
%So, it is very interesting to investigate if the similar pattern is also observed for chiral-odd GPDs in this  quark-diquark model in the longitudinal position space. 
The GPDs in longitudinal position space are defined as:
\be
H_T(x,\sigma,t)&=&{1\over 2 \pi} \int_0^{\zeta_f} d \zeta e^{i\zeta P^+b^-/2}H_T(x,\zeta,t),\nonumber\\
&=&{1\over 2 \pi} \int_0^{\zeta_f} d \zeta e^{i\zeta \sigma}H_T(x,\zeta,t),\\
E_T(x,\sigma,t)&=&{1\over 2 \pi} \int_0^{\zeta_f} d \zeta e^{i\zeta P^+b^-/2}E_T(x,\zeta,t),\nonumber\\
&=&{1\over 2 \pi} \int_0^{\zeta_f} d \zeta e^{i\zeta \sigma}E_T(x,\zeta,t).
\ee
Similarly one can obtain $\widetilde{H}_T(x,\sigma,t)$ and $\widetilde{E}_T(x,\sigma,t)$ as well.
Since we are considering the region $\zeta<x<1$, the upper limit of $\zeta$ integration $\zeta_f$ is given by $x$ if $x$ is smaller than $\zeta_{max}$, otherwise by $\zeta_{max}$ if $x$ is larger than $\zeta_{max}$ where the maximum value of $\zeta$ for a fixed $-t$ is given by
\be
%\zeta_{max}=\frac{-t}{2M_n^2}\bigg(\sqrt{1+\frac{4M_n^2}{(-t)}}-1\bigg).
\zeta_{max}=\sqrt{\frac{(-t)}{(-t+4M_n^2)}}.
\ee
We show the Fourier spectrum of all the chiral-odd GPDs for $u$ and $d$ quarks in longitudinal position space as a function of $\sigma$ for fixed $x=0.3$ and different values of $-t$ in Fig.\ref{glx6}. ${H}_T^q$, ${E}_T^q$ and $\widetilde{H}_T^q$ display a diffraction pattern in the $\sigma$ space as observed for the DVCS amplitude in~\cite{BDHAV},
%Similarly, the Fourier spectrum of $\widetilde{H}_T^q$ and $\widetilde{E}_T^q$ in longitudinal position space are shown in Fig.\ref{gtlx6}. Though $\widetilde{H}_T^q(x,\sigma,t)$ shows a diffraction pattern
but $\widetilde{E}_T^q(x,\sigma,t)$ does not show the same pattern. This is due to the fact that the distinctly different behavior of $\widetilde{E}_T^q(x,\zeta,t)$ with $\zeta$ compared to the other GPDs. This again shows that the diffraction pattern is not solely due to the finiteness of $\zeta$ integration, the functional form of the GPDs are also crucial.  For all the diffraction patterns the first minima appears at the same values of $\sigma$. We also show the chiral-odd GPDs in $\sigma$ space for fixed $-t=0.4$ $GeV^2$ and different values of $x$ in Fig.\ref{glt5}. In analogous to the single slit optical diffraction pattern, here $\zeta_{max}$ plays the role of the slit width.  Since the position of the minima are inversely proportional to the slit width, the minima move towards the center of the diffraction pattern as the slit width $\zeta_{max}$ increases. 
% In all cases, the primary maxima is followed by secondary maxima.
%%%%%%%%%%%%%%%%%%%..zeta dependent..zb..%%%%%%%%
%%%%%%%%%%%%%%%%%%%%%%%%%%%%%%%%%%%%%%%%%%%%%%%%%%%%%%%%%%%%%%%%%%%%%%%

%%%%%%%%%%%%%%%%%%%%%%%%%%%%%%%%%%%%%%%%%%%%%%%%%%%%%%%%%%%%%%%%%%%%%%%
\section{Summary}\label{summary}
%%%%%%%%%%%%%%%%%%%%%%%%%%%%%%%%%%%%%%%%%%%%%%%%%%%%%%%%%%%%%%%%%%%%%%%
we have investigated the chiral-odd GPDs for $u$ and $d$ quark in proton for both zero and nonzero skewness in the light front quark-diquark model predicted by the soft-wall AdS/QCD. We have found that the $\widetilde{E}^q_T(x,0,t)$ is zero in this model due to odd function in nature with $\zeta$. $\widetilde{H}^q_T$ shows opposite sign of ${H}^q_T$ for both $u$ and $d$ quark as expected from $SU(6)$. For zero skewness, all the chiral-odd GPDs for $u$ quark are opposite with respect to $d$ quark .
We have calculated the GPDs for nonzero skewness in the DGLAP region i.e., for ($x>\zeta$).  The peaks of the distributions move to higher values of $x$ for fixed $\zeta$ with increasing of $-t$ similar as the nature of $\zeta=0$. The height of the peaks increases and also shift to higher values of $x$ as $\zeta$ increases for fixed  $-t$. We observed markedly different behavior for $\widetilde{E}^q_T$ from the other chiral-odd GPDs when we plot the GPDs against $\zeta$ for fixed $x$ and different $-t$. It shows that with increasing $\zeta$, $\widetilde{E}^q_T$ started to increase smoothly from zero but other GPDs rise from different values at $\zeta=0$ for different values of $-t$. 

We have also presented all the chiral-odd GPDs in the transverse position or impact parameter($b$) as well as longitudinal position($\sigma$) spaces by taking FT of the GPDs with respect to transverse momentum transfer($\Delta_{\perp}$) and $\zeta$ respectively. The impact parameter $b$ gives a measure of  the transverse distance between the struck parton and the center of momentum of the hadron.
In this model, except $\widetilde{H}^q_T$, the behavior of the GPDs in the transverse impact parameter space for $u$ and $d$ quarks are quite different when plotted in $x$ and $b$. Except the magnitude, the nature of $H^q_T$, $E^q_T$ and $\widetilde{H}^q_T$ are more or less same when plotted against $\zeta$ and $b$ but $\widetilde{E}^q_T$ shows a different behavior. The width of the all distributions increase with increasing $\zeta$ as well as $x$ decreases. We found that the GPDs $H_T$, $E_T$ and $\widetilde{H}_T$ for $u$ and $d$ quarks in $\sigma$ space show diffraction patterns analogous to diffractive scattering of a wave in optics. A similar diffraction pattern also has been  observed  in some other models. The qualitative nature of the diffraction patterns for all three chiral-odd GPDs are same for both $u$ and $d$ quarks.
The general features of this pattern are mainly depends on the finiteness of $\zeta$ integration as well as the dependence of GPDs on $x$, $\zeta$ and $t$. Like other GPDs, $\widetilde{E}_T$ does not show the diffraction pattern. This is due to a different nature of $\widetilde{E}_T$ with $\zeta$ from the other GPDs.  It also indicates that the diffraction pattern is  not solely  due to finiteness of $\zeta$ integration and the functional behaviors of the GPDs are  important to have the diffraction pattern. 
%However, one needs to study the GPDs in $x<\zeta$ domain also in oder to get the full Lorentz invariant picture in the longitudinal position space.

%%%%%%%%%%%%%%%%%%%%%%%%%%%%%%%%%%%%%%%%%%%%%%%%%%%%%%%%%%%%%%%%%%%%%%%
%\section{Appendix}
%%%%%%%%%%%%%%%%%%%%%%%%%%%%%%%%%%%%%%%%%%%%%%%%%%%%%%%%%%%%%%%%%%%%%%%

%%%%%%%%%%%%%%%%%%%%%%%%%%%%%%%%%%%%

\end{document}